\documentclass{jfm}
\usepackage{graphicx}
\usepackage{epstopdf, epsfig}
\usepackage{amsmath}
\usepackage{mathrsfs}
\usepackage{booktabs}
\usepackage{subcaption}
\usepackage[percent]{overpic}

\shorttitle{RD decomposition in compressible rough-wall TBLs}
\shortauthor{Y. T. Su, L. H. Zhao and C. X. Xu}

\title{Extension of the Renard-Deck drag decomposition to compressible rough-wall turbulent boundary layers}

\author{Youtian Su\aff{1},
  Li-Hao Zhao\aff{1}
 \and Chunxiao Xu\aff{1}\corresp{\email{xucx@tsinghua.edu.cn}}}

\affiliation{\aff{1}AML, Department of Engineering Mechanics, Tsinghua University, Beijing 100084, PR China}

\begin{document}

\maketitle

\begin{abstract}
A generalized Renard–Deck (RD) drag decomposition is developed for spatially developing compressible turbulent boundary layers over rough surfaces. Derived from the locally resolved, time-averaged momentum equations, the formulation accounts for both viscous friction and pressure drag without requiring spatial averaging at the roughness-element scale. Eight direct numerical simulations are considered at freestream Mach numbers of 2.5 and 5.86, comparing smooth walls and two-dimensional transverse square ribs (\(\lambda_x/k=8\)) under adiabatic and cold-wall conditions. The decomposition reconstructs the total drag with errors below 5\%. Mean-flow viscous dissipation and turbulent kinetic-energy production together account for approximately 90\% of the drag work. While their contributions are comparable over smooth walls, turbulent production dominates over rough walls, contributing approximately 80\%. More importantly, roughness substantially alters the wall-normal distributions of both contributions. Viscous dissipation becomes concentrated near the roughness crest, whereas turbulent production peaks above the crest, with both contributions decaying more gradually into the outer layer than over smooth walls. The resulting differences in their cumulative contributions persist across a substantial fraction of the boundary layer, demonstrating that the energetic influence of surface roughness extends well beyond the roughness sublayer.
\end{abstract}

%

\section{Introduction}\label{sec:intro}

Rough-wall turbulent flows are commonly encountered in high-speed aerospace applications, where surface roughness can arise from manufacturing tolerances, thermal ablation and thermal protection system designs. Rough-wall turbulence has been extensively investigated, with the roughness function, virtual origin and outer-layer similarity providing a fundamental framework for characterizing roughness effects \citep{nikuradse1933stromungsgesetze,hama1954boundary,townsend1976structure,jimenez2004turbulent,chung2021predicting}. For compressible wall-bounded turbulence, various velocity transformations have been developed to account for density and viscosity variations in mean velocity profiles and improve their similarity under different wall thermal conditions \citep{van1951turbulent,volpiani2020data,griffin2021velocity,hasan2023incorporating}. More recently, experimental and numerical studies have examined roughness functions, outer-layer similarity and turbulence statistics in compressible rough-wall flows, with particular attention to the effects of roughness geometry, Mach number and wall thermal conditions \citep{latin2000flow,tyson2013numerical,modesti2022direct,aghaei2023supersonic,wang2024roughness,cogo2025surface,cogo2025development}. \cite{modesti2022direct} introduced a modified roughness height based on the ratio of wall to roughness-crest kinematic viscosities, yielding a weaker Mach-number dependence of the relationship between the transformed roughness function and the modified roughness height.
\cite{wang2024roughness} introduced a wall-temperature-dependent correction to the velocity transformation within the roughness layer to reconcile the roughness functions under adiabatic and cold-wall conditions, whereas \cite{braga2026influence} quantified their differences using the semi-locally scaled roughness height $k^*$. The temperature--velocity relationship is another important aspect of compressible turbulence, with theoretical developments progressing from the classical Crocco--Busemann and Walz relations to the generalized Reynolds analogy (GRA) \citep{busemann1931handbuch,crocco1932sulla,walz1962compressible,huang1994van,zhang2014generalized}. Since the pressure drag induced by roughness elements has no direct counterpart in heat transfer \citep{dipprey1963heat}, the similarity between momentum and heat transport may be disrupted, and previous studies have reported noticeable deviations from classical temperature--velocity relations near rough walls \citep{modesti2022direct,cogo2025surface,cogo2025development}. \cite{cogo2026reynolds} coupled a rough-wall drag model with a compressibility transformation and a temperature--velocity relation to predict wall shear stress and heat flux. \cite{su2026compressible} proposed a roughness-modified GRA by introducing effective boundary conditions at a reference height above the roughness elements. \cite{braga2026influence} introduced a linear correction to the GRA, with its coefficient determined by the difference in Reynolds analogy factors between smooth and rough walls, improving the prediction of mean temperature--velocity relationships.

To elucidate the physical mechanisms underlying skin-friction drag, several decomposition methods have been developed, including the Fukagata--Iwamoto--Kasagi (FIK) \citep{fukagata2002contribution}, angular-momentum-integral (AMI) \citep{elnahhas2022enhancement} and Renard--Deck (RD) \citep{renard2016theoretical} identities, based on streamwise momentum, angular-momentum and mean kinetic-energy balances, respectively. The RD identity provides an energy-based interpretation of skin-friction generation through mean-flow viscous dissipation, turbulent kinetic-energy production and spatial development effects. Extending this framework to compressible turbulent channel flows, \citet{li2019decomposition} found that the contribution of turbulent production increases with Reynolds number, while the relative contributions exhibit weak Mach-number dependence at comparable transformed friction Reynolds numbers. Subsequently, \citet{fan2019decomposition} investigated zero-pressure-gradient turbulent boundary layers, separating the viscous and turbulent contributions into incompressible and compressible components through compressibility transformations, and found that the relative contribution of spatial development exhibits weak Reynolds-number dependence. \cite{fan2022energy} found that wall cooling reduces near-wall production and shifts its pre-multiplied peak farther from the wall, while \cite{cogo2023assessment} showed that wall-cooling effects become more pronounced at higher Mach numbers.

\citet{bannier2015riblet} applied an extended FIK identity to flows over complex surfaces to investigate turbulent drag reduction mechanisms. Meanwhile, \citet{nikora2013spatially} established the double-averaged Navier--Stokes (DANS) framework through temporal and spatial averaging, introducing dispersive stresses, spatial porosity and fluid--solid interfacial forces to account for the effects of roughness geometry on mean flow and momentum transport. Based on the DANS framework, \citet{nikora2019friction} developed a FIK-type momentum-integral drag decomposition for rough-wall flows. They found that Reynolds stresses make substantial contributions to drag, while dispersive stresses can also play a significant role. Subsequently, \citet{zhang2024integral} reformulated the FIK, AMI and RD integral identities within the DANS framework and, by appropriately selecting the reference-frame velocity, proposed new decompositions that separately quantify bottom-wall skin friction and roughness-element drag. In addition, \citet{guo2022energy} derived an energy-based drag decomposition for incompressible periodic rough-wall channel flows from the kinetic-energy balance. By further employing a triple decomposition of the velocity field, they distinguished the contributions of plane-averaged and spatially heterogeneous motions to viscous dissipation and turbulent kinetic-energy production.

Despite the valuable insights provided by previous studies into drag generation over rough walls, extending existing decomposition approaches to spatially developing compressible rough-wall boundary layers remains challenging. First, double-averaged formulations generally rely on sufficient scale separation among the roughness elements, the spatial averaging region and the outer flow, a condition that may not be satisfied when investigating local flow structures and drag-generation mechanisms at the roughness-element scale. Second, for rough-wall flows with complex geometries, a decomposition based solely on the streamwise momentum equation is insufficient to fully characterize the connection between drag and flow energy, motivating the incorporation of the kinetic energy equation to establish a decomposition in terms of energy transport and dissipation. Furthermore, density variations, thermodynamic coupling and streamwise inhomogeneity in compressible rough-wall boundary layers introduce additional compressibility-related and spatial-development terms. A consistent mathematical treatment and clear physical interpretation of these contributions are therefore essential for understanding the underlying drag-generation mechanisms. Such a framework would enable systematic investigations of how Mach number, wall temperature and surface roughness influence the different contributions to drag power, thereby providing further insight into energy transport and drag generation in spatially developing compressible rough-wall boundary layers. 

The present study extends the Renard–Deck (RD) drag decomposition to spatially developing compressible turbulent boundary layers over rough surfaces. Starting from the time-averaged momentum equations, a generalized energy-based decomposition is derived by retaining the full three-dimensional stress and momentum-flux contributions, without requiring spatial averaging at the roughness-element scale. The resulting formulation relates the total surface drag, including both viscous friction and pressure drag, to mean-flow viscous dissipation, turbulent kinetic-energy production, convective transport and additional contributions associated with compressibility and spatial development. To investigate these mechanisms, eight direct numerical simulations are considered at freestream Mach numbers of \(Ma_\infty=2.5\) and \(5.86\), covering smooth and rough walls under adiabatic and cold-wall conditions. The roughness consists of two-dimensional transverse square ribs with a streamwise pitch-to-height ratio of \(\lambda_x/k=8\). The paper is organized as follows. Section 2 describes the computational methodology and statistical framework. Section 3 first examines the mean flow characteristics, temperature–velocity relationships and turbulence statistics, followed by the derivation and application of the generalized RD decomposition and a regional analysis of surface drag. Finally, the main conclusions are summarized in Section 4.

\section{Methodology}\label{sec:methodology}

The present study considers eight direct numerical simulation (DNS) cases of spatially developing compressible turbulent boundary layers, covering two free-stream Mach numbers ($Ma_\infty=2.5$ and $5.86$), two surface conditions (smooth and rough walls) and two wall thermal conditions (adiabatic and cooled walls). Four cases at $Ma_\infty=2.5$ are taken from our previous study \cite{su2026compressible}, while four additional cases at $Ma_\infty=5.86$ are introduced to extend the investigation of Mach-number effects on rough-wall turbulence.

\begin{figure}
    \centering
    \vspace{0.6cm}
    \includegraphics[width=0.85\textwidth]{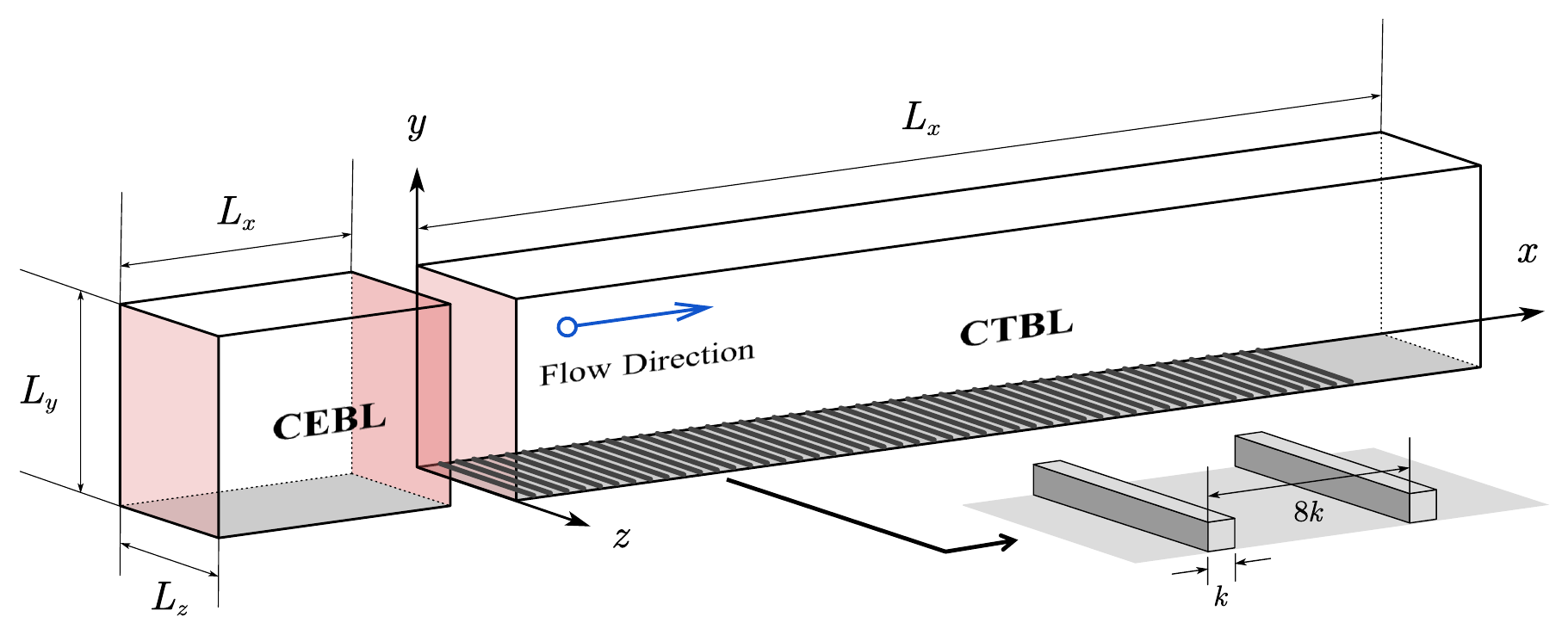}
    \vspace{0.6cm}

    \caption{Schematic of the computational domain configurations, comprising the upstream compressible equivalent boundary layer (CEBL) domain and the downstream main compressible turbulent boundary layer (CTBL) domain. The inset in the lower-right corner illustrates the geometric configuration of the roughness element.}
    \label{fig:computational_setup}
\end{figure}

The computational setup consists of an auxiliary compressible equivalent boundary layer (CEBL) domain and a main compressible turbulent boundary layer (CTBL) domain, as illustrated in figure \ref{fig:computational_setup}. Fully developed turbulent inflow conditions are generated using the CEBL method proposed by \cite{li2024inflow}. This method employs streamwise periodic boundary conditions and appropriate source terms to reproduce the mean mass, momentum and energy balances of a spatially developing boundary layer. The resulting turbulent fluctuations are subsequently imposed at the inlet of the main CTBL domain, where the three-dimensional compressible Navier--Stokes equations are solved. 

\begin{equation}
    \frac{\partial \rho}{\partial t} + \frac{\partial \rho u_j}{\partial x_j} = -M,
\end{equation}
\begin{equation}
    \frac{\partial \rho u_i}{\partial t} + \frac{\partial \rho u_i u_j}{\partial x_j} + \frac{\partial p}{\partial x_i} - \frac{\partial \sigma_{ij}}{\partial x_j} = -F \delta_{i1},
\end{equation}
\begin{equation}
    \frac{\partial \rho E}{\partial t} + \frac{\partial \rho H u_j}{\partial x_j} - \frac{\partial (\sigma_{ij}u_i - q_j)}{\partial x_j} = -Q
\end{equation}
where \mbox{$\rho$} is the fluid density, \mbox{$u_i$} is the velocity vector component, \mbox{$p$} is the pressure, and \mbox{$E$} and \mbox{$H$} represent the total energy and total enthalpy, respectively. $M$, $F$ and $Q$ denote the source terms introduced by the CEBL method. In the CEBL domain, their values are prescribed according to the target inlet Reynolds number, whereas they are identically zero in the CTBL domain. For a Newtonian fluid under Stokes' hypothesis, the viscous stress tensor \mbox{$\sigma_{ij}$} is given by
\begin{equation}
    \sigma_{ij} = \mu \left( \frac{\partial u_i}{\partial x_j} + \frac{\partial u_j}{\partial x_i} \right) - \frac{2}{3} \mu \frac{\partial u_k}{\partial x_k} \delta_{ij},
\end{equation}
where \mbox{$\mu$} is the dynamic viscosity, and \mbox{$\delta_{ij}$} is the Kronecker delta. And \mbox{$q_j$} is the heat flux vector based on Fourier's law. Throughout the paper, the Einstein summation convention over repeated indices is adopted. No-slip boundary conditions are imposed at the solid surfaces, while periodic boundary conditions are applied in the spanwise direction. Non-reflecting and supersonic outflow boundary conditions are prescribed at the upper and downstream boundaries, respectively.

The flow is assumed to be a calorically perfect gas with a constant specific-heat ratio of $\gamma=1.4$. The free-stream reference temperature is fixed at $T_\infty=270\,\mathrm{K}$, and the thermodynamic properties are determined using the ideal-gas equation of state. The dynamic viscosity is evaluated using Sutherland's law, with a constant molecular Prandtl number of $Pr=0.71$. Two wall thermal conditions are considered, including an adiabatic wall and an isothermal cooled wall with $T_w/T_r=0.5$, where $T_r$ denotes the recovery temperature. 

The inlet friction Reynolds number of the equivalent turbulent boundary layer (ETBL) is set to $Re_\tau=300$ for all cases at $Ma_\infty=2.5$. For the $Ma_\infty=5.86$ cases, the inlet values are $Re_\tau=325$ and $350$ for the adiabatic smooth- and rough-wall cases, respectively, while $Re_\tau=300$ is prescribed for both cooled-wall cases. The target friction Reynolds numbers at the analysis stations are summarized in table~\ref{tab:flow_parameters_split}. Here, the total drag stress $\tau_w$ used to define $Re_\tau$ accounts for both viscous friction and pressure drag.

\begin{table}
    \begin{center}
    \def~{\hphantom{0}}
    \setlength{\tabcolsep}{1.9pt}
    \begin{tabular}{lcccc c lcccc}

        \multicolumn{5}{c}{Present Simulations} && \multicolumn{5}{c}{Reference Literature} \\[3pt]

        Case & Surf. & $M_\infty$ & $T_w/T_r$ & $Re_{\tau}$ &\, \, \, \, \,& Case & Surf. & $M_\infty$ & $T_w/T_r$ & $Re_{\tau}$ \\[3pt]

        M2p5SA  & S & 2.5  & 1.0 & ~510 && \cite{zhang2018direct}         & S & 2.5  & 1.0 & ~510 \\
        M2p5SC  & S & 2.5  & 0.5 & ~510 && \cite{volpiani2018effects}     & S & 2.28 & 0.5 & ~511 \\
        M2p5RA  & R & 2.5  & 1.0 & 1050 && \cite{pirozzoli2011turbulence} & S & 2.0  & 1.0 & 1000 \\
        M2p5RC  & R & 2.5  & 0.5 & 1050 && \cite{schlatter2010assessment} & S & 0    & --  & 1047 \\[3pt]

        M5p86SA & S & 5.86 & 1.0 & ~450 && \cite{eitel2014simulation}     & S & 0    & --  & 2478 \\
        M5p86SC & S & 5.86 & 0.5 & ~450 && \cite{cogo2023assessment}\_A   & S & 6.0  & 1.0 & ~443 \\
        M5p86RA & R & 5.86 & 1.0 & ~960 && \cite{cogo2023assessment}\_C   & S & 6.0  & 0.57& ~443 \\
        M5p86RC & R & 5.86 & 0.5 & ~960 &&                                &   &      &     &      \\

    \end{tabular}
    \caption{Summary of the flow parameters for the present DNS cases (left) and the reference datasets used for validation and comparison (right). In the case labels, `S' and `R' identify smooth- and rough-wall configurations, respectively, while `A' and `C' distinguish adiabatic ($T_w/T_r=1.0$) and cooled ($T_w/T_r=0.5$) wall conditions. The same surface classifications are used in the `Surf.' column. For the present simulations, $Re_\tau$ denotes the target friction Reynolds number.}

    \label{tab:flow_parameters_split}
    \end{center}
\end{table}
The roughness consists of two-dimensional spanwise-aligned square bars with a fixed streamwise pitch-to-height ratio of $\lambda_x/k=8$, following the configurations considered by \citep{lee2007direct,lee2011direct}. The roughness height is determined using the inner scaling of the corresponding smooth-wall reference cases. Specifically, the wall-normal location corresponding to $y^+=24.3$ is identified from the smooth-wall statistics at the target friction Reynolds number and subsequently adopted as the roughness height $k$. For $Ma_\infty=2.5$, this procedure yields $k=0.0859\delta_{in}$ and $0.0803\delta_{in}$ for the adiabatic and cooled-wall cases, respectively. Applying the same criterion to the $Ma_\infty=5.86$ cases gives $k=0.0804\delta_{in}$ and $0.0856\delta_{in}$ for the corresponding thermal conditions. For all rough-wall cases, 116 roughness elements are arranged periodically in the streamwise direction.

The dimensions of the computational domains are determined following the procedure adopted in our previous study. The auxiliary CEBL domain has a fixed streamwise extent of $L_x=10\delta_{in}$, while its wall-normal and spanwise dimensions are identical to those of the corresponding main CTBL domain. The latter is divided into an effective computational region of length $L_e$ and a downstream buffer region of length $L_b=30\delta_{in}$. The buffer region, where a smooth bottom wall is imposed for all cases, is introduced to attenuate outgoing disturbances and minimize spurious numerical reflections. To accommodate the thicker boundary layers and larger turbulent structures associated with surface roughness, the spanwise domain width is increased from $L_z=5.9\delta_{in}$ for smooth-wall cases to $9.0\delta_{in}$ for rough-wall cases. The same domain-design strategy is applied to both Mach numbers, with the specific domain dimensions summarized in table~\ref{tab:computational_details}.

\begin{table}
    \begin{center}
    \def~{\hphantom{0}}
    \setlength{\tabcolsep}{4.0pt} 
    \begin{tabular}{l c c c c c}

        Case & $L_x \times L_y \times L_z \, / \delta_{in}^3$ & $N_x \times N_y \times N_z$  & $k/\delta_{in}$ & $x_t/\delta_{in}$ & $\delta/\delta_{in}$ \\[5pt]
        
        M2p5SA  & $~95 \times 12.5 \times 5.9$  & $2270 \times 300 \times 310$ & --     & 59.0 & 1.84 \\
        M2p5SC  & $~95 \times 12.5 \times 5.9$  & $2270 \times 300 \times 310$ & --     & 40.2 & 1.71 \\
        M2p5RA  & $109 \times 12.5 \times 9.0$ & $6980 \times 300 \times 473$ & 0.0859 & 48.5 & 2.49 \\
        M2p5RC  & $103 \times 12.5 \times 9.0$ & $6980 \times 300 \times 473$ & 0.0803 & 40.2 & 2.41 \\[3pt]
        
        M5p86SA & $~95 \times 12.5 \times 5.9$  & $2270 \times 360 \times 310$ & --     & 55.7 & 1.57 \\
        M5p86SC & $~95 \times 12.5 \times 5.9$  & $2270 \times 360 \times 310$ & --     & 46.7 & 1.60 \\
        M5p86RA & $104 \times 12.5 \times 9.0$ & $6980 \times 360 \times 473$ & 0.0804 & 41.5 & 2.04 \\
        M5p86RC & $109 \times 12.5 \times 9.0$ & $6980 \times 360 \times 473$ & 0.0856 & 70.3 & 2.67 \\
   
    \end{tabular}
    \caption{Computational domain dimensions, grid resolutions, and \textit{a posteriori} boundary-layer parameters for the simulated cases. The variables $L_{x,y,z}$ and $N_{x,y,z}$ denote the physical domain sizes and grid point counts in the streamwise, wall-normal, and spanwise directions, respectively. Here, $x_t$ defines the streamwise station where the target friction Reynolds number is evaluated, and $\delta$ corresponds to the local boundary-layer thickness at $x_t$.}
    \label{tab:computational_details}
    \end{center}
\end{table}

Consistent with our previous methodology \citep{su2026compressible}, the complex solid boundaries are resolved using a sharp-interface immersed boundary method (IBM). The physical dimensions of the computational domains, global grid sizes, and the specific locations of the target analysis stations are summarized in table \ref{tab:computational_details}. To ensure the high-fidelity resolution of the roughness elements and near-wall turbulence, the specific grid distribution strategy, its physical rationale, and the \textit{a posteriori} validation of the computational resolution are detailed in Appendix \ref{app:grid_resolution}. All flow statistics are obtained by averaging over a time interval of $342\delta_/U_\infty$ to ensure statistical convergence.

\subsection{Statistical framework and double-averaging method}

The statistical treatment follows a double-averaging procedure to account for the spatial heterogeneity introduced by the two-dimensional roughness. Reynolds and Favre averages are first introduced to describe the turbulent statistics of the compressible flow. For an instantaneous quantity $\phi(x,y,z,t)$, the Reynolds decomposition is written as
\begin{equation}
\phi(x,y,z,t)=\overline{\phi}(x,y)+\phi'(x,y,z,t)
\end{equation}
where $\overline{\phi}(x,y)$ denotes the Reynolds-averaged quantity, additionally averaged along the homogeneous spanwise direction, and $\phi'$ is the corresponding turbulent fluctuation. The Favre decomposition is defined as
\begin{equation}
\phi(x,y,z,t)=\tilde{\phi}(x,y)+\phi''(x,y,z,t)
\end{equation}
where $\tilde{\phi}=\overline{\rho\phi}/\overline{\rho}$ denotes the Favre mean and $\phi''$ the associated Favre fluctuation.

For the rough-wall cases, the time-averaged quantities retain a periodic streamwise variation associated with the roughness geometry. In particular, the separation and recirculation between successive roughness elements produce spatial variations that remain after temporal and spanwise averaging. These roughness-induced variations are separated from the spatially averaged field by introducing an intrinsic spatial average over each roughness period. A phase indicator function $\Gamma(x,y)$ is used to exclude the solid region from the averaging operation. Owing to the two-dimensional roughness geometry, $\Gamma$ is independent of $z$, with $\Gamma=1$ in the fluid and $\Gamma=0$ inside the solid. The intrinsic spatial average of a time-averaged quantity is then defined as
\begin{equation}
\langle \overline{\phi }\rangle (x_n+\frac{\lambda _x}{2},y)=\frac{\int_{x_n}^{x_n+\lambda _x}{\overline{\phi }(x,y)\Gamma (x,y)dx}}{\int_{x_n}^{x_n+\lambda _x}{\Gamma (x,y)dx}}
\label{eq:spatial_average}
\end{equation}
where $x_n$ denotes the streamwise location of the beginning of the $n$th roughness unit and $\lambda_x$ is the streamwise roughness period. The resulting profile is assigned to the centre of the corresponding roughness period, $x_n+\lambda_x/2$.

With this definition, the instantaneous quantity can be expressed through the double-averaging, or triple-decomposition, framework as
\begin{equation}
\phi(x,y,z,t)=\langle \overline{\phi} \rangle(x_n+\frac{\lambda _x}{2},y)+\overline{\phi}^d(x,y)+\phi'(x,y,z,t)
\end{equation}
where
\begin{equation}
\overline{\phi}^d(x,y)=\overline{\phi}(x,y)-\langle \overline{\phi} \rangle(x_n+\lambda_x/2,y)
\end{equation}
is the spatial fluctuation, or dispersive component. It represents the stationary spatial variation of the mean field induced by the roughness elements and is therefore distinct from the turbulent fluctuation $\phi'$.

For the analysis of mean-flow quantities in the Results and discussion section, the wall-normal profiles are spatially averaged over the corresponding roughness period according to equation \eqref{eq:spatial_average}. For compactness, the operator $\langle\cdot\rangle$ is omitted in the following discussion, such that quantities written as $\overline{\phi}$ should be interpreted as the corresponding spatially averaged values $\langle\overline{\phi}\rangle(x_n+\lambda_x/2,y)$. In section \ref{sec:RD_decomposition}, the drag decomposition is instead performed directly on the locally resolved mean fields within each roughness period, so that the streamwise variations associated with the intra-period flow structures are retained. The resulting decomposition therefore preserves the detailed spatial organization of the individual drag contributions within a roughness unit.

\section{Results and discussion}\label{sec:results_discussion}
\subsection{Overall flow characterization} \label{sec:flow_characterization}
\begin{figure}
    \centering

    \begin{minipage}{\textwidth}
        \raggedright
        (\textit{a})\\[-0.15em]
        \includegraphics[width=\textwidth]{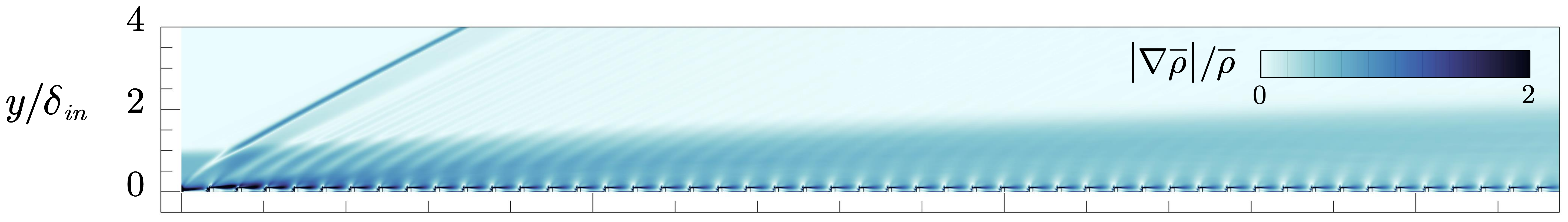}
    \end{minipage}

    \par\vspace{0.3cm}

    \begin{minipage}{\textwidth}
        \raggedright
        (\textit{b})\\[-0.15em]
        \includegraphics[width=\textwidth]{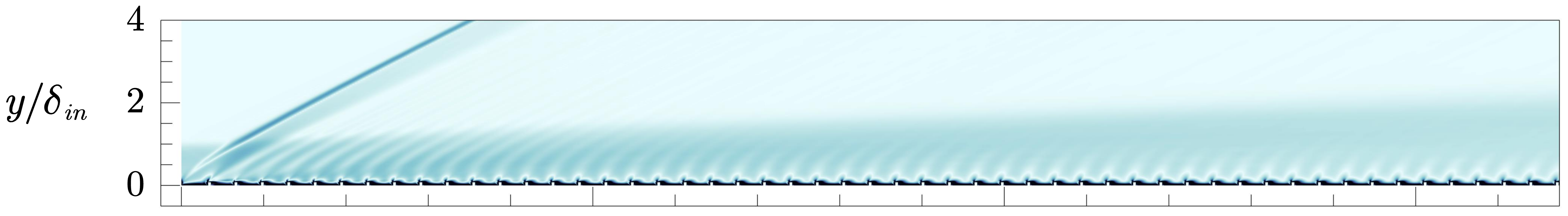}
    \end{minipage}

    \par\vspace{0.3cm}

    \begin{minipage}{\textwidth}
        \raggedright
        (\textit{c})\\[-0.15em]
        \includegraphics[width=\textwidth]{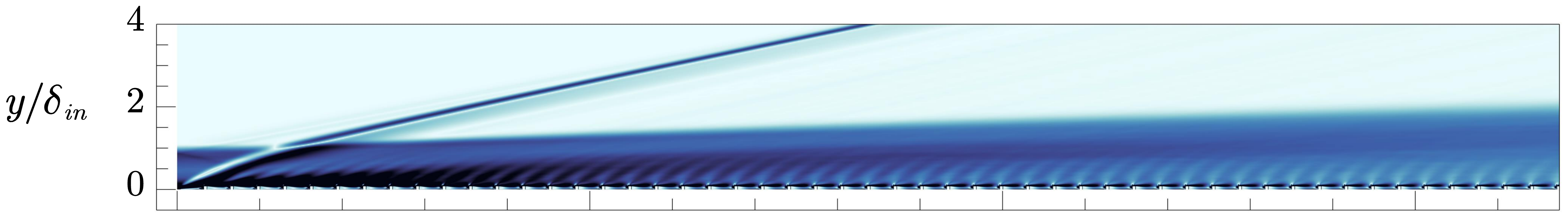}
    \end{minipage}

    \par\vspace{0.3cm}

    \begin{minipage}{\textwidth}
        \raggedright
        (\textit{d})\\[-0.15em]
        \includegraphics[width=\textwidth]{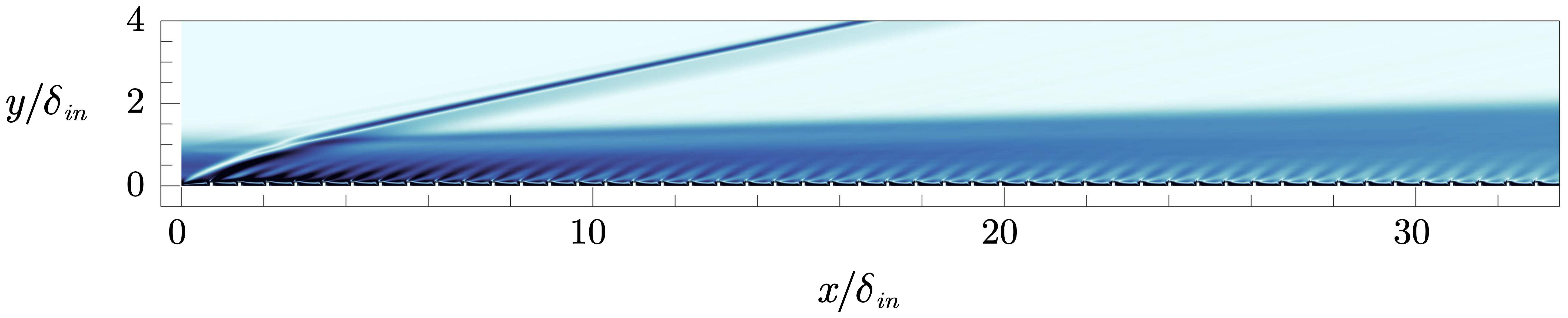}
    \end{minipage}

    \caption{Contours of the normalized density-gradient magnitude,
    $\left|\nabla\bar{\rho}\right|/\bar{\rho}$,
    in the rough-wall turbulent boundary layers for
    (a) M2p5RA, (b) M2p5RC, (c) M5p86RA and (d) M5p86RC.}
    \label{fig:schlieren_contours}
\end{figure}

The interaction between the incoming turbulent boundary layer and the rough surface generates a series of compression waves and oblique shocks, as illustrated in figure~\ref{fig:schlieren_contours}. Upon encountering the first roughness element, the flow undergoes a pronounced compression, producing a strong oblique shock. Further downstream, successive roughness elements induce additional shocks as the boundary layer progressively adjusts to the roughness geometry. The shock structures exhibit distinct characteristics at different Mach numbers. At $Ma_\infty=5.86$, the shocks are generally stronger than those at $Ma_\infty=2.5$, yet the shocks generated in the downstream adjustment region do not extend appreciably beyond the boundary-layer edge. In contrast, at $Ma_\infty=2.5$, some of the roughness-induced shocks penetrate the boundary layer and propagate into the free stream. This difference can be partly attributed to the decrease in Mach angle with increasing Mach number, which causes the shock waves to propagate at a smaller inclination to the streamwise direction and thus requires a longer streamwise distance to reach the boundary-layer edge. Despite these differences, all cases eventually reach a fully developed rough-wall flow state after an initial adjustment region, characterized by a spatially periodic pattern of roughness-induced flow structures.

\begin{figure}
    \centering
    \begin{minipage}[t]{0.49\textwidth}
        \raggedright 
        (\textit{a}) \\[0.48em] 
        \includegraphics[width=\textwidth]{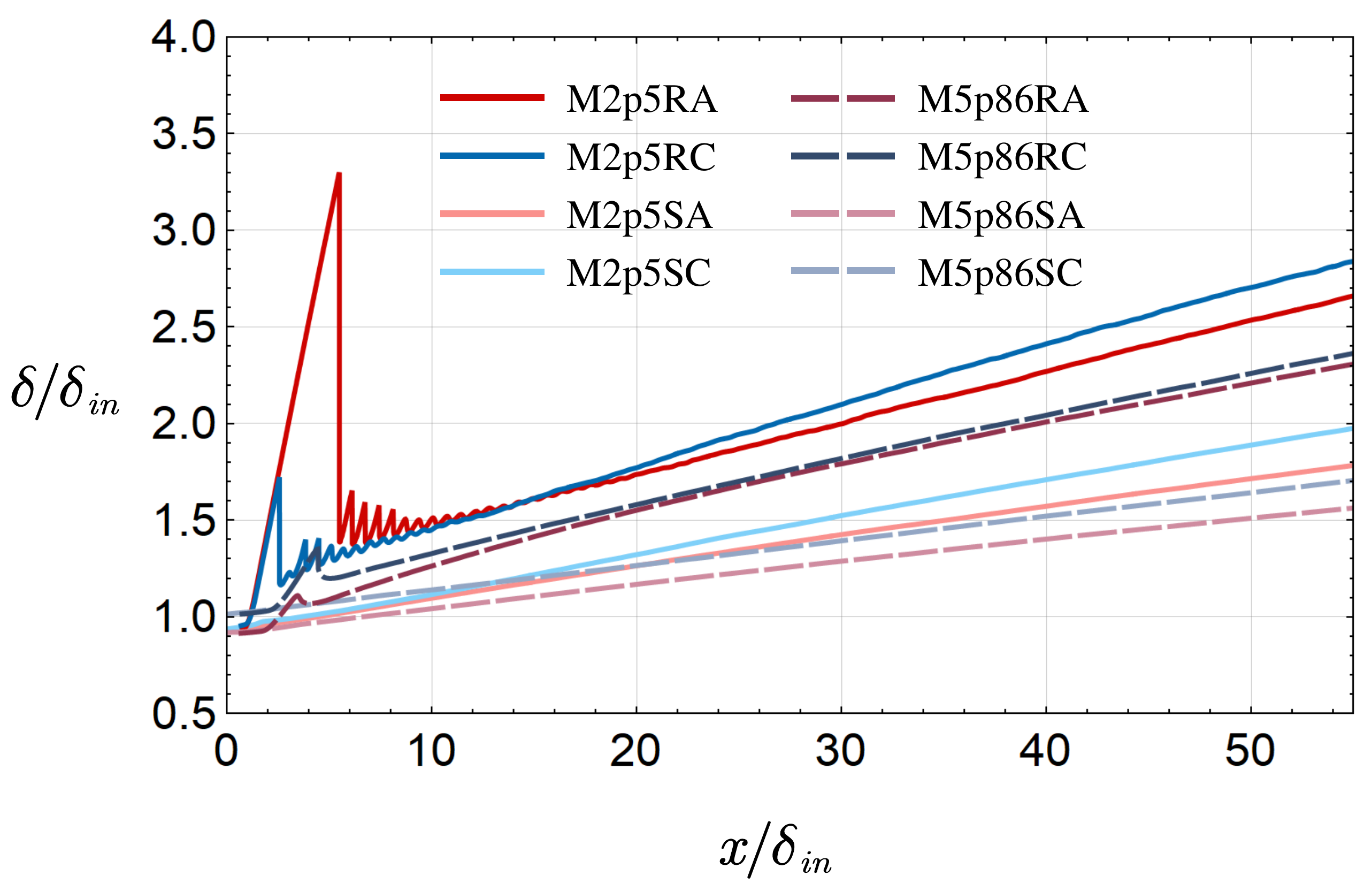}
        \label{fig:x_delta}
    \end{minipage}
    \hfill 
    \begin{minipage}[t]{0.493\textwidth}
        \raggedright 
        (\textit{b}) \\[0.67em] 
        \includegraphics[width=\textwidth]{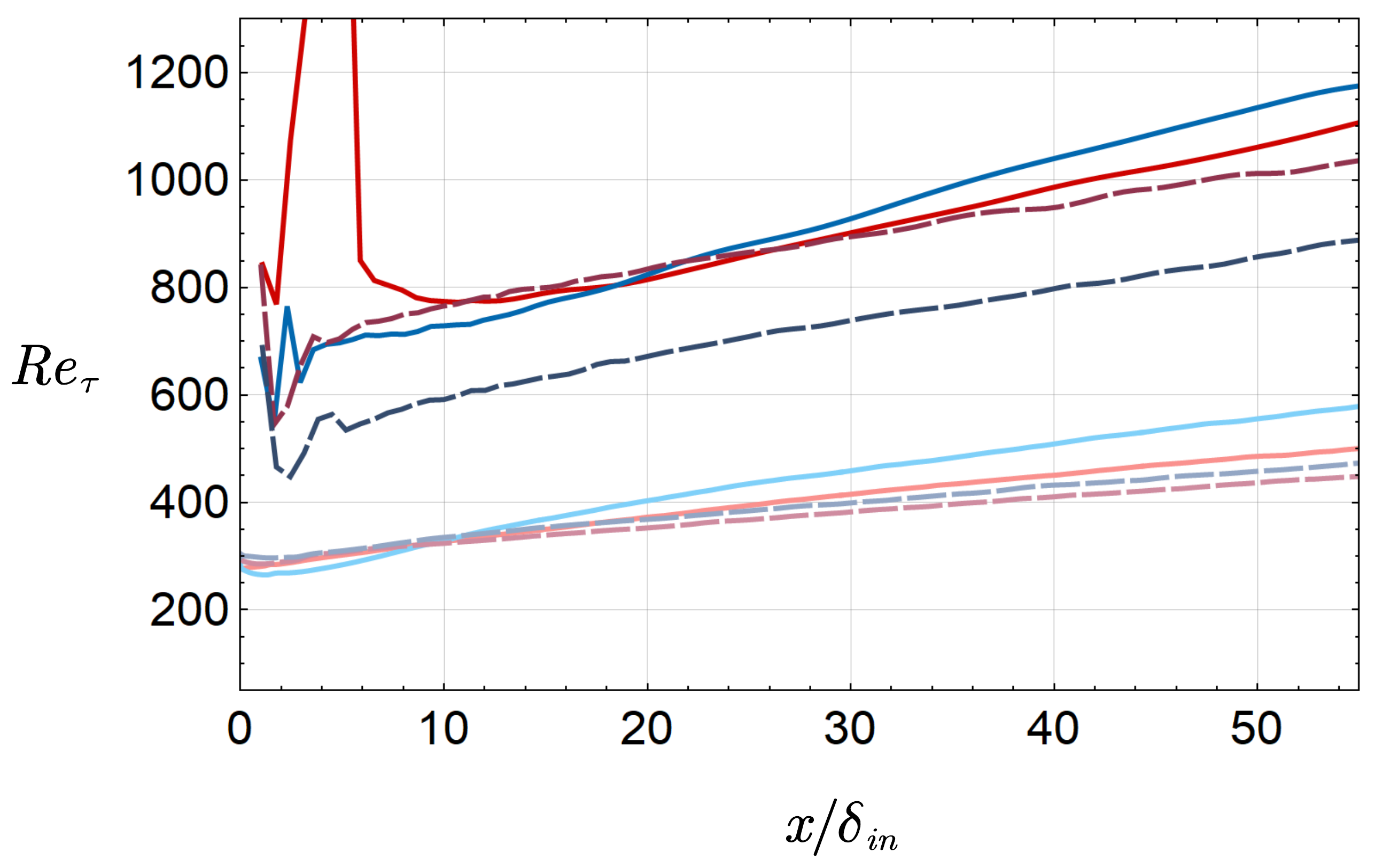} 
        \label{fig:x_ReTau}
    \end{minipage}
    
    \caption{
        Streamwise evolution of (a) boundary layer thickness $\delta$ and (b) friction Reynolds number $Re_\tau$.}
    \label{fig:x_BDL}
\end{figure}

This behavior is further corroborated by the streamwise evolution of the boundary-layer thickness and friction Reynolds number shown in figure~\ref{fig:x_BDL}. As shown in figure~\ref{fig:x_BDL}(a), pronounced fluctuations in the boundary-layer thickness can be detected over the streamwise interval $0<x/\delta_{in}<20$ for the $Ma_\infty=2.5$ rough-wall cases. In contrast, for the $Ma_\infty=5.86$ rough-wall cases, only the strong disturbance induced by the first roughness element produces a noticeable response at the boundary-layer edge. Regarding the streamwise growth of the boundary layer, the cooled-wall cases generally exhibit a higher growth rate than their adiabatic counterparts, and this difference is more pronounced at the lower Mach number. The introduction of roughness leads to a thicker boundary layer. In addition, the drag enhancement associated with the rough surface results in a higher friction Reynolds number $Re_\tau$.

\subsection{Mean velocity profiles}\label{sec:Mean_velocity_profiles}

For the rough-wall cases, the zero-plane displacement and roughness function are determined following the procedure established in our previous study \cite{su2026compressible}. Briefly, the zero-plane displacement is obtained through a kinematic optimization that maximizes the extent of the continuous logarithmic region in which the diagnostic function deviates by no more than 10\% from the theoretical value of $1/\kappa$. Once the optimal displacement is identified, the logarithmic intercept is evaluated by averaging over this region. The roughness function $\Delta U^+$ is then calculated as the difference between the prescribed smooth-wall intercept $B$ and the fitted rough-wall intercept. The resulting values are summarized in table \ref{tab:zero_plane}. The zero-plane displacement relative to the roughness height increases with increasing Mach number and decreasing wall temperature. This trend is consistently observed for both the kinematic optimization based on the VD and GFM transformations and the zero-moment method. Notably, the optimal zero-plane displacement obtained from the kinematic optimization exhibits a particularly pronounced sensitivity to increasing Mach number, as evidenced by the cold-wall cases, where $d^*/k^*$ increases from less than half of $d_m/k$ in case M2p5RC to a value comparable to $d_m/k$ in case M5p86RC.

\begin{table}
    \centering
    \setlength{\tabcolsep}{4pt}
    \begin{tabular}{lccccccccc}
        Case & $d^+$ & $d^*$ & $k^+$ & $k^*$ &
        $d^+/k^+$ & $d^*/k^*$ & $d_m/k$ &
        $\Delta U_{\mathrm{VD}}^+$ & $\Delta U_{\mathrm{GFM}}^+$ \\[3pt]
        M2p5RA  & 8.4   & 5.7   & 36.1 & 36.7 & 0.23 & 0.15 & 0.60 & 10.6 & 10.8 \\
        M2p5RC  & 10.5  & 5.7   & 34.8 & 27.0 & 0.30 & 0.21 & 0.64 & 9.9  & 9.5  \\
        M5p86RA & 17.15 & 12.1  & 37.7 & 39.1 & 0.45 & 0.31 & 0.62 & 10.6 & 10.6 \\
        M5p86RC & 20.28 & 13.2  & 30.8 & 25.4 & 0.65 & 0.52 & 0.68 & 8.82 & 8.18 \\
    \end{tabular}
    \caption{Summary of the roughness parameters obtained for the different Mach numbers and thermal boundary conditions at the target location $x_t$. The zero-plane displacements $d^+$ and $d^*$, together with the roughness heights $k^+$ and $k^*$, are normalized by the inner and semi-local inner scales, respectively. The quantity $d_m/k$ gives the displacement estimated from the zero-moment method, while $\Delta U_{\mathrm{VD}}^+$ and $\Delta U_{\mathrm{GFM}}^+$ denote the roughness functions evaluated using the corresponding velocity transformations.}
    \label{tab:zero_plane}
\end{table}

\begin{figure}
    \centering

    \begin{minipage}[t]{0.49\textwidth}
        \raggedright
        (\textit{a}) \\[-0.15em]
        \includegraphics[width=\textwidth]{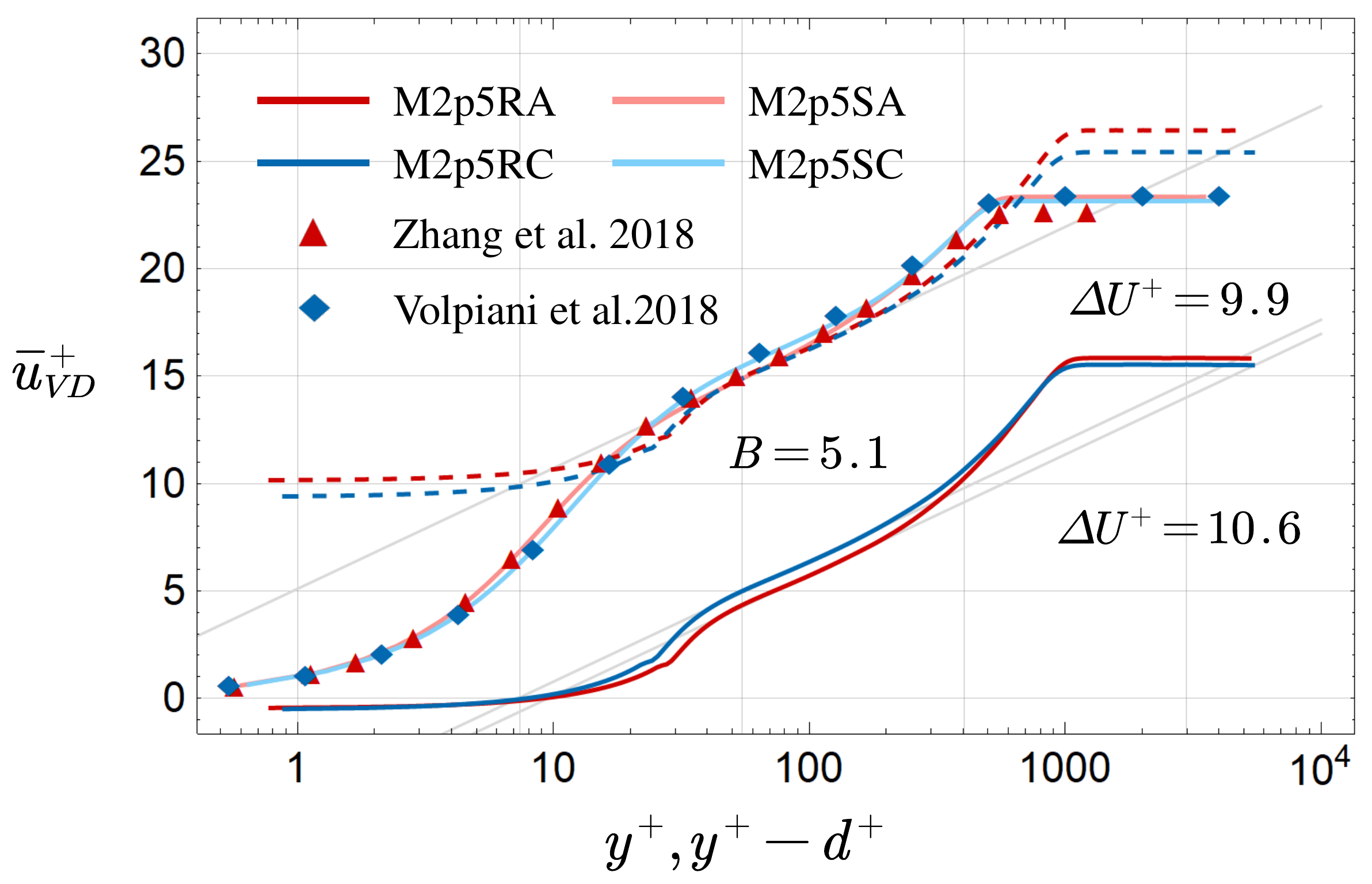}
    \end{minipage}
    \hfill
    \begin{minipage}[t]{0.495\textwidth}
        \raggedright
        (\textit{b}) \\[-0.15em]
        \includegraphics[width=\textwidth]{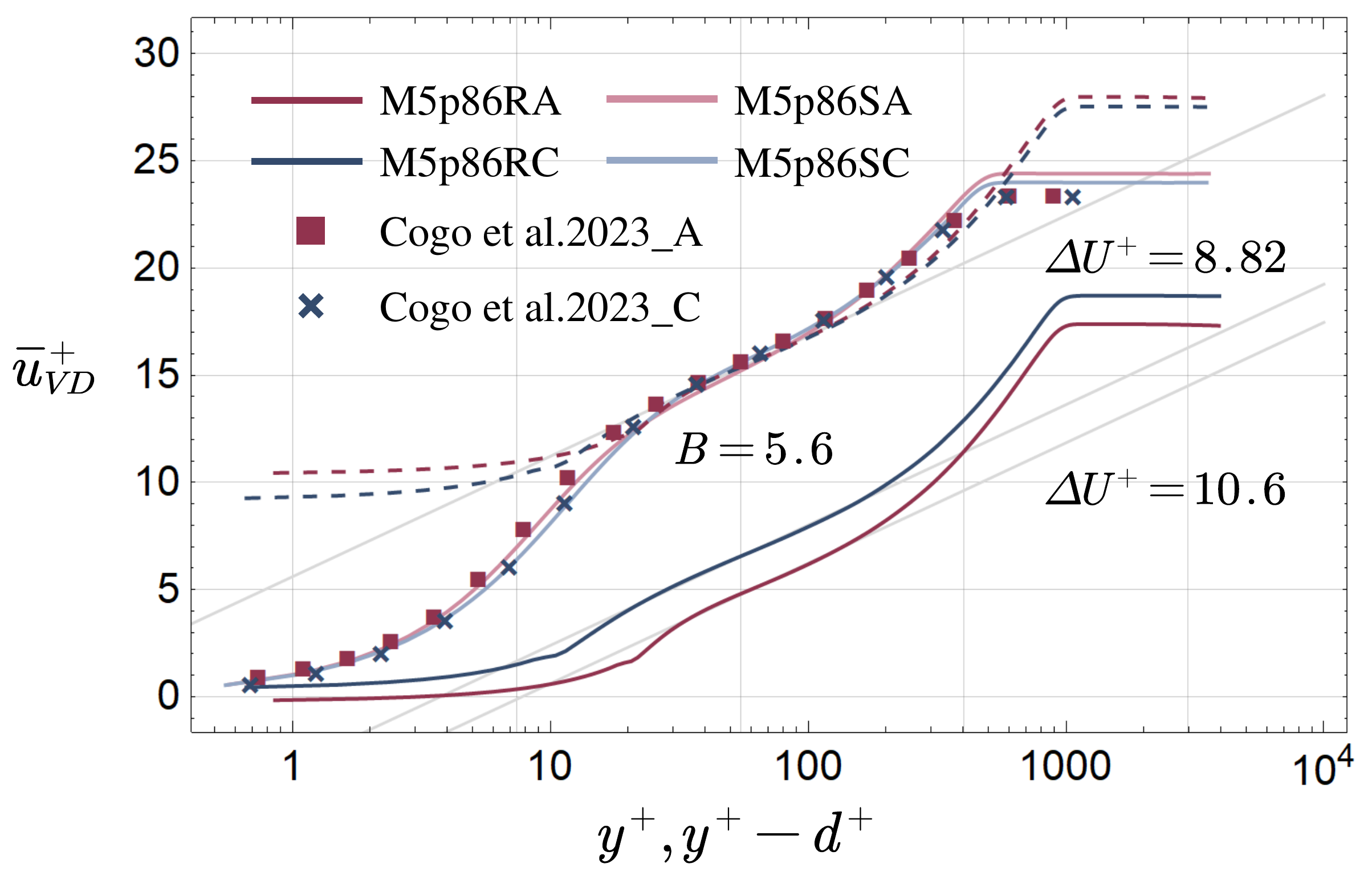}
    \end{minipage}

    \vspace{0.25cm}

    \begin{minipage}[t]{0.49\textwidth}
        \raggedright
        (\textit{c}) \\[-0.15em]
        \includegraphics[width=\textwidth]{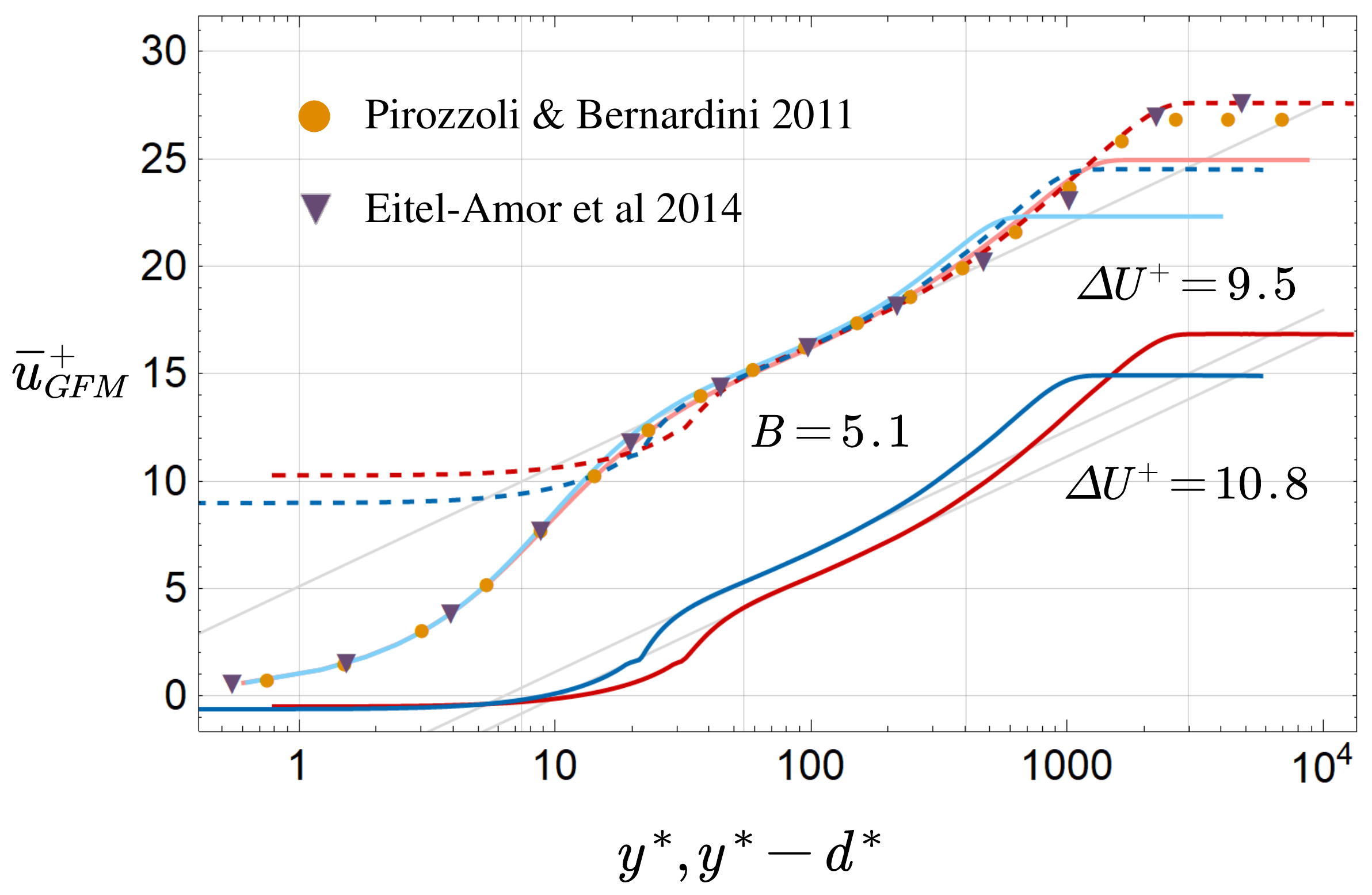}
    \end{minipage}
    \hfill
    \begin{minipage}[t]{0.495\textwidth}
        \raggedright
        (\textit{d}) \\[-0.15em]
        \includegraphics[width=\textwidth]{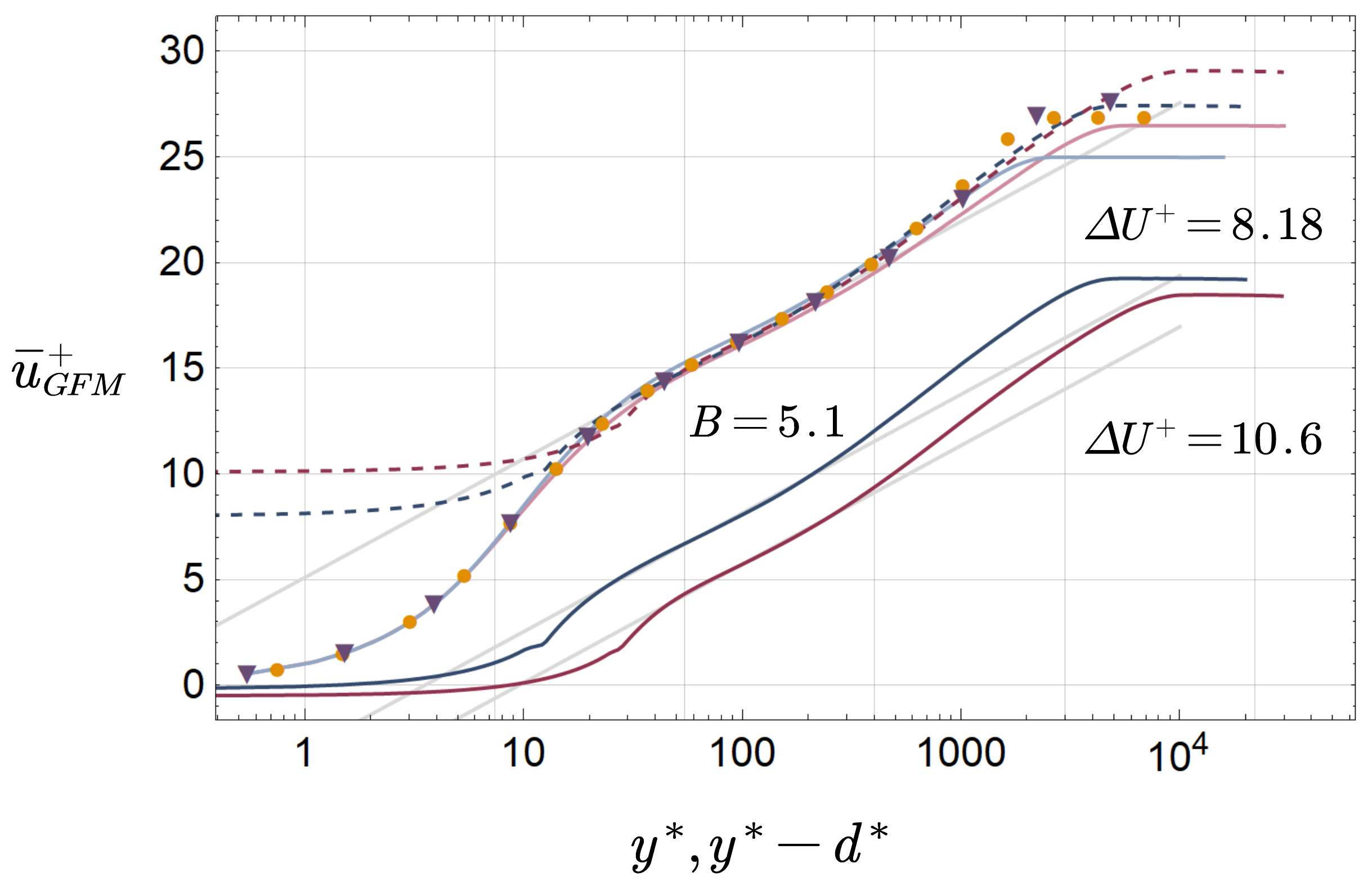}
    \end{minipage}

    \caption{Transformed mean streamwise velocity profiles: (a) van Driest (VD) transformation at $Ma=2.5$; (b) VD transformation at $Ma=5.86$; (c) Griffin--Fu--Moin (GFM) transformation at $Ma=2.5$; (d) GFM transformation at $Ma=5.86$. Gray lines: logarithmic law with $\kappa=0.41$ and  intercept $B$. The roughness function $\Delta U^+$ indicates the downward shift relative to the corresponding smooth-wall profile. Dashed lines of corresponding colors: rough-wall velocity profiles corrected by adding $\Delta U^+$.}
    \label{fig:mean_velocity_transformation}
\end{figure}

Figure \ref{fig:mean_velocity_transformation} presents the transformed mean streamwise velocity profiles. The smooth-wall cases show good agreement with the corresponding reference data, providing further validation of the present simulations. For the van Driest (VD) transformation, the logarithmic intercept $B$ exhibits a Mach-number dependence, with values of 5.1 and 5.6 at $Ma=2.5$ and $5.86$, respectively. In contrast, the Griffin--Fu--Moin (GFM) transformation yields a consistent intercept of $B=5.1$ for both Mach numbers. At $Ma=2.5$, the adiabatic- and cold-wall cases exhibit comparable roughness functions $\Delta U^+$ under the VD transformation, resulting in nearly overlapping rough-wall velocity profiles. At $Ma=5.86$, however, the pronounced difference in $\Delta U^+$ between the two thermal conditions leads to a clear separation of the profiles, with the cold-wall case exhibiting a smaller roughness function than its adiabatic-wall counterpart. For the adiabatic-wall cases, $\Delta U^+$ is relatively insensitive to both Mach number and the choice of velocity transformation. By contrast, for the cold-wall cases, $\Delta U^+$ decreases with increasing Mach number, and the VD transformation consistently yields larger roughness functions than the GFM transformation at a given Mach number.

\begin{figure}
    \centering

    \begin{minipage}[t]{0.4\textwidth}
        \raggedright
        (\textit{a}) \\[-0.15em]
        \includegraphics[width=\textwidth]{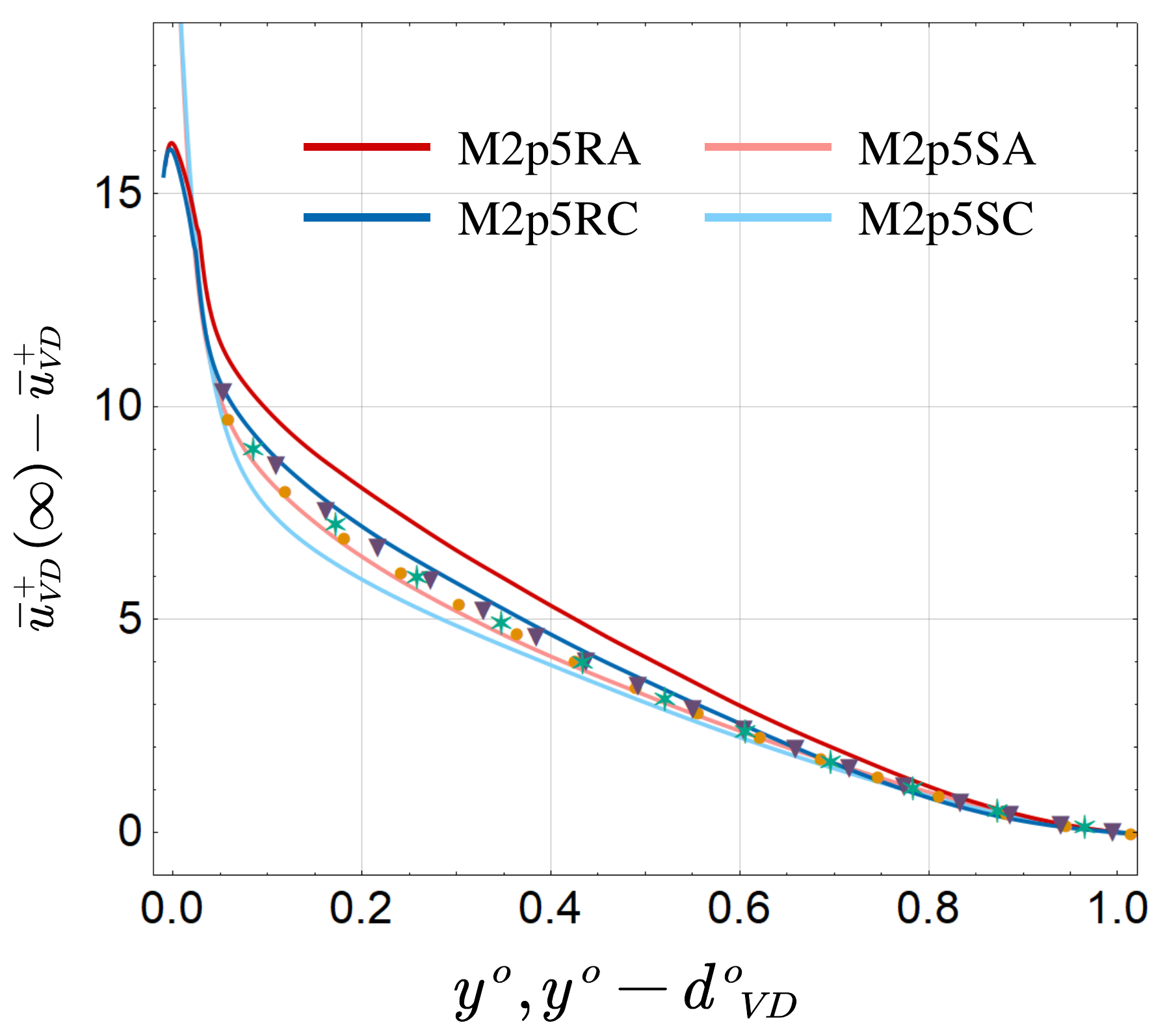}
    \end{minipage}
    \hspace{0.04\textwidth}
    \begin{minipage}[t]{0.4\textwidth}
        \raggedright
        (\textit{b}) \\[-0.15em]
        \includegraphics[width=\textwidth]{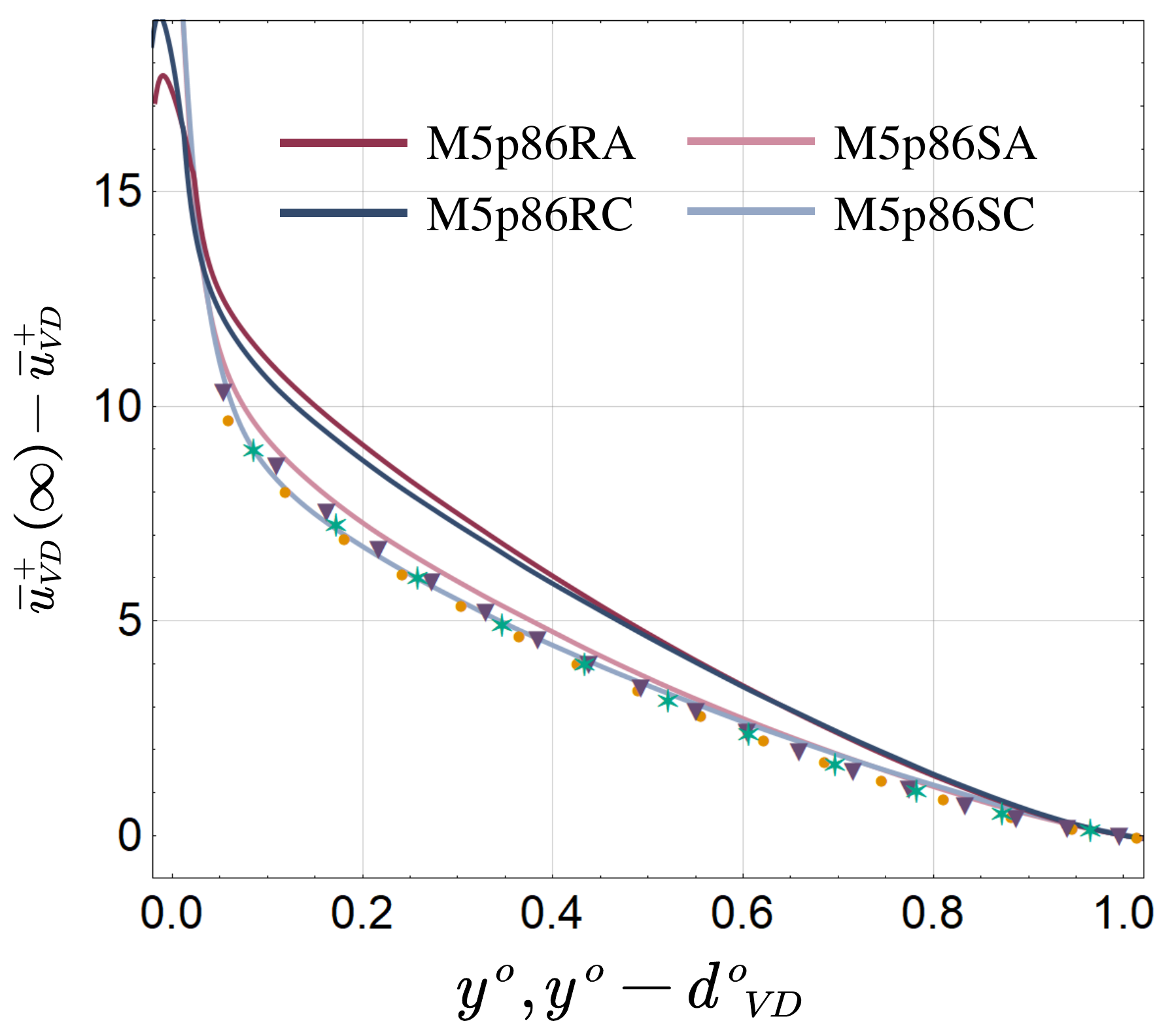}
    \end{minipage}

    \vspace{0.25cm}

    \begin{minipage}[t]{0.4\textwidth}
        \raggedright
        (\textit{c}) \\[-0.15em]
        \includegraphics[width=\textwidth]{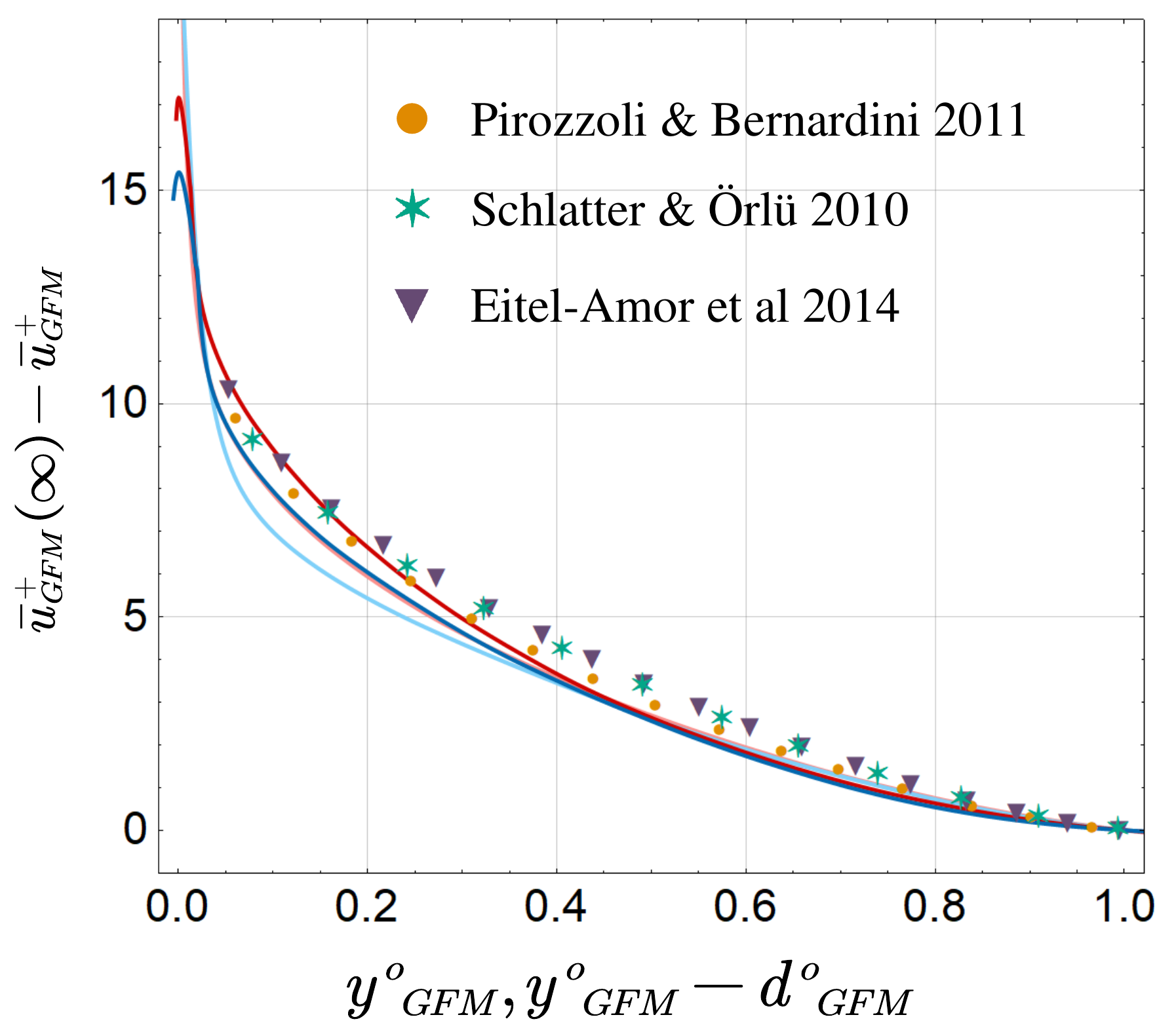}
    \end{minipage}
    \hspace{0.04\textwidth}
    \begin{minipage}[t]{0.4\textwidth}
        \raggedright
        (\textit{d}) \\[-0.15em]
        \includegraphics[width=\textwidth]{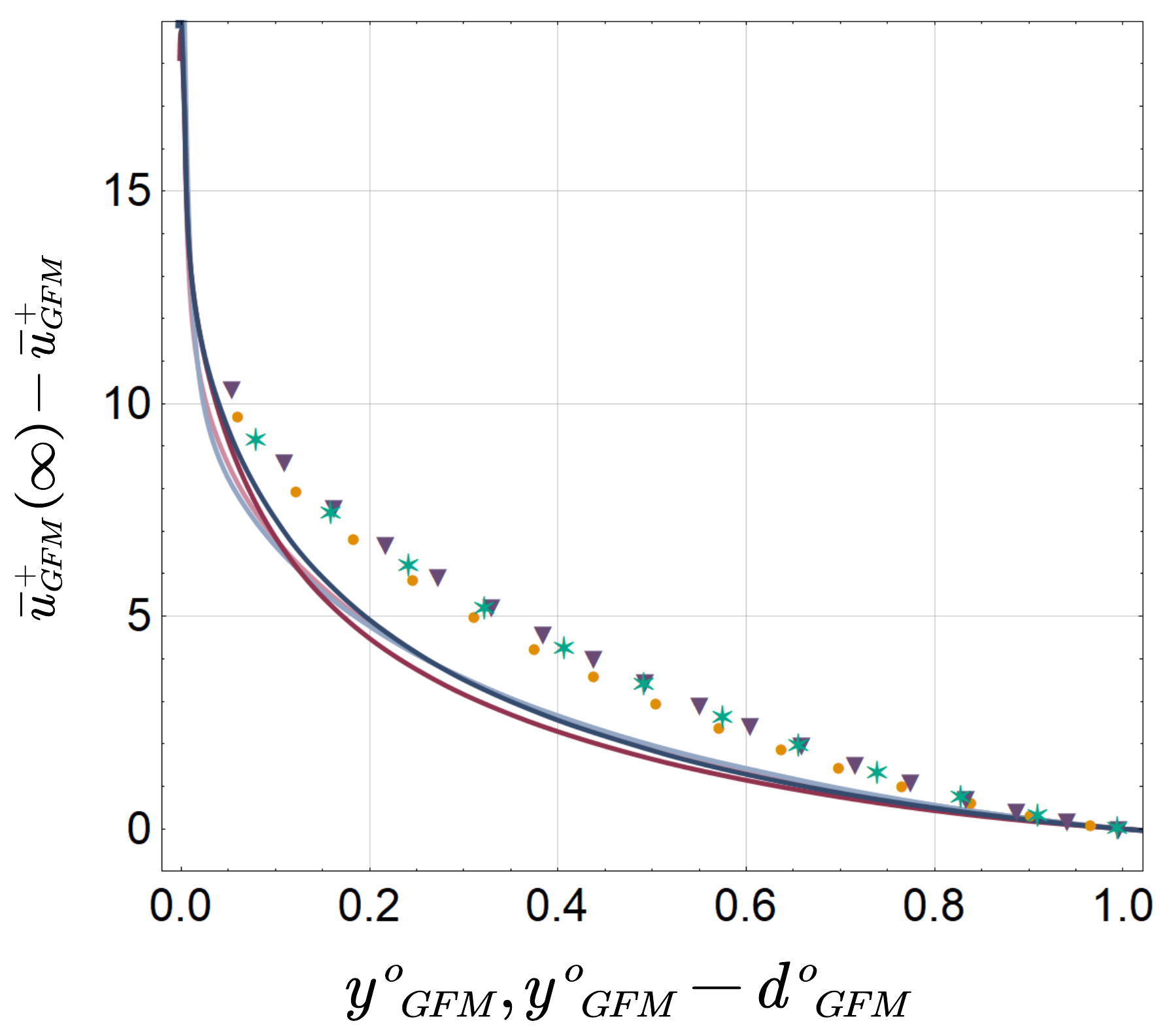}
    \end{minipage}

    \caption{Transformed mean streamwise velocity-defect profiles in outer scaling: (a) van Driest (VD) transformation at $Ma=2.5$; (b) VD transformation at $Ma=5.86$; (c) Griffin--Fu--Moin (GFM) transformation at $Ma=2.5$; (d) GFM transformation at $Ma=5.86$. The superscript $o$ denotes outer-scaled quantities, with both the height and the boundary-layer thickness evaluated using the corresponding transformation.}
    \label{fig:mean_outer_similarity}
\end{figure}

The outer-layer behaviour of the mean velocity profiles is examined in terms of the velocity defect, as shown in figure \ref{fig:mean_outer_similarity}. In the original coordinates, the differences among the smooth-wall cases are relatively small. For the rough-wall cases, however, accounting only for the zero-plane displacement in the van Driest (VD) transformation results in substantial deviations from the corresponding smooth-wall profiles, particularly at the higher Mach number. When the height is transformed according to the Griffin--Fu--Moin (GFM) transformation, a clear Mach-number dependence emerges in the velocity-defect profiles. Specifically, the profiles shift downwards with increasing Mach number over the outer-scaled height range $0.4 \lesssim y^o \lesssim 1.0$. More importantly, at a given Mach number, the GFM transformation leads to a close collapse of the velocity-defect profiles between the rough- and smooth-wall cases, demonstrating good outer-layer similarity.

\subsection{General Reynolds analogy}\label{sec:General_Reynolds_analogy}

\begin{figure}
    \centering

    \begin{minipage}[t]{0.47\textwidth}
        \raggedright
        (\textit{a}) \\[-0.15em]
        \includegraphics[width=\textwidth]{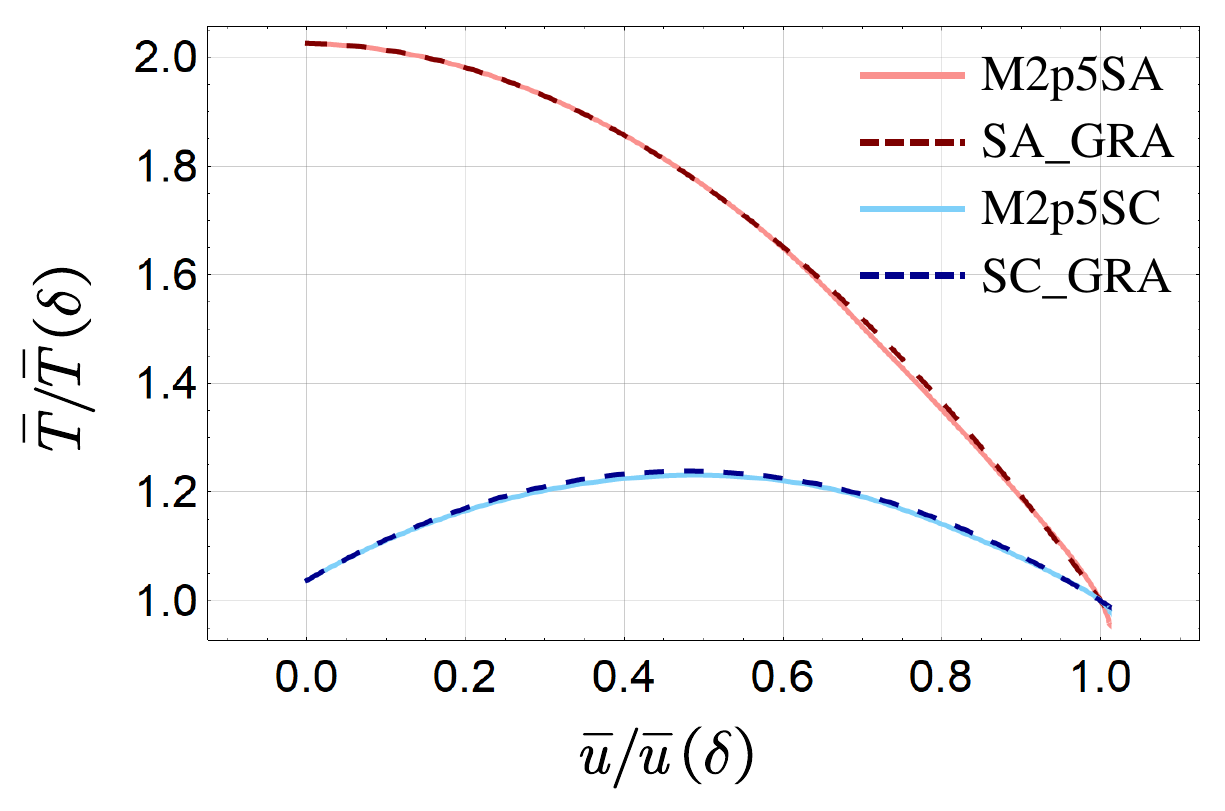}
    \end{minipage}
    \hspace{0.04\textwidth}
    \begin{minipage}[t]{0.47\textwidth}
        \raggedright
        (\textit{b}) \\[-0.15em]
        \includegraphics[width=\textwidth]{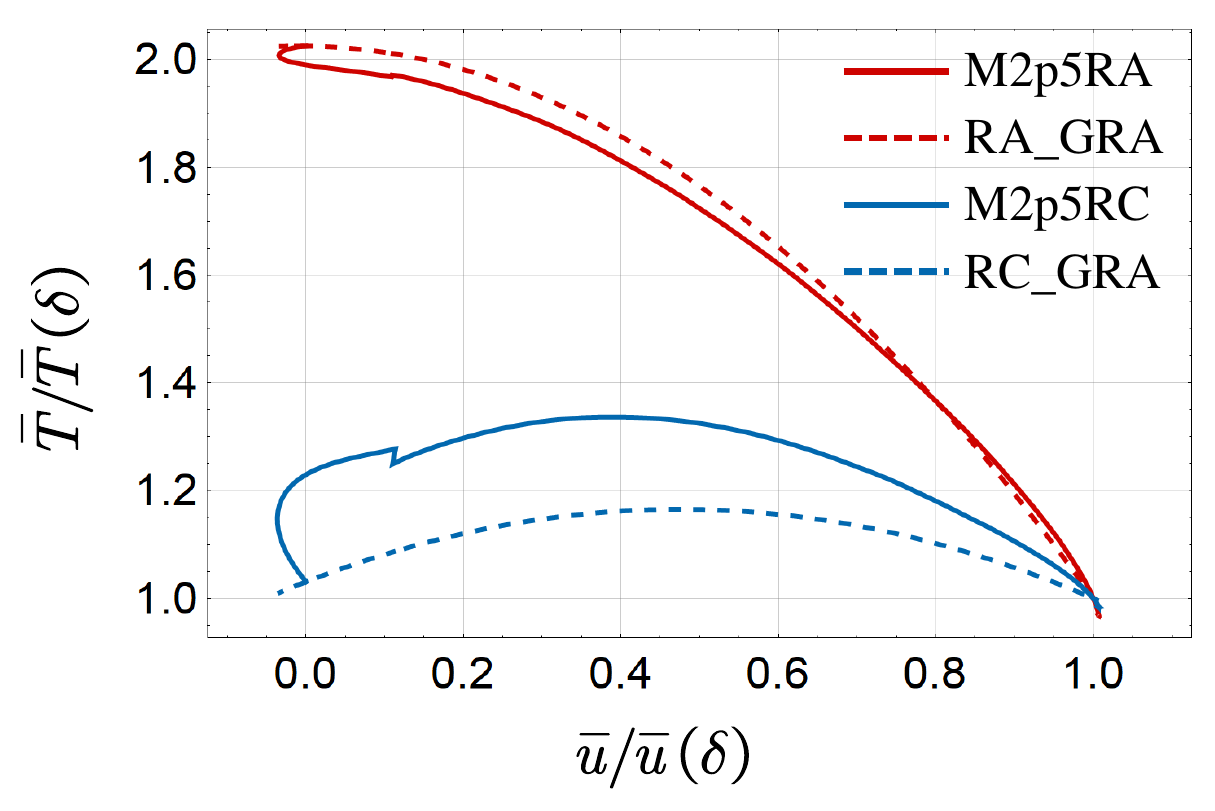}
    \end{minipage}

    \par\vspace{0.25cm}

    \begin{minipage}[t]{0.47\textwidth}
        \raggedright
        (\textit{c}) \\[-0.15em]
        \includegraphics[width=\textwidth]{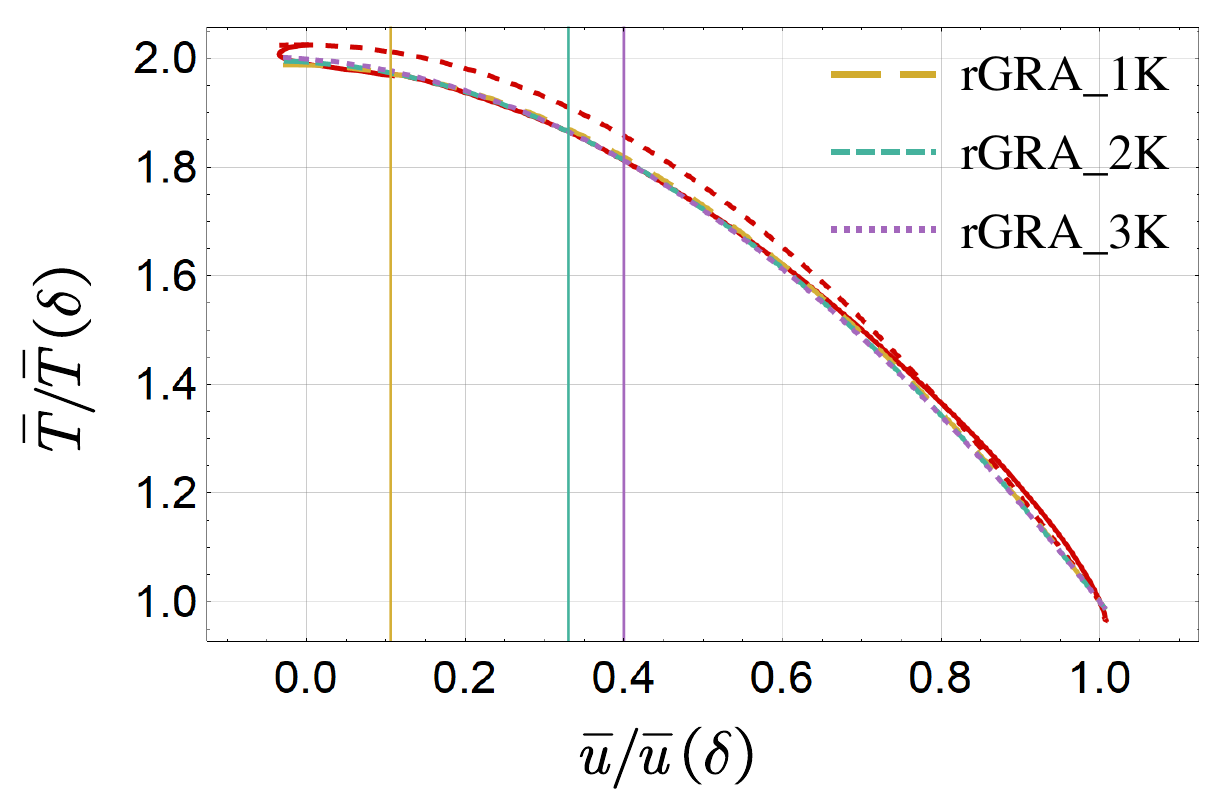}
    \end{minipage}
    \hspace{0.04\textwidth}
    \begin{minipage}[t]{0.47\textwidth}
        \raggedright
        (\textit{d}) \\[-0.15em]
        \includegraphics[width=\textwidth]{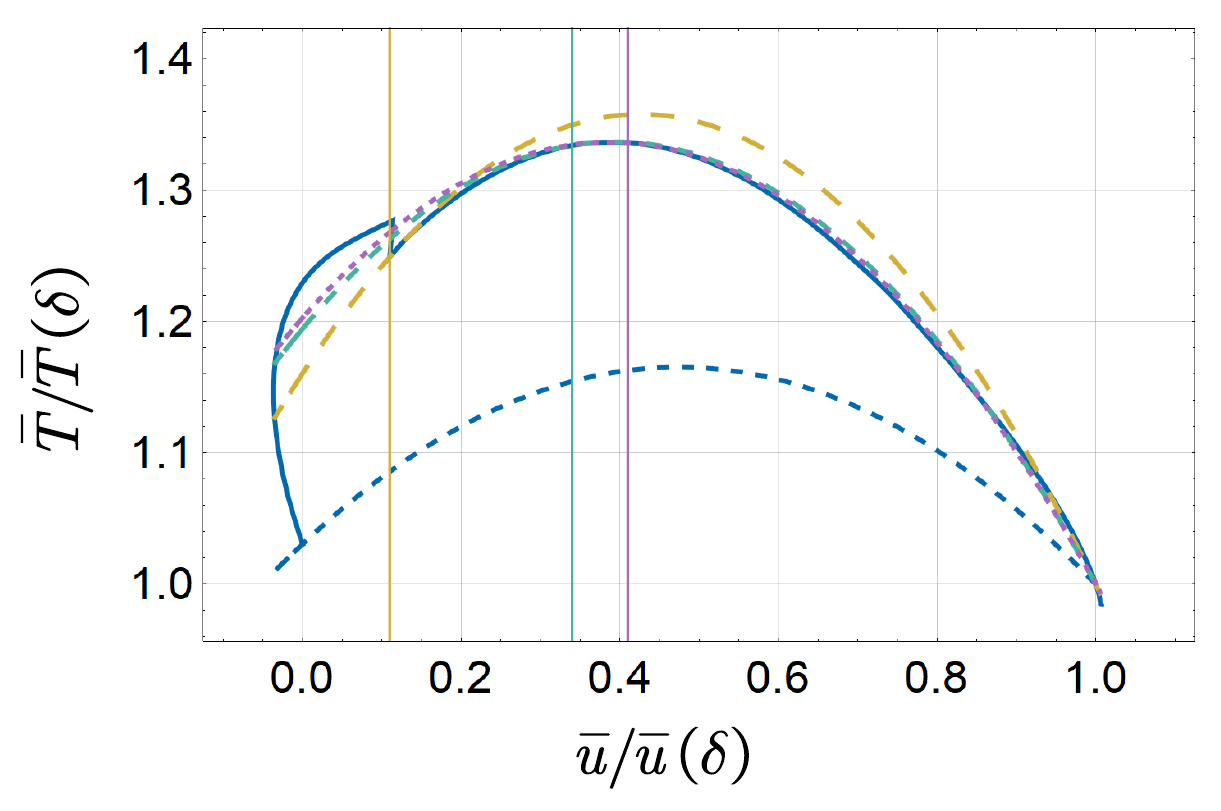}
    \end{minipage}

    \caption{Mean temperature--velocity relations in outer scaling at $Ma=2.5$: (a) smooth-wall cases; (b) rough-wall cases; (c) adiabatic rough-wall case; (d) cold rough-wall case. Solid lines: DNS results. Dashed lines: classical GRA predictions. Coloured dashed lines in (c,d): rGRA predictions with reference heights $k$, $2k$ and $3k$. Vertical solid lines: mean velocities at the corresponding reference heights.}
    \label{fig:GRA_M2p5}
\end{figure}

\begin{figure}
    \centering

    \begin{minipage}[t]{0.47\textwidth}
        \raggedright
        (\textit{a}) \\[-0.15em]
        \includegraphics[width=\textwidth]{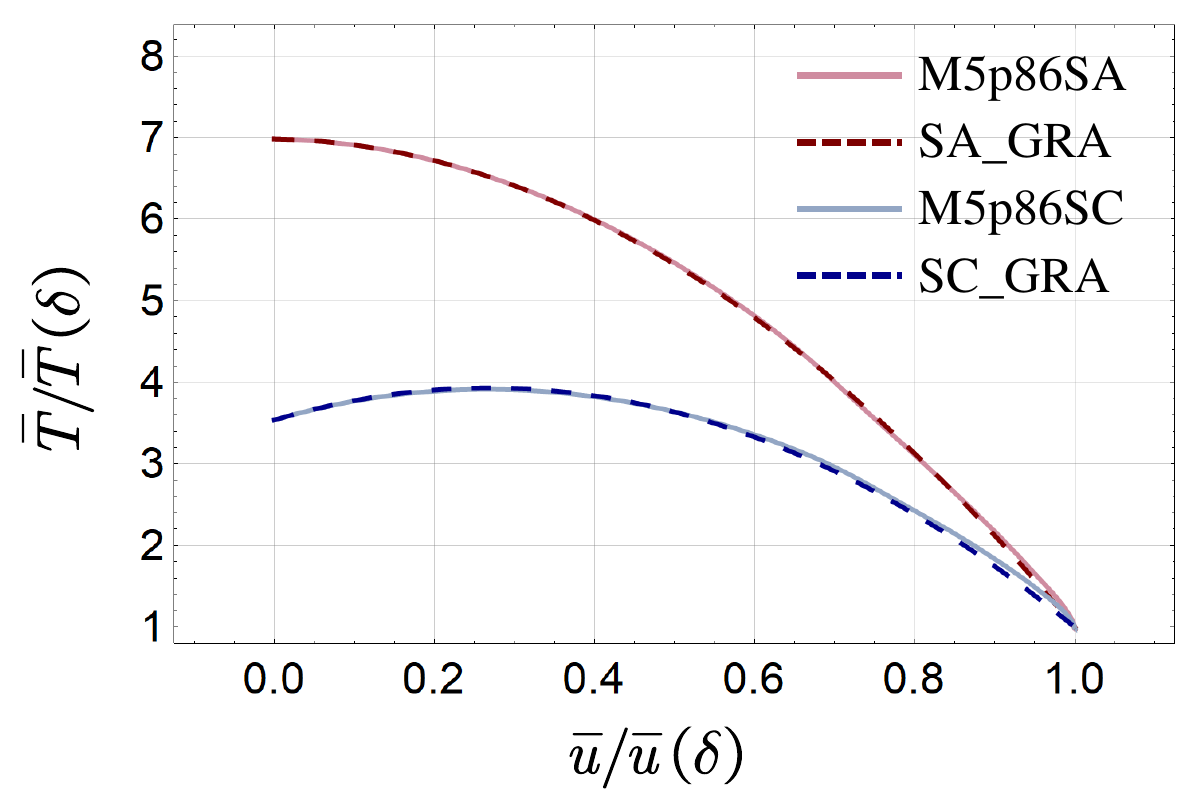}
    \end{minipage}
    \hspace{0.04\textwidth}
    \begin{minipage}[t]{0.47\textwidth}
        \raggedright
        (\textit{b}) \\[-0.15em]
        \includegraphics[width=\textwidth]{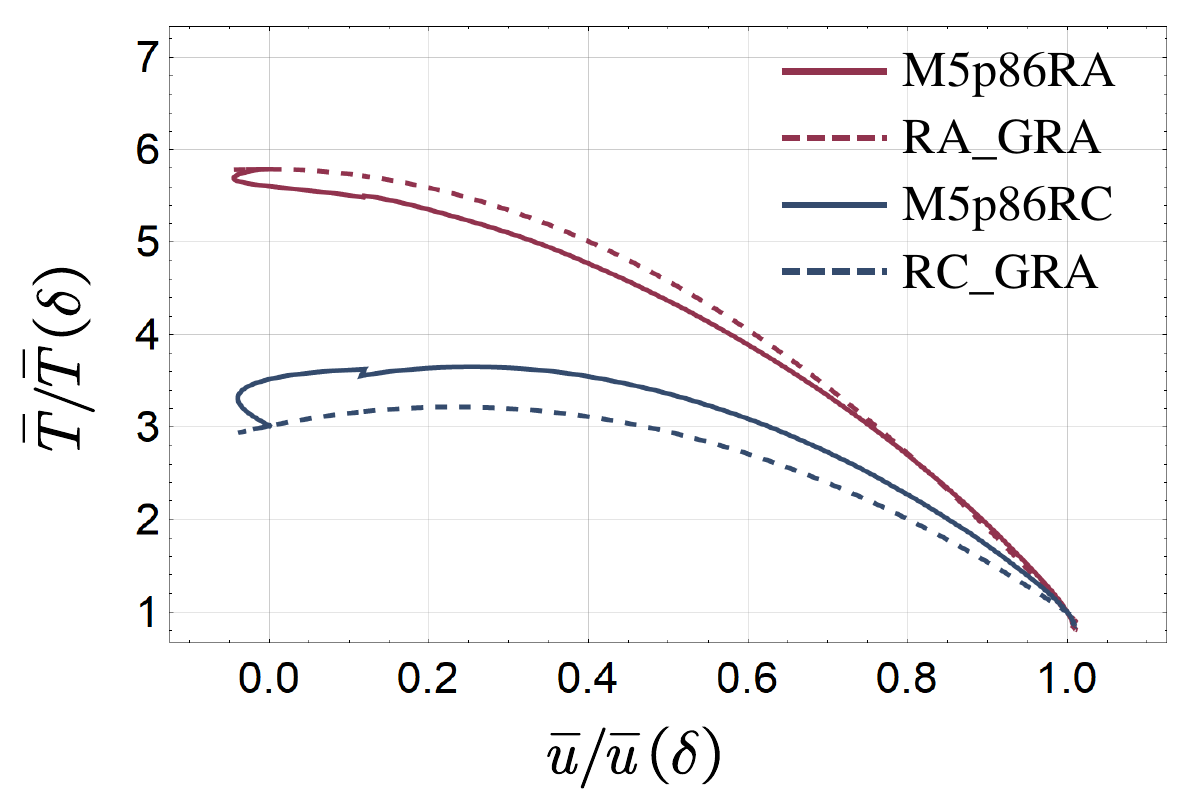}
    \end{minipage}

    \par\vspace{0.25cm}

    \begin{minipage}[t]{0.47\textwidth}
        \raggedright
        (\textit{c}) \\[-0.15em]
        \includegraphics[width=\textwidth]{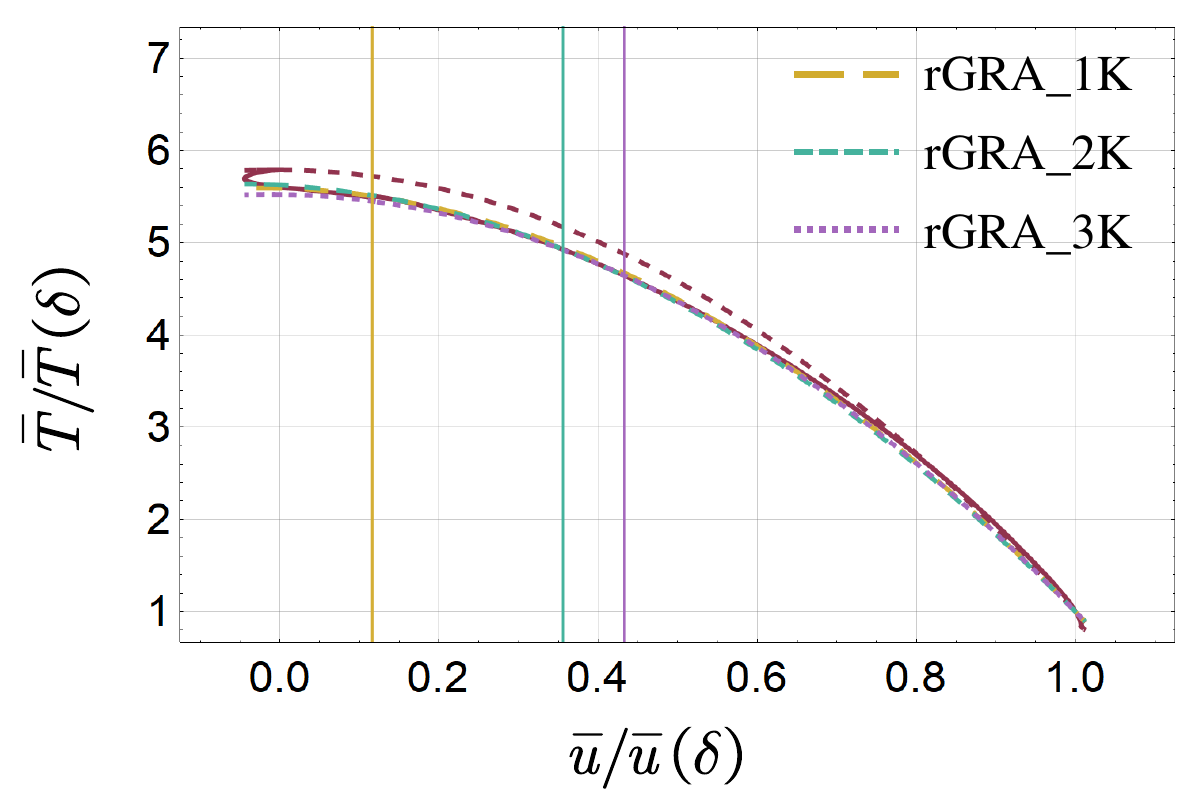}
    \end{minipage}
    \hspace{0.04\textwidth}
    \begin{minipage}[t]{0.47\textwidth}
        \raggedright
        (\textit{d}) \\[-0.15em]
        \includegraphics[width=\textwidth]{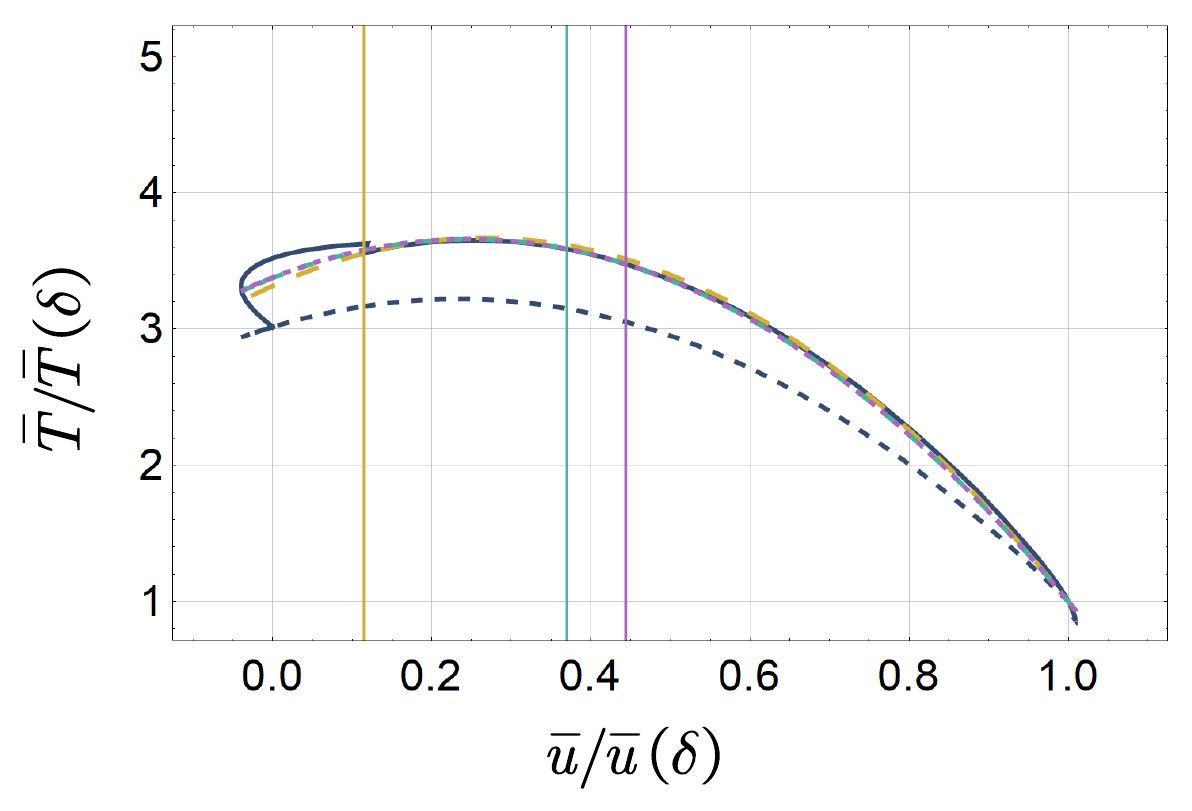}
    \end{minipage}

    \caption{Mean temperature--velocity relations in outer scaling at $Ma=5.86$: (a) smooth-wall cases; (b) rough-wall cases; (c) adiabatic rough-wall case; (d) cold rough-wall case. Solid lines: DNS results. Dashed lines: classical GRA predictions. Colored dashed lines in (c,d): rGRA predictions with reference heights $k$, $2k$ and $3k$. Vertical solid lines: mean velocities at the corresponding reference heights.}
    \label{fig:GRA_M5p86}
\end{figure}

For smooth-wall turbulent boundary layers, the relationship between the mean velocity and temperature fields can be described by the general Reynolds analogy (GRA) proposed by \citet{zhang2014generalized}. Based on the assumption that the effective turbulent Prandtl number is approximately unity, the GRA establishes a quadratic relationship between the mean temperature and streamwise velocity, expressed as
\begin{equation}
\frac{\bar{T}_{\mathrm{GRA}}}{\bar{T}_{\delta}}=\frac{\bar{T}_w}{\bar{T}_{\delta}}+\frac{\bar{u}_{\delta}\Gamma_w}{\bar{T}_{\delta}}\left(\frac{\bar{u}}{\bar{u}_{\delta}}\right)+\left(\frac{\bar{T}_{\delta}-\bar{T}_w-\bar{u}_{\delta}\Gamma_w}{\bar{T}_{\delta}}\right)\left(\frac{\bar{u}}{\bar{u}_{\delta}}\right)^2
\label{eq:GRA}
\end{equation}
where $\Gamma_w=-(Pr/C_p)(\bar{q}_w/\bar{\tau}_w)$ characterizes the coupling between wall heat flux and shear stress, and the subscript $\delta$ denotes quantities evaluated at the boundary-layer edge. As shown in figures \ref{fig:GRA_M2p5}(a) and \ref{fig:GRA_M5p86}(a), the GRA predictions agree well with the DNS results for the smooth-wall cases at both Mach numbers. In contrast, figures \ref{fig:GRA_M2p5}(b) and \ref{fig:GRA_M5p86}(b) reveal substantial discrepancies between the GRA predictions and DNS results in the near-wall region of the rough-wall cases, particularly under cold-wall conditions.

To overcome this limitation, \citet{su2026compressible} developed a rough-wall general Reynolds analogy (rGRA) by introducing a reference point above the roughness crest, where the strong near-wall disturbances are sufficiently weakened. By replacing the physical wall conditions with the local temperature, velocity and temperature--velocity gradient at a reference height $h$, the rGRA takes the form
\begin{equation}
\begin{aligned}
\bar{T}_{\mathrm{rGRA}}(h,\bar{u})={}&\bar{T}_h+\left.\frac{\partial\bar{T}}{\partial\bar{u}}\right|_h(\bar{u}-\bar{u}_h)
&+\left[\bar{T}_{\delta}-\bar{T}_h-(\bar{u}_{\delta}-\bar{u}_h)\left.\frac{\partial\bar{T}}{\partial\bar{u}}\right|_h\right]\left(\frac{\bar{u}-\bar{u}_h}{\bar{u}_{\delta}-\bar{u}_h}\right)^2
\end{aligned}
\label{eq:rGRA}
\end{equation}
where the subscript $h$ denotes quantities evaluated at the reference height (the labels $1k$, $2k$ and $3k$ in figures \ref{fig:GRA_M2p5} and \ref{fig:GRA_M5p86} indicate reference heights of one, two and three times the roughness height, respectively). By introducing a reference point sufficiently above the roughness crest, the rGRA avoids the strong spatial heterogeneity of the thermal field within the roughness sublayer, thereby providing a more accurate reconstruction of the temperature-velocity relationship.

\begin{figure}
    \centering

    \begin{minipage}[t]{0.47\textwidth}
        \raggedright
        (\textit{a}) \\[-0.15em]
        \includegraphics[width=\textwidth]{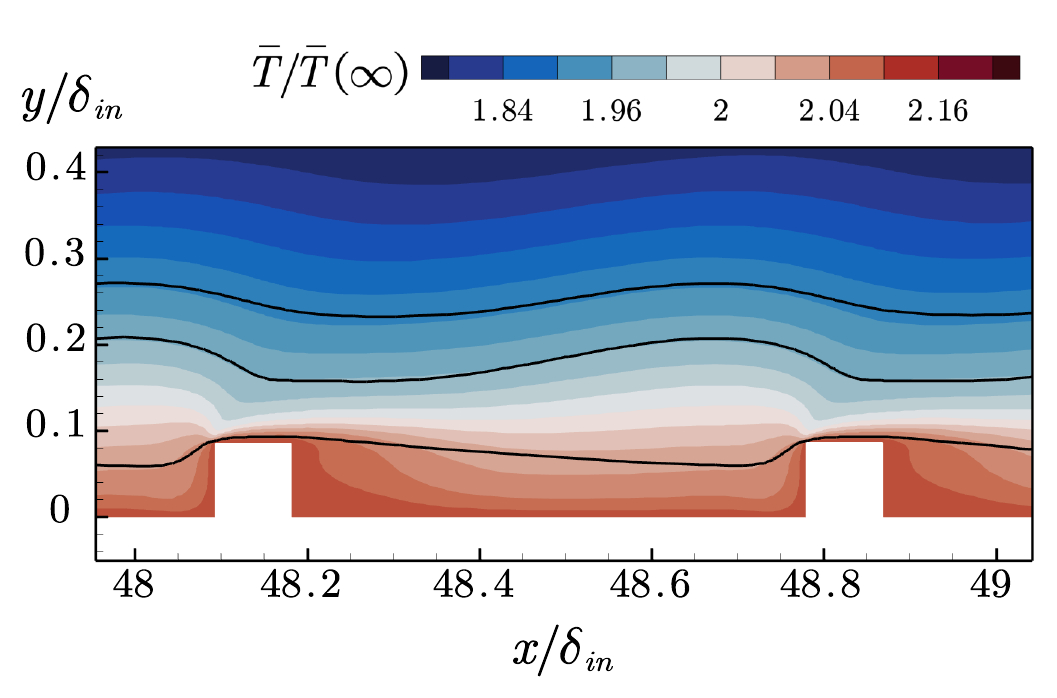}
    \end{minipage}
    \hspace{0.04\textwidth}
    \begin{minipage}[t]{0.47\textwidth}
        \raggedright
        (\textit{b}) \\[-0.15em]
        \includegraphics[width=\textwidth]{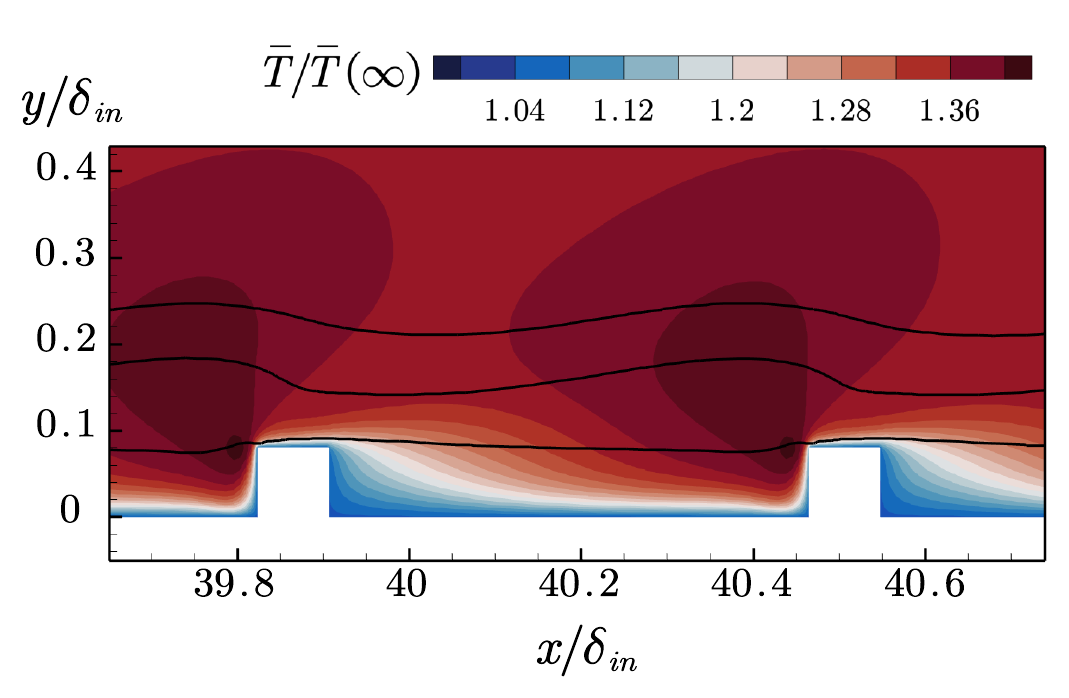}
    \end{minipage}

    \par\vspace{0.25cm}

    \begin{minipage}[t]{0.47\textwidth}
        \raggedright
        (\textit{c}) \\[-0.15em]
        \includegraphics[width=\textwidth]{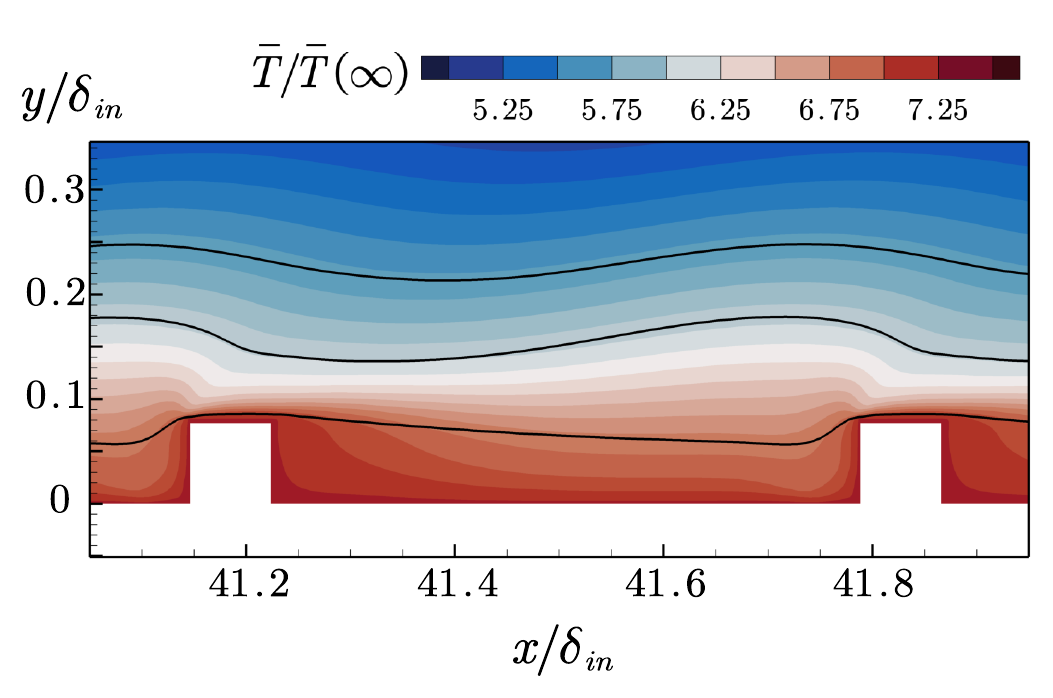}
    \end{minipage}
    \hspace{0.04\textwidth}
    \begin{minipage}[t]{0.47\textwidth}
        \raggedright
        (\textit{d}) \\[-0.15em]
        \includegraphics[width=\textwidth]{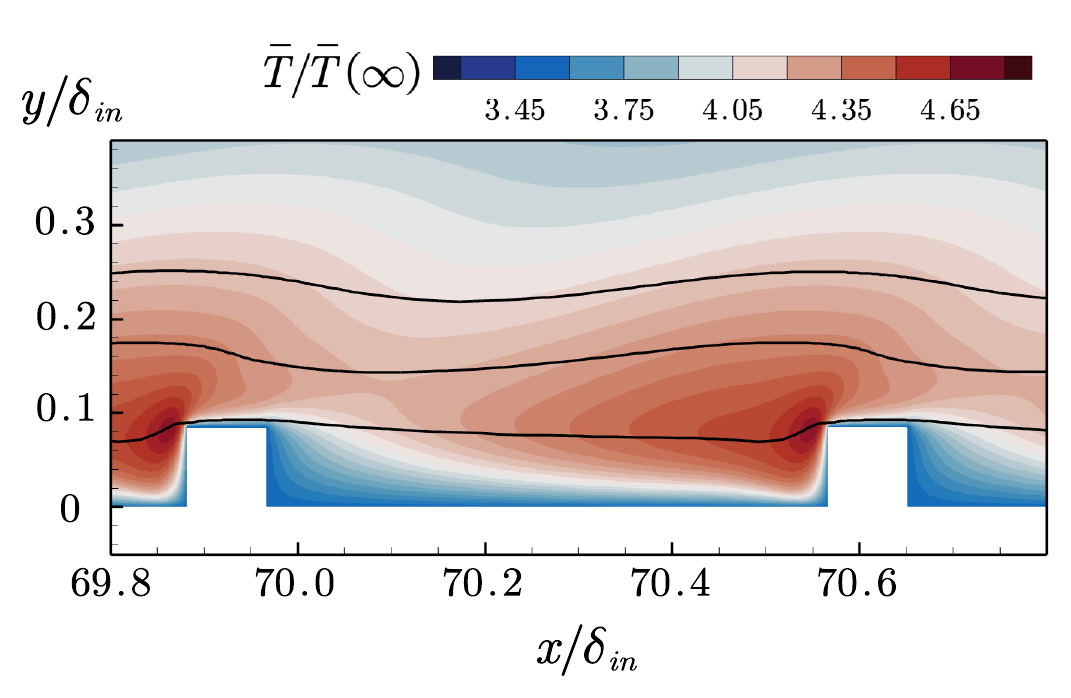}
    \end{minipage}

    \caption{Contours of the mean temperature overlaid with isolines of the mean streamwise velocity for (a) M2p5RA, (b) M2p5RC, (c) M5p86RA and (d) M5p86RC.}
    \label{fig:temperature_contours}
\end{figure}

For the adiabatic-wall cases, accurate temperature predictions can be obtained with a reference height of $h=k$. For the cold-wall case at $Ma=2.5$, however, the reference height must be increased to $h=2k$ to achieve good agreement with the DNS results, whereas $h=k$ remains sufficient at $Ma=5.86$. The physical origin of these differences can be further elucidated by examining the mean temperature contours and the corresponding streamwise velocity isolines in figure \ref{fig:temperature_contours}. For the adiabatic-wall cases, figures \ref{fig:temperature_contours}(a,c) show that the temperature contours and streamwise velocity isolines are nearly parallel above the roughness crests, consistent with the assumption that the flow progressively recovers a quasi-parallel state away from the wall. In the cold-wall cases, wall cooling leads to a substantial reduction in temperature within the recirculation regions, while localized high-temperature regions are observed near the windward faces of the roughness elements. These features give rise to pronounced streamwise-periodic temperature variations, resulting in substantial differences between the temperature contours and velocity isolines in the lower roughness sublayer, as illustrated in figures \ref{fig:temperature_contours}(b,d). This discrepancy becomes less pronounced at the higher Mach number, where the spatial extent of the high-temperature regions near the windward faces is considerably reduced, allowing the temperature field to approach a quasi-parallel state over a shorter wall-normal distance.

\subsection{Velocity fluctuations}\label{sec:Velocity_fluctuations}

To account for the mean density variations induced by compressibility, the root-mean-square (r.m.s.) velocity fluctuations are scaled following the density-weighted transformation of \citet{morkovin1962effects}
\begin{equation}
u_{\mathrm{r.m.s.},i}^{*}=\left(\frac{\bar{\rho}}{\bar{\rho}_w}\right)^{1/2}\frac{\sqrt{\widetilde{u_i''u_i''}}}{u_\tau}
\label{eq:rms_scaling}
\end{equation}
where $u_i''$ denotes the Favre velocity fluctuation, $\bar{\rho}$ and $\bar{\rho}_w$ are the local mean and wall densities, respectively, and $u_\tau$ is the friction velocity. This transformation accounts for the effects of mean density variations on the turbulent velocity fluctuations, facilitating comparisons across different Mach numbers.

\begin{figure}
    \centering

    \def\myfigheight{5cm}
    \def\widthA{0.48\textwidth}
    \def\widthB{0.48\textwidth}


    \begin{minipage}[t]{\widthA}
        {\raggedright (\textit{a})\par}
        \vspace{-1.2em}
        \centering
        \includegraphics[width=\linewidth,height=\myfigheight,keepaspectratio]
        {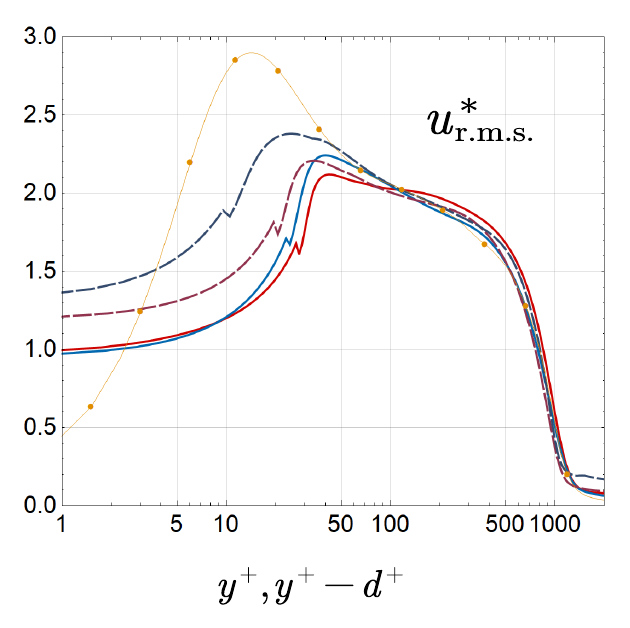}
    \end{minipage}%
    \hspace{0.01\textwidth}%
    \begin{minipage}[t]{\widthB}
        {\raggedright (\textit{b})\par}
        \vspace{-1.2em}
        \centering
        \includegraphics[width=\linewidth,height=\myfigheight,keepaspectratio]
        {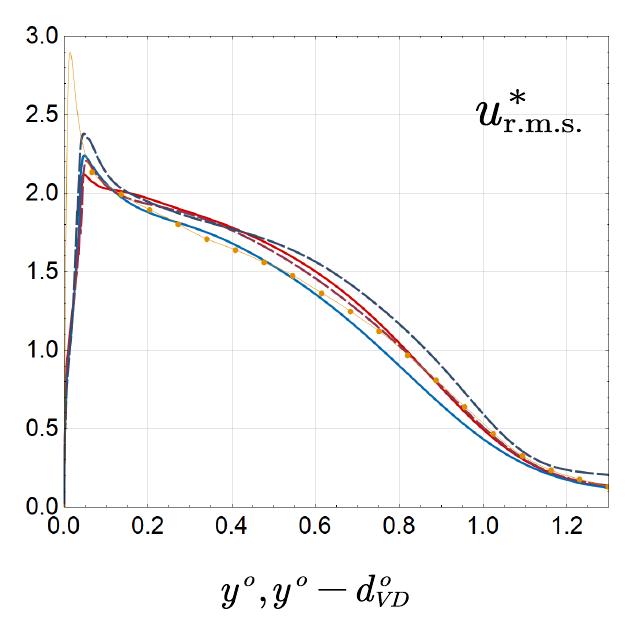}
    \end{minipage}

    \par\vspace{0.15cm}


    \begin{minipage}[t]{\widthA}
        {\raggedright (\textit{c})\par}
        \vspace{-1.2em}
        \centering
        \includegraphics[width=\linewidth,height=\myfigheight,keepaspectratio]
        {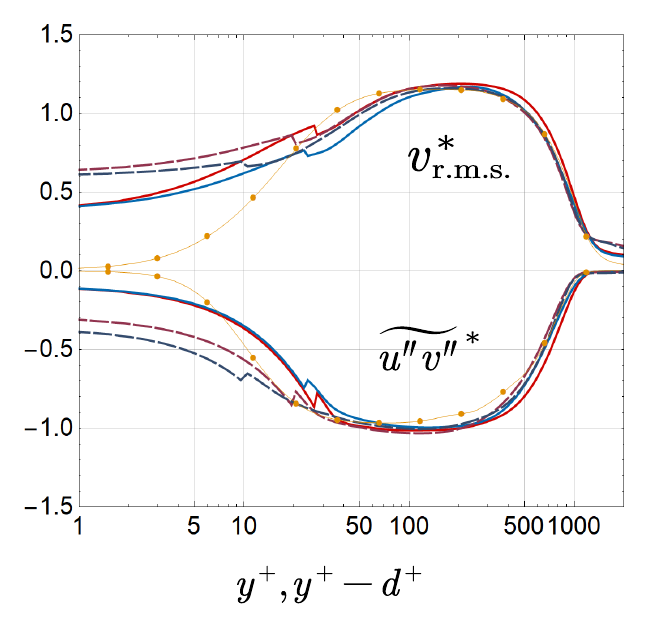}
    \end{minipage}%
    \hspace{0.01\textwidth}%
    \begin{minipage}[t]{\widthB}
        {\raggedright (\textit{d})\par}
        \vspace{-1.2em}
        \centering
        \includegraphics[width=\linewidth,height=\myfigheight,keepaspectratio]
        {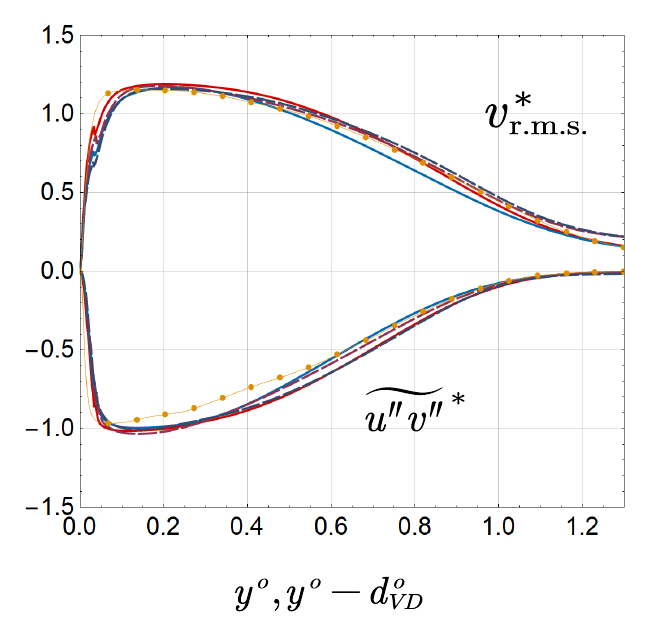}
    \end{minipage}

    \par\vspace{0.15cm}


    \begin{minipage}[t]{\widthA}
        {\raggedright (\textit{e})\par}
        \vspace{-1.2em}
        \centering
        \includegraphics[width=\linewidth,height=\myfigheight,keepaspectratio]
        {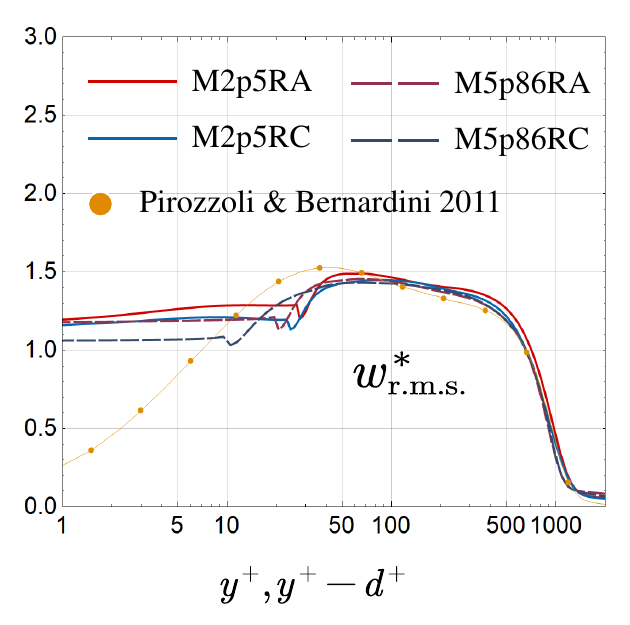}
    \end{minipage}%
    \hspace{0.01\textwidth}%
    \begin{minipage}[t]{\widthB}
        {\raggedright (\textit{f})\par}
        \vspace{-1.2em}
        \centering
        \includegraphics[width=\linewidth,height=\myfigheight,keepaspectratio]
        {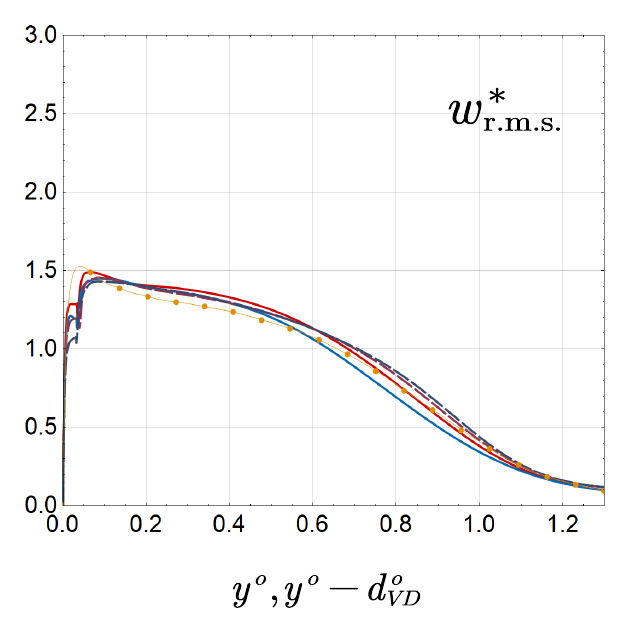}
    \end{minipage}

    \caption{Profiles of the turbulent velocity fluctuations obtained using the density-weighted transformation of Morkovin: (a,b) streamwise velocity fluctuation, (c,d) wall-normal velocity fluctuation and Reynolds shear stress, and (e,f) spanwise velocity fluctuation. The left column shows the inner-scaled profiles, whereas the right column shows the corresponding outer-scaled profiles. The superscript $o$ denotes outer-scaled quantities, with both the height and the boundary-layer thickness evaluated using the corresponding transformation.}
    \label{fig:RMS_ALL}
\end{figure}

As shown in figure \ref{fig:RMS_ALL}(a), the presence of roughness elements suppresses the peak in the streamwise velocity fluctuations observed near the roughness height in the smooth-wall cases, giving rise instead to a weaker peak within the roughness sublayer above the roughness crests. In comparison, the cold-wall cases exhibit higher fluctuation peaks than their adiabatic-wall counterparts at the same Mach number, and the peak intensity increases with increasing Mach number.

The influence of roughness on the wall-normal velocity fluctuations is relatively weak, with the fluctuation intensities remaining close to those of the corresponding smooth-wall cases and showing little sensitivity to wall temperature (figure \ref{fig:RMS_ALL}c). By comparison, the Reynolds shear stress and spanwise velocity fluctuations exhibit peak magnitudes comparable to their smooth-wall counterparts, but are slightly reduced towards the upper edge of the logarithmic region, as shown in figures \ref{fig:RMS_ALL}(c,e). This reduction is also evident in the outer-scaled profiles over the range $0.15 \lesssim y^o \lesssim 0.45$, as illustrated in figures \ref{fig:RMS_ALL}(d,f). Under outer scaling, the velocity fluctuation profiles of the adiabatic rough-wall cases exhibit a weak Mach-number dependence, while remaining distinct from their smooth-wall counterparts for $y^o \lesssim 0.7$, followed by a rapid decay farther from the wall. The effect of wall cooling is not monotonic with Mach number: at $Ma=2.5$, the cold-wall case exhibits lower outer-layer fluctuation intensities than its adiabatic-wall counterpart, whereas the opposite trend is observed at $Ma=5.86$.

\subsection{Generalized Renard--Deck drag decomposition for rough-wall boundary layers}\label{sec:RD_decomposition}

In the present study, we derive a generalized Renard–Deck (RD) drag decomposition for compressible turbulent boundary layers over rough surfaces. Building on the energy-based framework of \cite, the proposed formulation incorporates compressibility effects and accounts for both viscous friction and pressure drag exerted on the rough surface. The derivation begins with the statistically stationary Reynolds-averaged momentum equation, which can be expressed in tensor form as
\begin{equation}
\nabla\cdot\left(\overline{\boldsymbol{\mathcal T}}-\overline{\rho}\widetilde{\boldsymbol U}\widetilde{\boldsymbol U}\right)=0,\qquad \overline{\boldsymbol{\mathcal T}}=\overline{\boldsymbol T}-\overline{\rho}\widetilde{\boldsymbol U''\boldsymbol U''}
\label{eq:momentum_vec_identity}
\end{equation}
where $\overline{\boldsymbol{\mathcal T}}$ is the total stress tensor incorporating the Reynolds stress, and $\overline{\boldsymbol T} = -\overline{p}\boldsymbol{I}+\boldsymbol{\overline{\sigma}}$ denotes the mean pressure and viscous stresses. Following the energy-based approach of \cite{renard2016theoretical}, the momentum equation is multiplied by the mean velocity relative to the free stream, $\overline{\boldsymbol U}-\boldsymbol U_\infty$, and integrated over the fluid domain $\Omega_f$. Applying the product rule yields
\begin{equation}
\begin{aligned}
0={}&\iiint_{\Omega_f}\nabla\cdot\left[(\overline{\boldsymbol U}-\boldsymbol U_\infty)\cdot\overline{\boldsymbol{\mathcal T}}\right]\,\mathrm{d}V\\
&-\iiint_{\Omega_f}\nabla\overline{\boldsymbol U}:\overline{\boldsymbol{\mathcal T}}\,\mathrm{d}V\\
&-\iiint_{\Omega_f}(\overline{\boldsymbol U}-\boldsymbol U_\infty)\cdot\nabla\cdot\left(\overline{\rho}\widetilde{\boldsymbol U}\widetilde{\boldsymbol U}\right)\,\mathrm{d}V
\end{aligned}
\label{RD_original_three}
\end{equation}

\begin{figure}
    \centering
    \begin{minipage}[t]{0.208\textwidth}
        \raggedright
        (\textit{a}) \\[0.208em]
        \includegraphics[width=\textwidth]{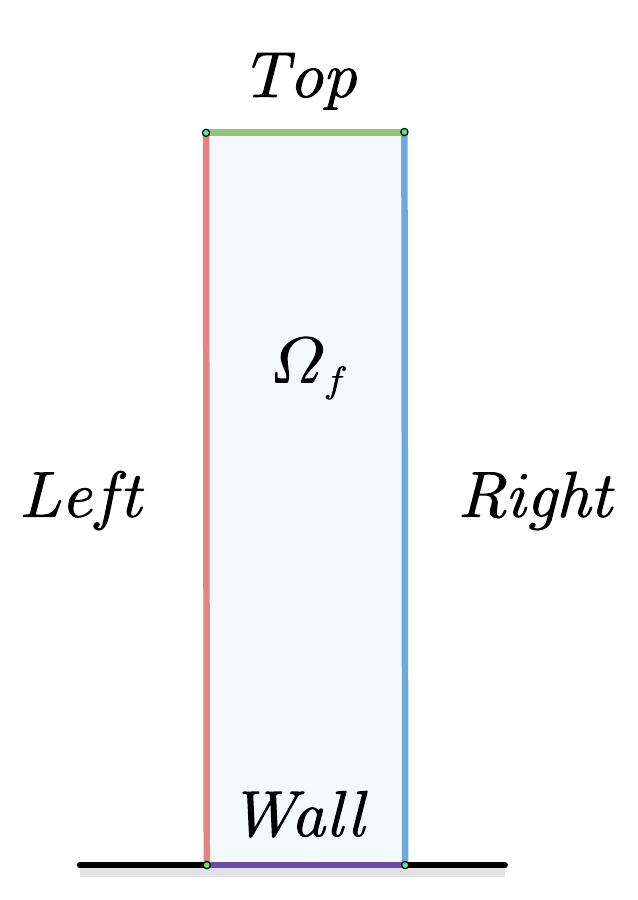}
    \end{minipage}
    \hfill
    \begin{minipage}[t]{0.409\textwidth}
        \raggedright
        (\textit{b}) \\[0.409em]
        \includegraphics[width=\textwidth]{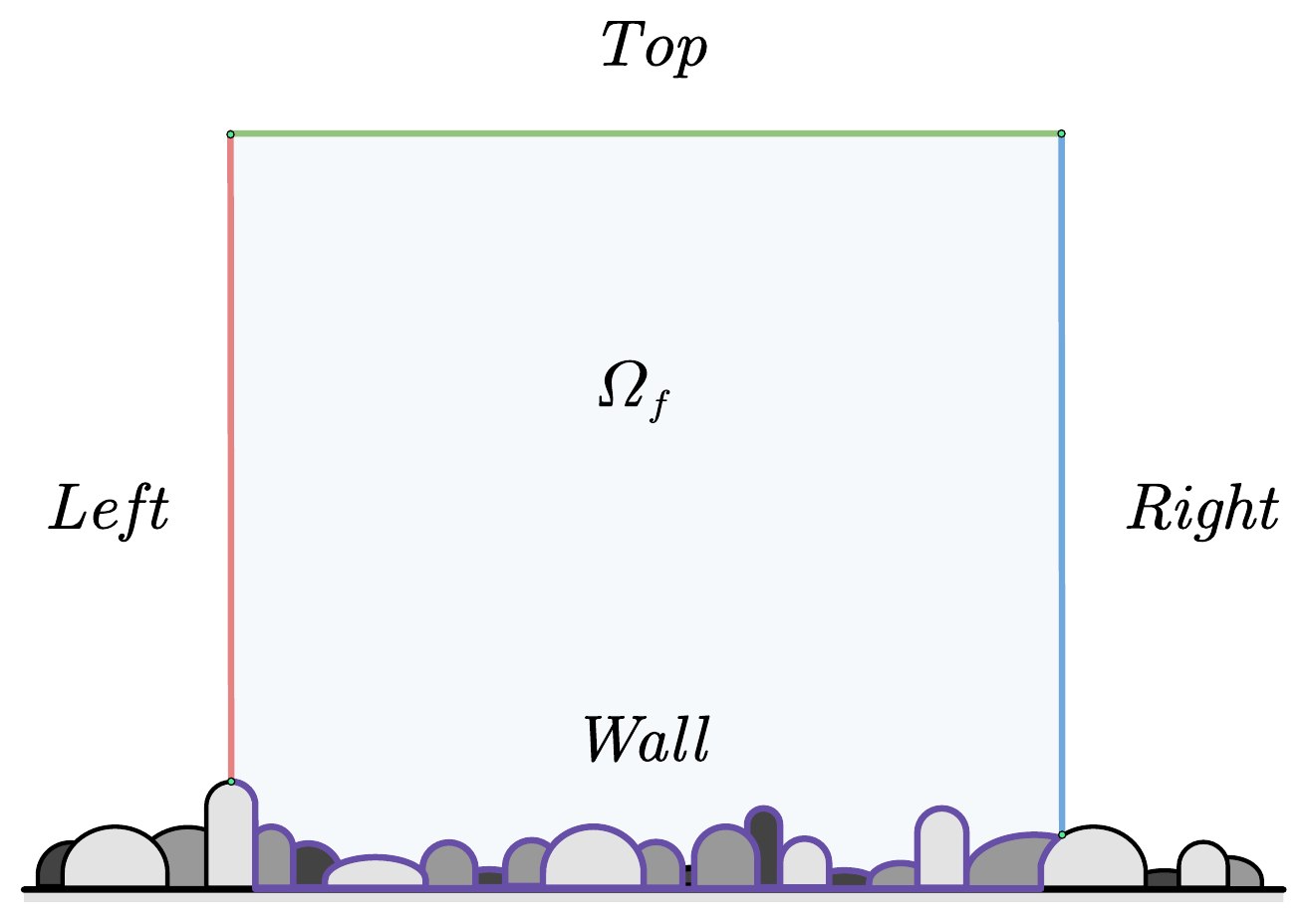}
    \end{minipage}
    \hfill
    \begin{minipage}[t]{0.355\textwidth}
        \raggedright
        (\textit{c}) \\[0.355em]
        \includegraphics[width=\textwidth]{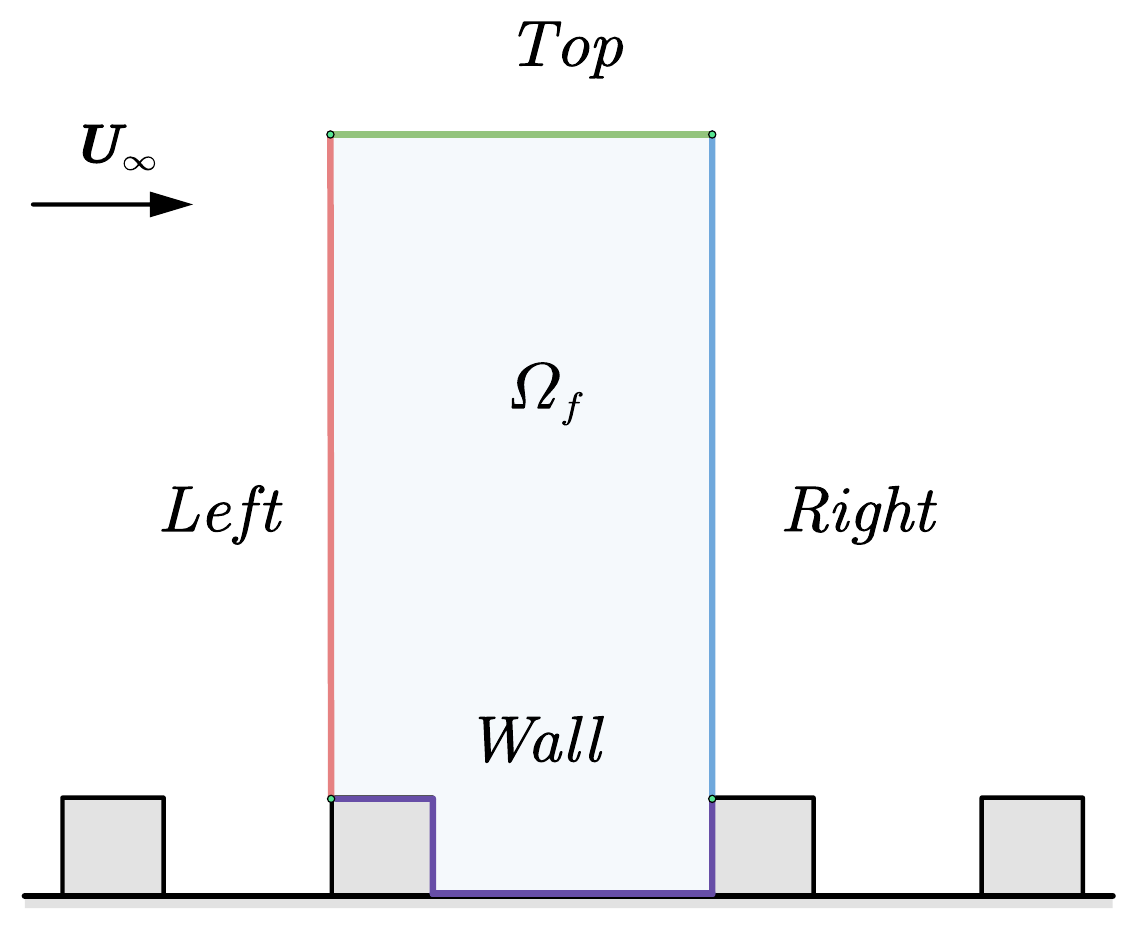}
    \end{minipage}

    \caption{Schematic illustration of the integration domains for the generalized Renard--Deck (RD) decomposition: (a) smooth wall, (b) random roughness and (c) span-wise square-bar roughness.}
    \label{fig:RDField}
\end{figure}

The integration domain is bounded by the upstream ($L$, Left), downstream ($R$, Right), upper ($T$, Top) and solid-wall ($W$, Wall) boundaries, as labelled in figure \ref{fig:RDField}, which illustrates the corresponding integration domains for smooth and rough surfaces.

Applying the divergence theorem to the first term isolates the total drag $F_D$, including both viscous friction and pressure drag, through the work of the solid boundary in the free-stream reference frame. The second term represents the work of the mean velocity gradients against the total stress, while the third term describes the transport of mean kinetic energy. Expanding these contributions and applying the corresponding boundary conditions leads to the generalized RD decomposition

\begingroup
\small
\setlength{\jot}{3pt}
\begin{equation}
\begin{aligned}
U_{\infty}F_D
=&\underbrace{
\left.\iint_S-\left[\left(\overline{\boldsymbol{U}}-\boldsymbol{U}_{\infty}\right)\cdot\overline{\boldsymbol{\mathcal{T}}}\right]_x\,\mathrm{d}A\right|_L^R
+\iiint_{\Omega}\gamma\left[-\overline{p}\frac{\partial\overline{u}}{\partial x}\right]\,\mathrm{d}V
+\left.\iint_S\mu\left[(\overline{\boldsymbol{U}}\cdot\nabla)\overline{\boldsymbol{U}}\right]_x\,\mathrm{d}A\right|_L^R
}_{\text{\scriptsize Stream-wise development (SD)}}\\
&+\underbrace{\iiint_{\Omega}\gamma\overline{v}\frac{\partial\overline{p}}{\partial y}\,\mathrm{d}V}_{\text{\scriptsize Normal displacement (ND)}}
+\underbrace{\iiint_{\Omega}-\gamma\overline{p}\frac{\partial\overline{w}}{\partial z}\,\mathrm{d}V}_{\text{\scriptsize Spanwise inhomogeneity (SI)}}\\
&+\underbrace{
\iiint_{\Omega}\gamma\left[
-\frac{2}{3}\mu\overline{\Theta}^{2}
-\nabla\mu\cdot\left((\overline{\boldsymbol{U}}\cdot\nabla)\overline{\boldsymbol{U}}\right)
-\mu\overline{\boldsymbol{U}}\cdot\nabla\overline{\Theta}
+\left(\overline{\boldsymbol{U}}-\widetilde{\boldsymbol{U}}\right)\cdot\overline{\rho}\left(\widetilde{\boldsymbol{U}}\cdot\nabla\right)\widetilde{\boldsymbol{U}}
\right]\,\mathrm{d}V
}_{\substack{\text{\scriptsize Additional compressible term (AC)}}}\\
&+\underbrace{\iiint_{\Omega}\gamma\left(\Phi_{\nu}+\Phi_R\right)\,\mathrm{d}V}_{\text{\scriptsize Dissipation term (DT)}}
+\underbrace{
\left.\iint_S\frac{1}{2}\overline{\rho}\,\widetilde{U}_x\left|\widetilde{\boldsymbol{U}}-\boldsymbol{U}_{\infty}\right|^2\,\mathrm{d}A\right|_L^R
}_{\text{\scriptsize Convection term (CT)}}
\end{aligned}
\label{eq:RD_generalized_final}
\end{equation}
\endgroup

Here, $\overline{\Theta}=\nabla\cdot\overline{\boldsymbol U}$ denotes the mean velocity dilatation. The viscous and turbulent contributions entering the dissipation term are defined as
\begin{equation}
\Phi_\nu=\mu\frac{\partial\overline U_j}{\partial x_k}\frac{\partial\overline U_j}{\partial x_k},
\qquad
\Phi_R=-\overline{\rho}\widetilde{U_j''U_k''}\frac{\partial\overline U_j}{\partial x_k}
\end{equation}
The term $SD$ represents the contribution associated with streamwise development of the boundary layer. It contains the streamwise variation of the stress work between the upstream and downstream cross-sections, the streamwise pressure contribution, and the corresponding viscous transport term. The $ND$ term accounts for the pressure work associated with wall-normal displacement of the mean flow, whereas $SI$ represents the contribution arising from spanwise inhomogeneity induced by the surface geometry. The $AC$ term collects the additional contributions associated with compressibility, including dilatational effects, viscosity-gradient effects, the coupling between the mean velocity and dilatation gradients, and the contribution arising from the difference between Reynolds- and Favre-averaged velocities. The $DT$ term comprises the mean-flow viscous dissipation $\Phi_\nu$ and turbulent kinetic-energy production $\Phi_R$, corresponding to the direct dissipation of mean-flow kinetic energy and the transfer of energy from the mean flow to turbulent fluctuations, respectively. Finally, $CT$ represents the convection of mean kinetic energy through the upstream and downstream cross-sections in the reference frame where the free stream is at rest. The detailed derivation of these terms is provided in the Appendix.

\begin{figure}
\centering

\begin{minipage}[t]{0.48\textwidth}
\hspace*{-1.2em}(\textit{a})\par
\vspace{-0.15em}
\centering
\includegraphics[width=\linewidth]{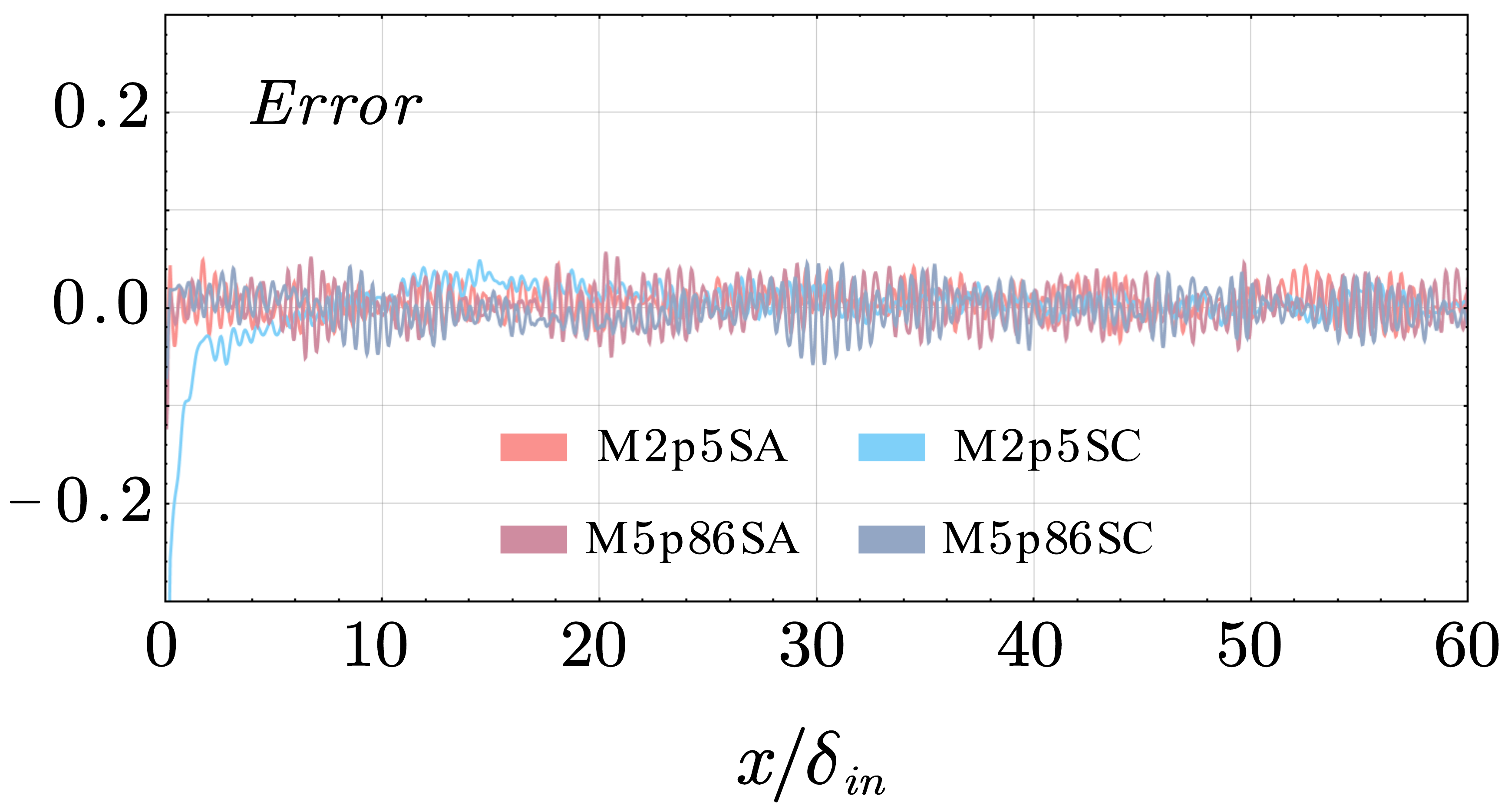}
\end{minipage}%
\hspace{0.04\textwidth}%
\begin{minipage}[t]{0.475\textwidth}
\hspace*{-1.2em}(\textit{b})\par
\vspace{-0.15em}
\centering
\includegraphics[width=\linewidth]{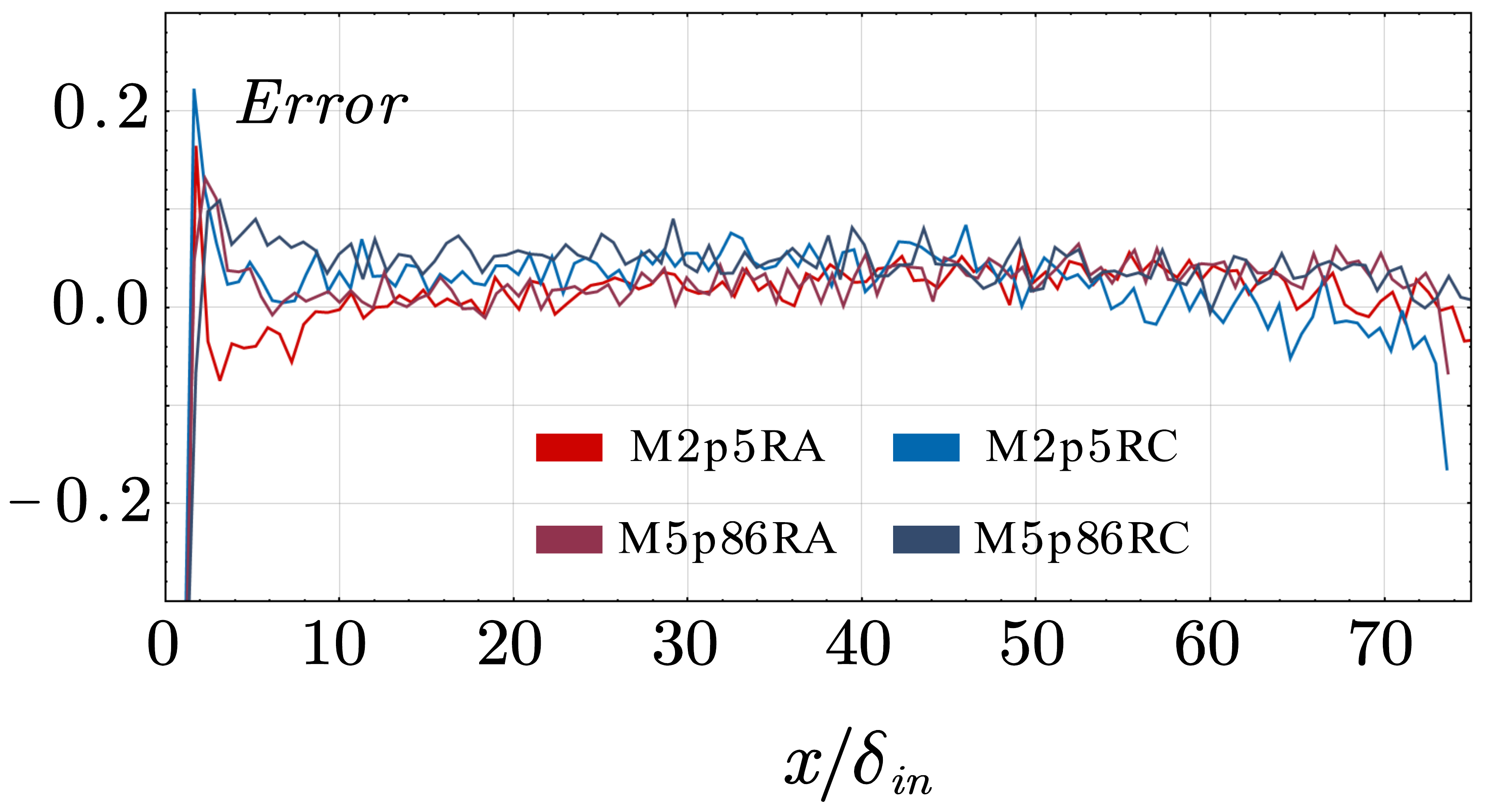}
\end{minipage}

\caption{Relative errors of the generalized Renard--Deck (RD) decomposition for (a) smooth-wall cases and (b) rough-wall cases.}
\label{fig:GRD_Error}
\end{figure}

For both the smooth- and rough-wall cases, the drag is evaluated from the right-hand side of equation \eqref{eq:RD_generalized_final}. For the smooth-wall cases, the integration domain has a streamwise extent of $0.3\delta_{in}$, as illustrated in figure \ref{fig:RDField}(a), whereas for the rough-wall cases the integration domain spans one roughness unit. To assess the accuracy of the generalized RD decomposition, the relative error, denoted by $\mathrm{Error}$, is defined between the drag obtained from equation \eqref{eq:RD_generalized_final} and that calculated directly by integrating the surface traction over the wall. As shown in figure \ref{fig:GRD_Error}, the maximum error for the smooth-wall cases is of the order of $5\%$, while that for the rough-wall cases remains below $10\%$. Moreover, the streamwise-averaged error over the target region is below $5\%$ for all cases. The close agreement between the reconstructed and directly evaluated drag demonstrates that the generalized RD decomposition provides a reliable evaluation of the total drag in the present flows. The remaining discrepancies may arise from two sources. First, higher-order correlations involving temperature-induced viscosity fluctuations are neglected in the derivation to retain the dominant contributions to the mean-flow energy balance. \cite{li2019decomposition} reported that neglecting these terms introduces errors of approximately $0.3$--$0.9\%$. Second, the relatively larger residual errors in the rough-wall cases may arise from differences between the numerical operators used in the DNS and those employed in post-processing. Specifically, nonlinear reconstruction is applied to the instantaneous convective fluxes in the DNS, whereas the derivative terms in the RD decomposition are evaluated using linear finite differences on time-averaged fields. Since nonlinear reconstruction and temporal averaging do not generally commute, discrepancies may arise in the discrete balances, particularly near roughness elements where strong velocity gradients and recirculating flows are present.

For the subsequent analysis of the individual drag contributions and their wall-normal distributions, the generalized RD decomposition is first evaluated separately for each roughness unit. To increase the number of statistical samples and obtain smoother and more reliable results, the resulting fractional contributions and outer-normalized wall-normal distributions, to be defined in equation \eqref{eq:def_ciStar}, are then averaged over 15 roughness units distributed symmetrically about the target location. In addition, since the present configurations are homogeneous in the spanwise direction, the spanwise-inhomogeneity term, $SI$, vanishes and is therefore omitted from the following analysis.

\begin{table}
    \centering
    \setlength{\tabcolsep}{5pt}
    \begin{tabular}{lcccccccc}

        Case & $SD$ & $ND$ & $AC$ & $DT$ & $CT$ & $Error$ & $DV$ & $DR$ \\[3pt]

        M2p5SA  & -0.2 &  0.3 &  0.3 & 89.1 & 10.7 & 0.2 & 40.8 & 48.3 \\
        M2p5SC  & -0.3 &  0.3 &  0.0 & 90.1 & 10.3 & 0.5 & 43.0 & 47.1 \\
        M2p5RA  &  1.9 & -0.1 &  0.2 & 88.4 & 13.7 & 3.3 &  6.6 & 81.7 \\
        M2p5RC  &  1.6 & -0.2 & -0.3 & 87.2 & 16.5 & 4.8 & 10.2 & 77.0 \\[3pt]
        M5p86SA & -0.3 &  0.5 &  0.5 & 90.8 &  8.8 & 0.4 & 43.3 & 47.5 \\
        M5p86SC & -0.5 &  0.5 &  0.5 & 89.7 & 10.0 & 0.1 & 44.6 & 45.1 \\
        M5p86RA &  1.1 &  0.8 &  0.3 & 87.7 & 13.2 & 3.1 &  7.0 & 80.7 \\
        M5p86RC &  2.3 &  1.5 & -0.2 & 89.5 &  9.7 & 2.7 & 11.8 & 77.7 \\

    \end{tabular}
    \caption{Relative contributions of the terms in the generalized Renard--Deck drag decomposition at target streamwise locations $x_t$. All values are expressed as percentages of the total drag. The contributions of $D_{\nu}$ and $D_R$ correspond to the two components of the dissipation term $DT$.}
    \label{tab:GRD_decomposition}
\end{table}

As summarized in table \ref{tab:GRD_decomposition}, the contributions of $SD$, $ND$ and $AC$ remain small for both the smooth- and rough-wall cases, indicating that streamwise development, wall-normal displacement and the additional compressibility effects make only minor contributions to the total drag. The decomposition is instead dominated by the dissipation term $DT$ and the convection term $CT$, whose relative contributions are only moderately affected by surface roughness. The $DT$ term accounts for approximately $90\%$ of the total drag, establishing a direct connection, within the generalized RD framework, between boundary-layer drag and the combined effects of mean-flow viscous dissipation and turbulent kinetic-energy production. To further compare the rough-wall cases with a smooth-wall flow at a comparable friction Reynolds number, the $Ma=2.0$, $Re_\tau\approx1000$ case of \cite{pirozzoli2011turbulence} is also considered. To perform the drag decomposition from a single wall-normal profile, an approximate treatment is required. Since the contributions associated with streamwise derivatives in the dissipation term are negligible for the present smooth-wall cases, these contributions are neglected for the reference case. The resulting $DT$ contribution is $89.7\%$ of the total drag, with $37.6\%$ arising from mean-flow viscous dissipation and $52.2\%$ from turbulent kinetic-energy production. This partition is consistent with the dominant role of $DT$ observed in the present cases. The $CT$ term contributes approximately $10\%$ and represents the streamwise change in the convective transport of mean kinetic energy associated with the development of the boundary layer. At $Ma=2.5$, roughness leads to an increase in the relative contribution of $CT$ compared with the corresponding smooth-wall cases. At $Ma=5.86$, however, no consistent roughness-induced trend is observed, with the adiabatic rough-wall case exhibiting a larger $CT$ contribution than its smooth-wall counterpart, whereas the two cold-wall cases remain comparable.

The dominant energetic pathway associated with the drag work can be further examined by decomposing the $DT$ term into the mean-flow viscous dissipation, $DV$, and the turbulent kinetic-energy production, $D_R$. For the smooth-wall cases at $Re_\tau\approx450$ and $510$, the two contributions are of comparable magnitude, each accounting for approximately $40$--$50\%$ of the total drag. The contribution of $D_V$ is slightly smaller than that of $D_R$, and the difference between them decreases as the wall temperature is reduced. Comparison with the $Re_\tau=1000$ smooth-wall case of \cite{pirozzoli2011turbulence} further indicates that, for smooth-wall flows, increasing the friction Reynolds number reduces the relative contribution of mean-flow viscous dissipation while increasing that of turbulent kinetic-energy production, consistent with the trend reported by \cite{li2019decomposition}. A similar Reynolds-number trend is observed for the rough-wall cases at $Re_\tau\approx960$--$1050$, for which $DV$ decreases and $DR$ increases. However, the partition is markedly shifted towards turbulent kinetic-energy production. In all rough-wall cases, $DR$ accounts for approximately $80\%$ of the total drag, substantially exceeding the corresponding smooth-wall values. This effect is particularly evident for the adiabatic rough-wall cases, where the contribution of $DV$ is only about $7\%$. These results indicate that, for the present spanwise square-bar roughness, most of the drag work is transferred from the mean flow to the roughness-induced turbulent motions rather than being directly dissipated through the mean velocity gradients.

The wall-normal distributions of the individual contributions in the generalized RD decomposition can be obtained by integrating the corresponding integrands in equation \eqref{eq:RD_generalized_final} over the plane at each wall-normal location. Denoting the wall-normal distribution of the $i$th contribution by $\mathcal{C}_i(y)$, the volume-integral terms (e.g. the $ND$, $AC$ and $DT$ contributions) are evaluated as
\begin{equation}
\mathcal{C}_i(y)=\iint_{A_y}\mathcal{Q}_i(x,y,z)\,\mathrm{d}x\,\mathrm{d}z
\end{equation}
where $\mathcal{Q}_i$ represents the corresponding local integrand and $A_y$ denotes the fluid cross-sectional area in the $x$--$z$ plane at a given wall-normal location $y$. For contributions associated with the upstream and downstream boundaries (e.g. the $CT$ contribution), the wall-normal distribution evaluated from
\begin{equation}
\mathcal{C}_i(y)=\left.\int_{S_y}\mathcal{Q}_{i}(y,z)\,\mathrm{d}z\right|_L^R
\end{equation}
where $S_y$ denotes the spanwise fluid extent at the corresponding upstream and downstream sections. The generalized RD relation satisfies
\begin{equation}
U_\infty F_D=\sum_i \int \mathcal{C}_i(y)\,\mathrm{d}y
\end{equation}

This formulation allows the wall-normal locations responsible for the different drag-work contributions to be examined directly. To facilitate comparison among cases with different drag levels and boundary-layer thicknesses, the wall-normal distributions are normalized by the characteristic drag-work density $U_\infty F_D/\delta$. Accordingly, the normalized distribution of each term is defined as
\begin{equation}
\mathcal{C}_i^{*}(y)=\frac{\delta\,\mathcal{C}_i(y)}{U_\infty F_D}
\label{eq:def_ciStar}
\end{equation}
consequently
\begin{equation}
1=\sum_i\int \mathcal{C}_i^{*}(y)\,\mathrm{d}\left(\frac{y}{\delta}\right)
\end{equation}
The area under each normalized profile gives the relative contribution of the corresponding term to the total drag work. Its distribution along $y^o=y/\delta$ indicates where this contribution is concentrated in the wall-normal direction.

\begin{figure}
    \centering

    \begin{minipage}[t]{0.318\textwidth}
        \makebox[0pt][l]{\hspace*{-1em}(\textit{a})}\par\vspace{-0.15em}
        \makebox[\linewidth][c]{%
            \includegraphics[width=\textwidth]{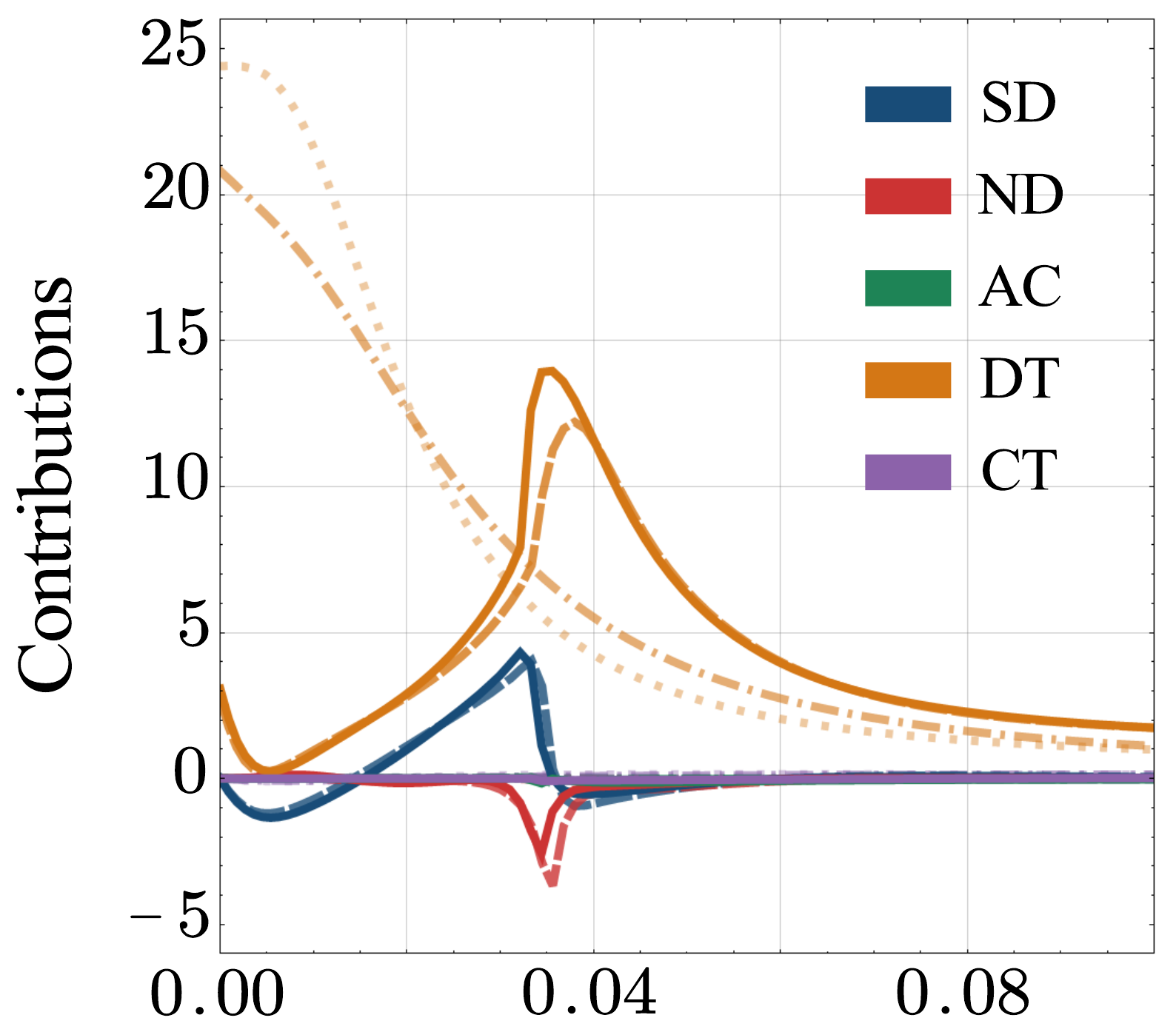}%
        }
    \end{minipage}
    \hspace{0.01\textwidth}
    \begin{minipage}[t]{0.581\textwidth}
        \makebox[0pt][l]{\hspace*{-0.4em}(\textit{b})}\par\vspace{-0.15em}
        \makebox[\linewidth][c]{%
            \includegraphics[width=\textwidth]{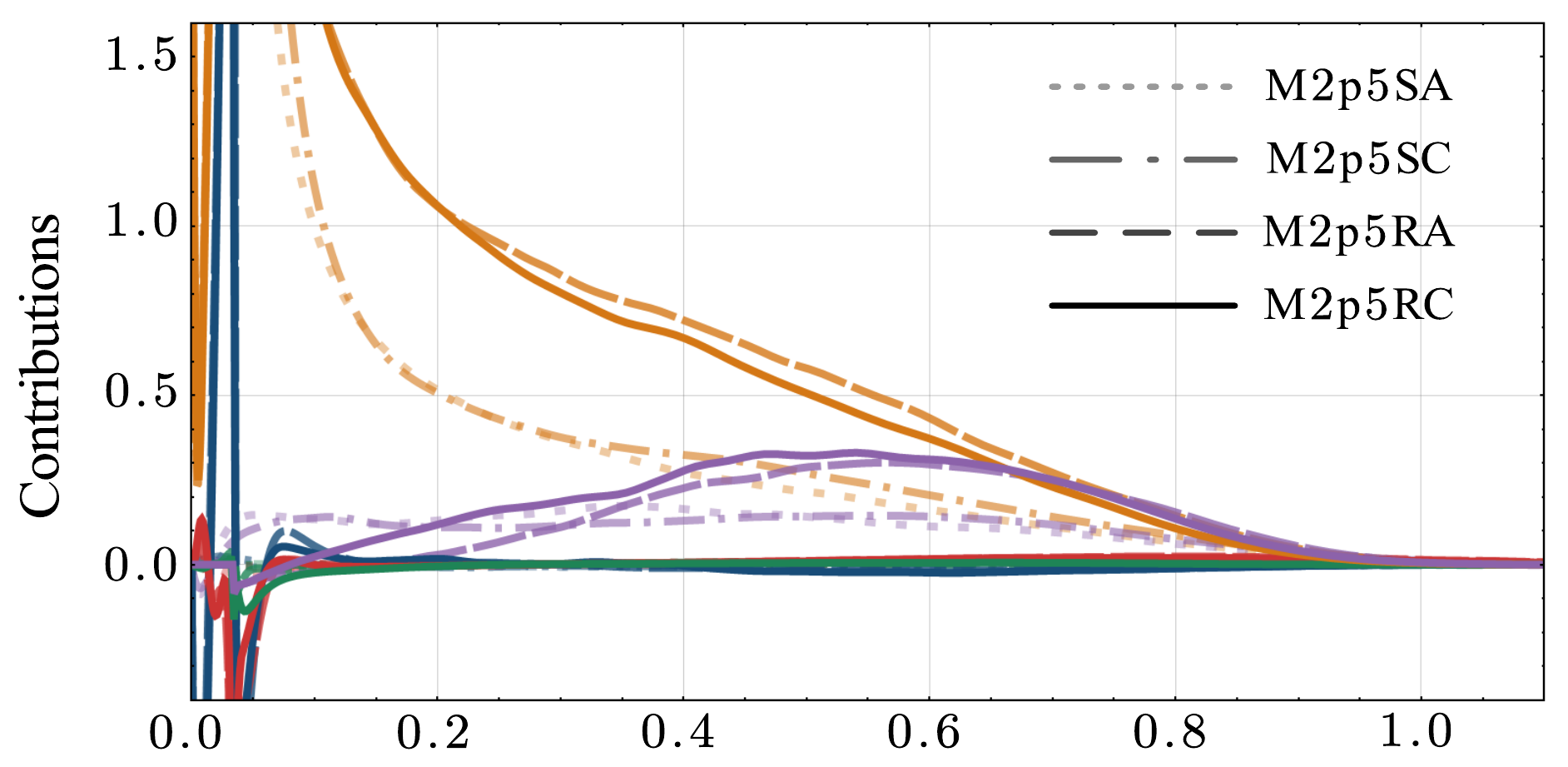}%
        }
    \end{minipage}

    \par\vspace{0.2cm}

    \begin{minipage}[t]{0.318\textwidth}
        \makebox[0pt][l]{\hspace*{-1em}(\textit{c})}\par\vspace{-0.15em}
        \makebox[\linewidth][c]{%
            \includegraphics[width=\textwidth]{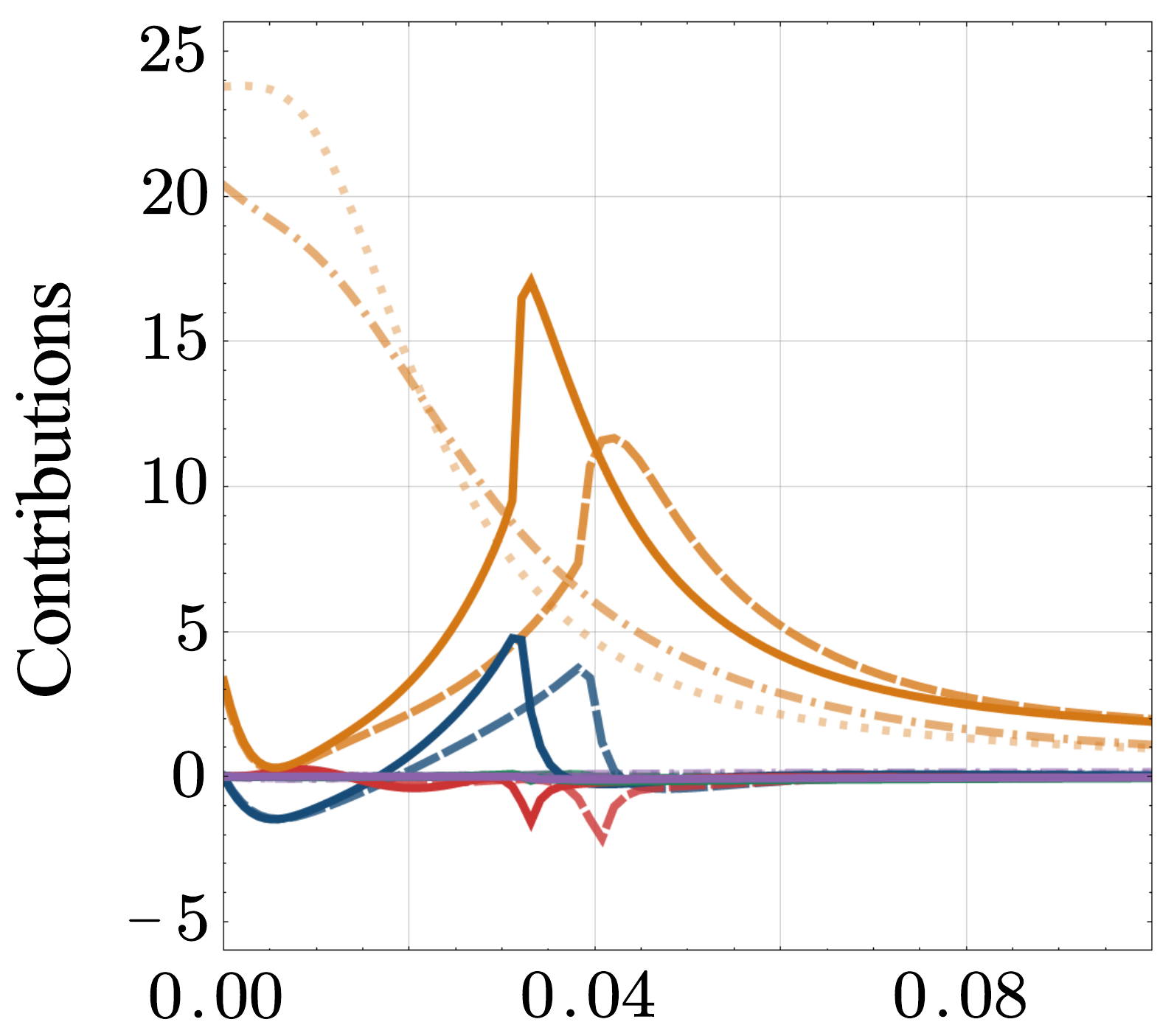}%
        }
    \end{minipage}
    \hspace{0.01\textwidth}
    \begin{minipage}[t]{0.581\textwidth}
        \makebox[0pt][l]{\hspace*{-0.4em}(\textit{d})}\par\vspace{-0.15em}
        \makebox[\linewidth][c]{%
            \includegraphics[width=\textwidth]{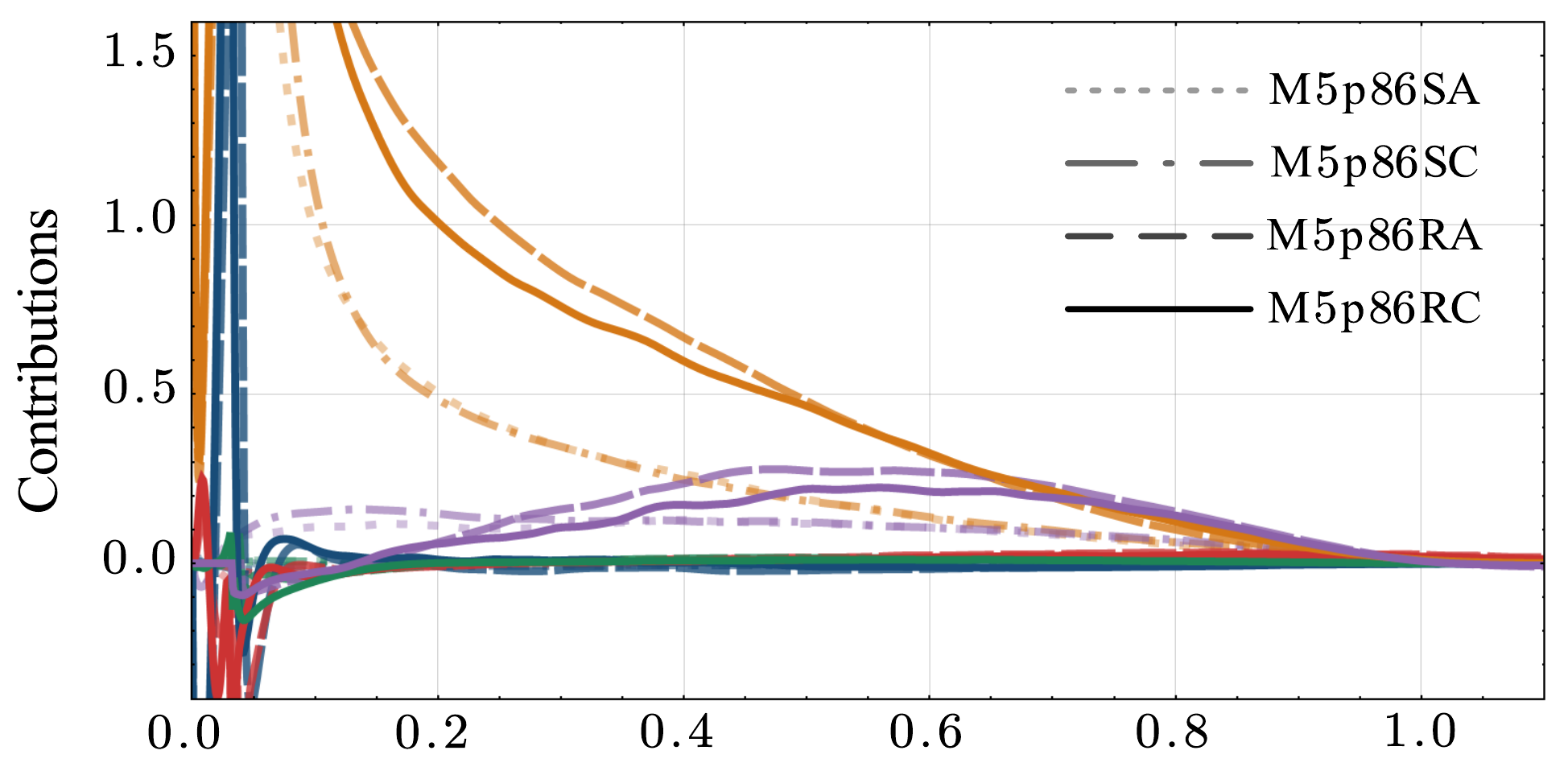}%
        }
    \end{minipage}

    \par\vspace{0.2cm}

    \begin{minipage}[t]{0.318\textwidth}
        \makebox[0pt][l]{\hspace*{-1em}(\textit{e})}\par\vspace{-0.15em}
        \makebox[\linewidth][c]{%
            \includegraphics[width=\textwidth]{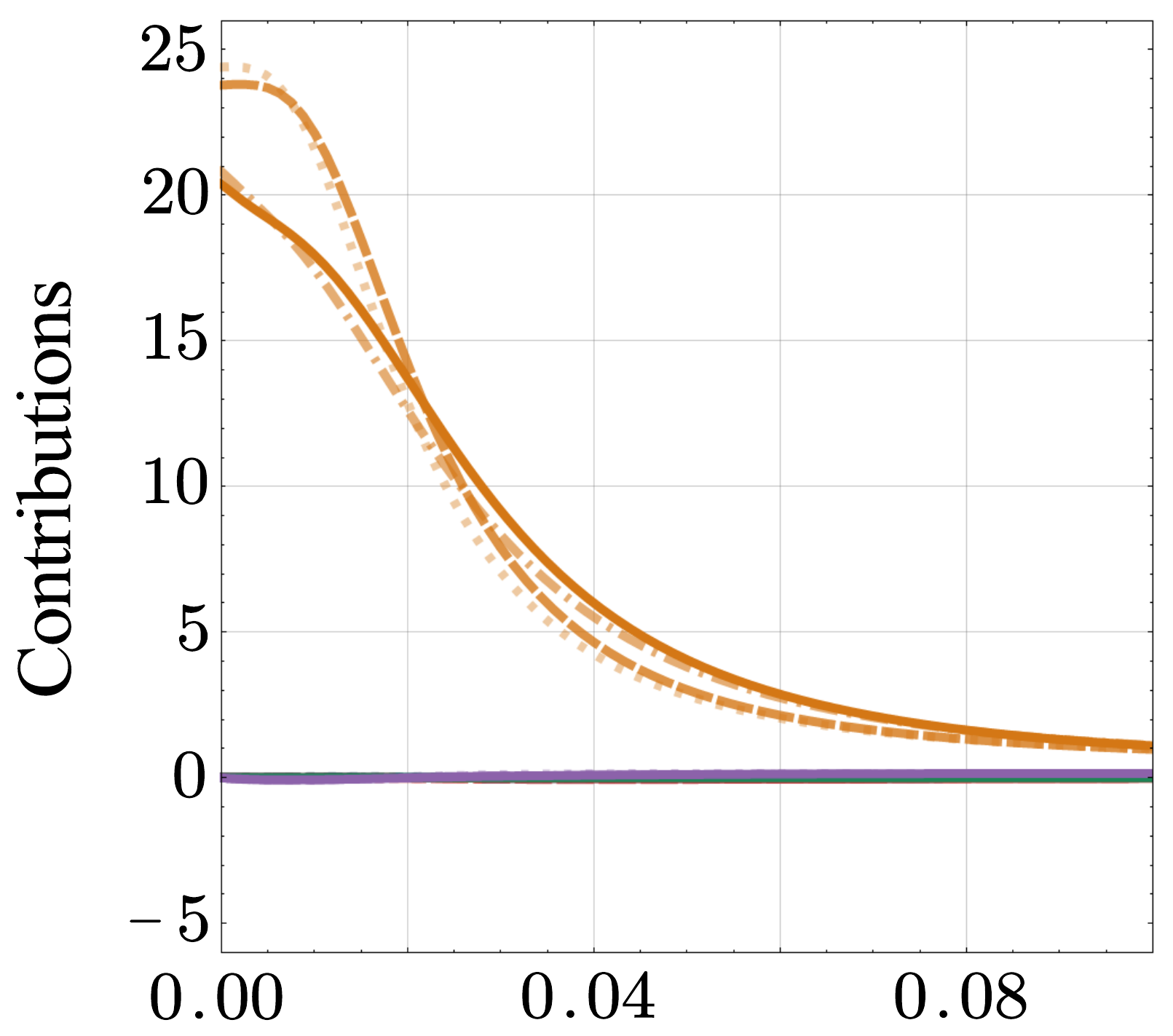}%
        }
    \end{minipage}
    \hspace{0.01\textwidth}
    \begin{minipage}[t]{0.581\textwidth}
        \makebox[0pt][l]{\hspace*{-0.4em}(\textit{f})}\par\vspace{-0.15em}
        \makebox[\linewidth][c]{%
            \includegraphics[width=\textwidth]{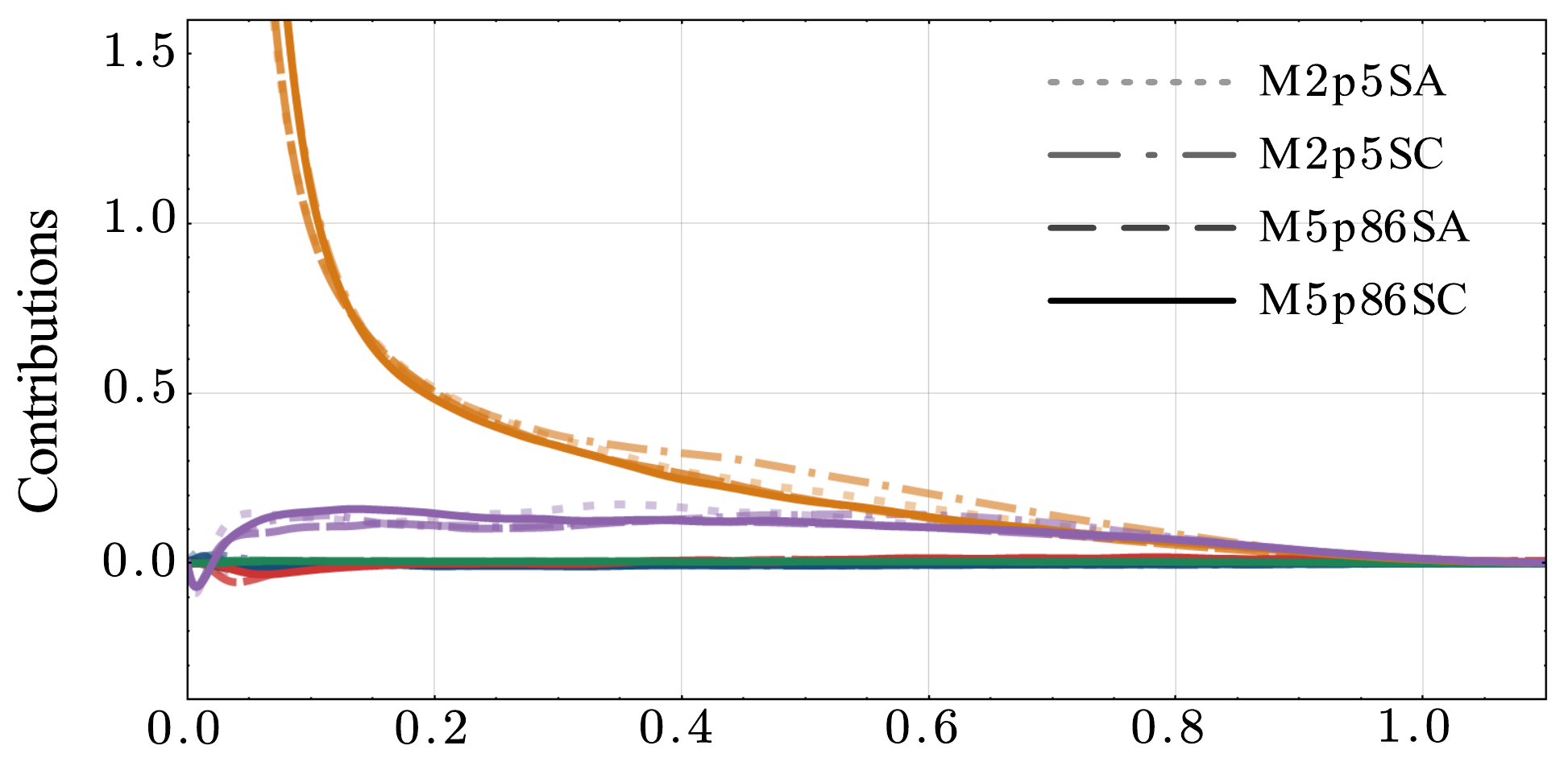}%
        }
    \end{minipage}

    \par\vspace{0.2cm}

    \begin{minipage}[t]{0.318\textwidth}
        \makebox[0pt][l]{\hspace*{-1em}(\textit{g})}\par\vspace{-0.15em}
        \makebox[\linewidth][c]{%
            \includegraphics[width=\textwidth]{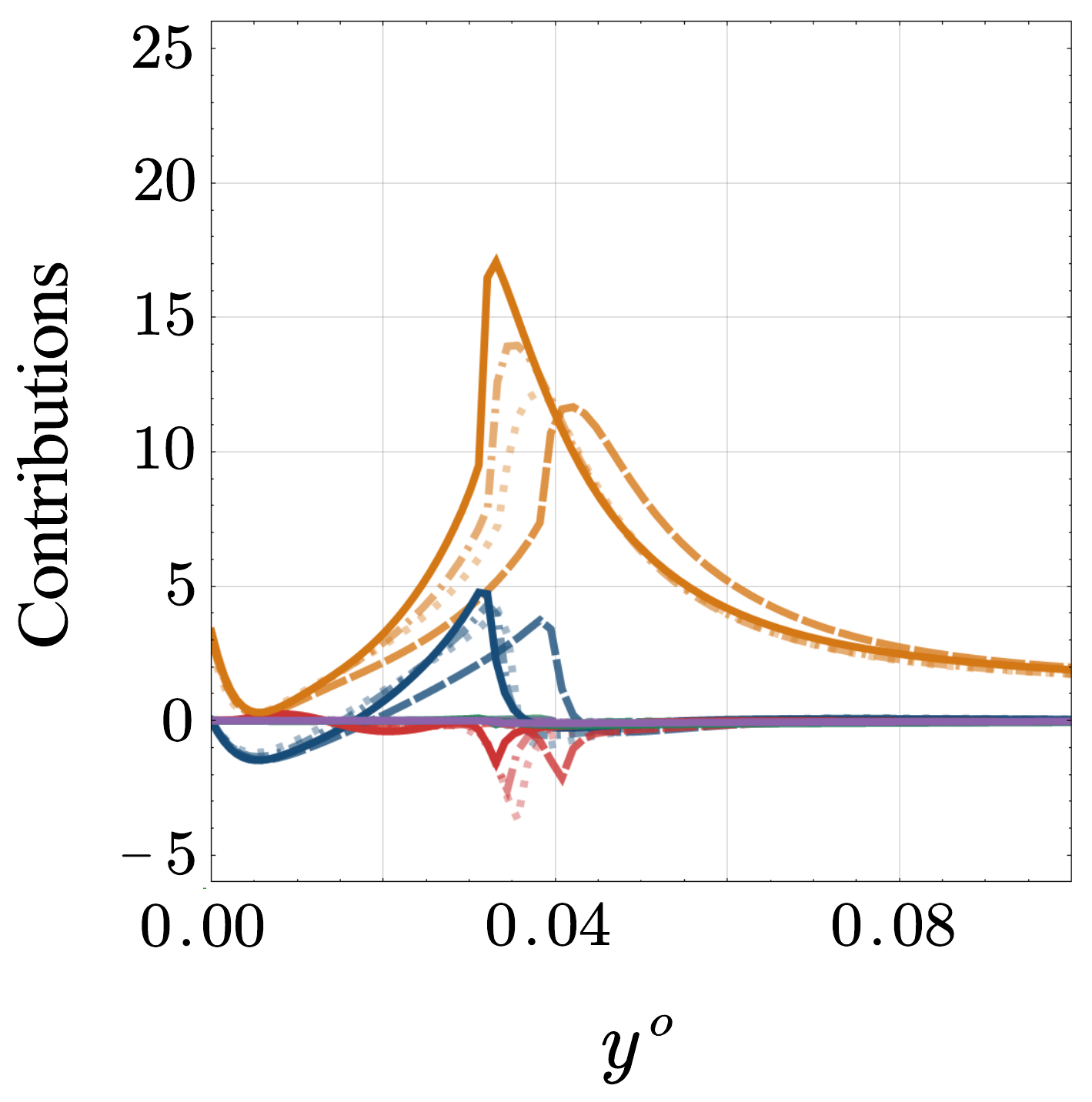}%
        }
    \end{minipage}
    \hspace{0.01\textwidth}
    \begin{minipage}[t]{0.581\textwidth}
        \makebox[0pt][l]{\hspace*{-0.4em}(\textit{h})}\par\vspace{-0.15em}
        \makebox[\linewidth][c]{%
            \includegraphics[width=\textwidth]{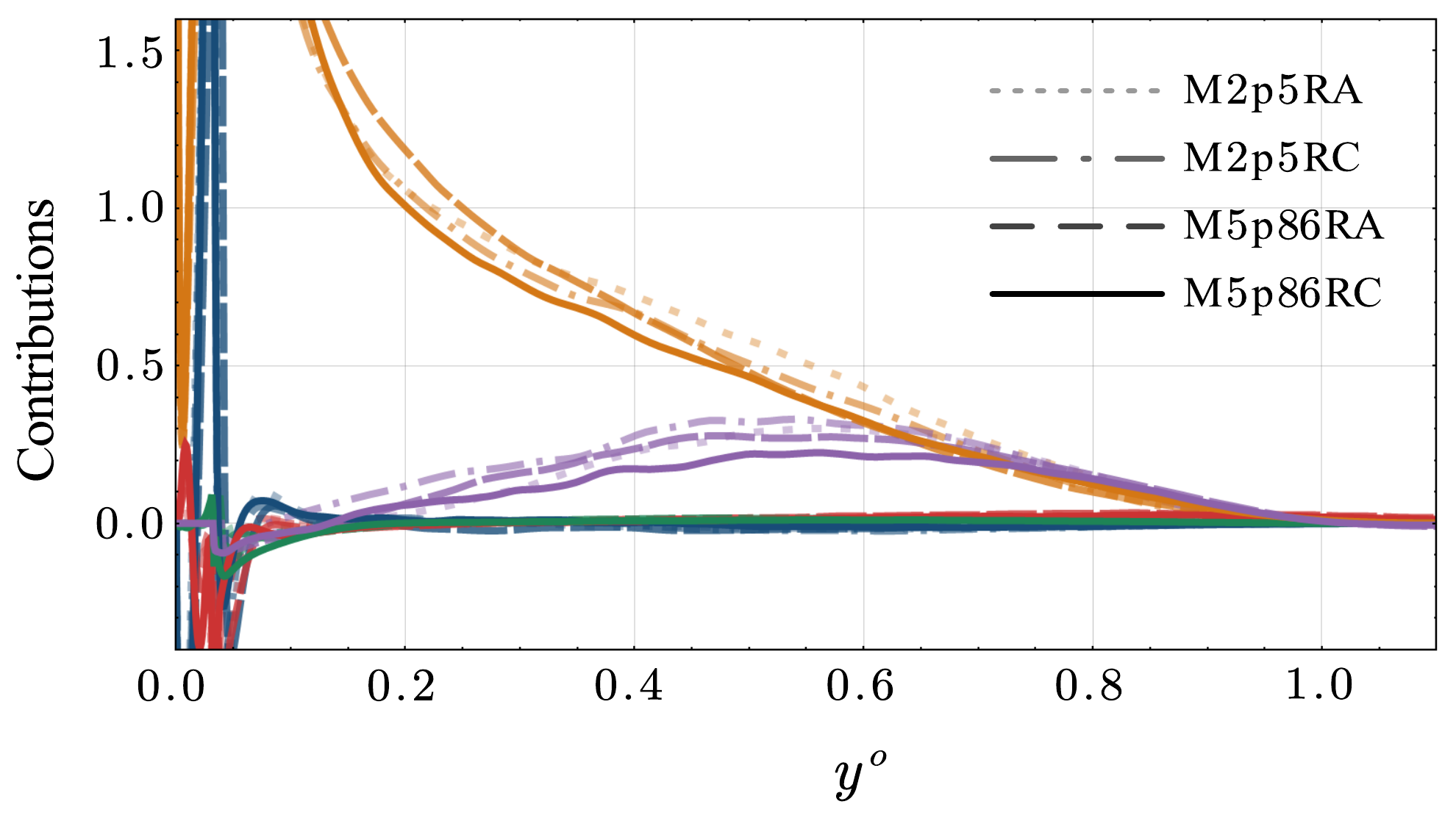}%
        }
    \end{minipage}

    \caption{Spatial distributions of the drag-decomposition terms, where ``Contributions'' denotes the normalized contribution $\mathcal{C}_i^*$ defined by equation \eqref{eq:def_ciStar}. Each row compares the same four cases, with the left panel showing the near-wall distribution and the right panel showing the corresponding outer-layer distribution.}
    \label{fig:GRD_ALL}
\end{figure}

We first examine the near-wall distributions of the individual drag-decomposition terms. For the smooth-wall cases, the dissipation term $DT$ dominates in the near-wall region, as shown in figure \ref{fig:GRD_ALL}(e). Its maximum occurs at the wall and the contribution decreases progressively with increasing wall-normal distance. The near-wall behaviour of the rough-wall cases is more complex, as shown in figure \ref{fig:GRD_ALL}(g). Similar to the smooth-wall cases, $DT$ remains the dominant contribution near the wall. However, only a relatively weak dissipation peak appears at the bottom plane, $y=0$. Within the roughness layer, the dissipation first decreases and then increases with increasing $y$, forming a second and much stronger peak near the roughness crest. For the smooth-wall cases, the remaining terms are negligible in the near-wall region compared with $DT$. In contrast, the rough-wall cases exhibit distinct structures of the streamwise-development term $SD$ below the roughness crest. This contribution mainly originates from the second term in the $SD$ component of equation \eqref{eq:RD_generalized_final}. As illustrated in figure \ref{fig:dragField}, the recirculating flow within the roughness layer produces strong streamwise variations in the mean velocity, thereby enhancing this contribution. The resulting $SD$ distribution decreases initially with increasing height and subsequently increases, reaching its maximum near the roughness crest. The normal-displacement term $ND$ also becomes appreciable in the rough-wall cases. As the flow is deflected around the roughness elements, substantial wall-normal motions are generated, and the corresponding $ND$ contribution is therefore concentrated primarily near the roughness crest. For the $DT$ term, the peak near the roughness crest is lower in the rough-wall cases than the corresponding near-wall peak in the smooth-wall cases, as shown in figure \ref{fig:GRD_ALL}(a,c). For the smooth-wall cases, the adiabatic condition exhibits a higher $DT$ peak than the cold-wall condition, whereas the opposite trend is observed for the rough-wall cases, where the peak near the roughness crest is higher under cold-wall conditions. Increasing the Mach number has little influence on the peak value for the smooth-wall cases, as shown in figure \ref{fig:GRD_ALL}(e), while the rough-wall cases exhibit a higher peak contribution at the higher Mach number, as shown in figure \ref{fig:GRD_ALL}(g).

In the outer layer, the drag-work contributions are primarily associated with the $DT$ and $CT$ terms. The $DT$ contribution decreases with increasing wall-normal distance for all cases. As shown in figure \ref{fig:GRD_ALL}(b, d), the decay is considerably faster for the smooth-wall cases than for the rough-wall cases, while the difference between the adiabatic- and cold-wall conditions remains small in the outer region. For the rough-wall cases, the $DT$ contribution decays slightly more slowly under adiabatic conditions than under cold-wall conditions. In addition, for both smooth- and rough-wall cases, the higher-Mach-number cases exhibit a slightly faster decay than their lower-Mach-number counterparts. This behavior is consistent with the observation that the overall $DT$ contributions remain comparable among the cases, whereas the near-wall peak is higher at the higher Mach number. The wall-normal distributions of the $CT$ term differ substantially between the smooth- and rough-wall cases. For the smooth-wall cases, $CT$ reaches its maximum in the inner region and then decays gradually through the outer layer. In contrast, for the rough-wall cases, the peak is shifted markedly away from the wall, occurring at approximately half of the boundary-layer thickness, and its magnitude is substantially larger than that observed for the smooth-wall cases, as shown in figure \ref{fig:GRD_ALL}(b, d).

\begin{figure}
\centering

\begin{minipage}[t]{0.428\textwidth}
\hspace*{-1.2em}(\textit{a})\par
\vspace{-0.15em}
\centering
\includegraphics[width=\linewidth]{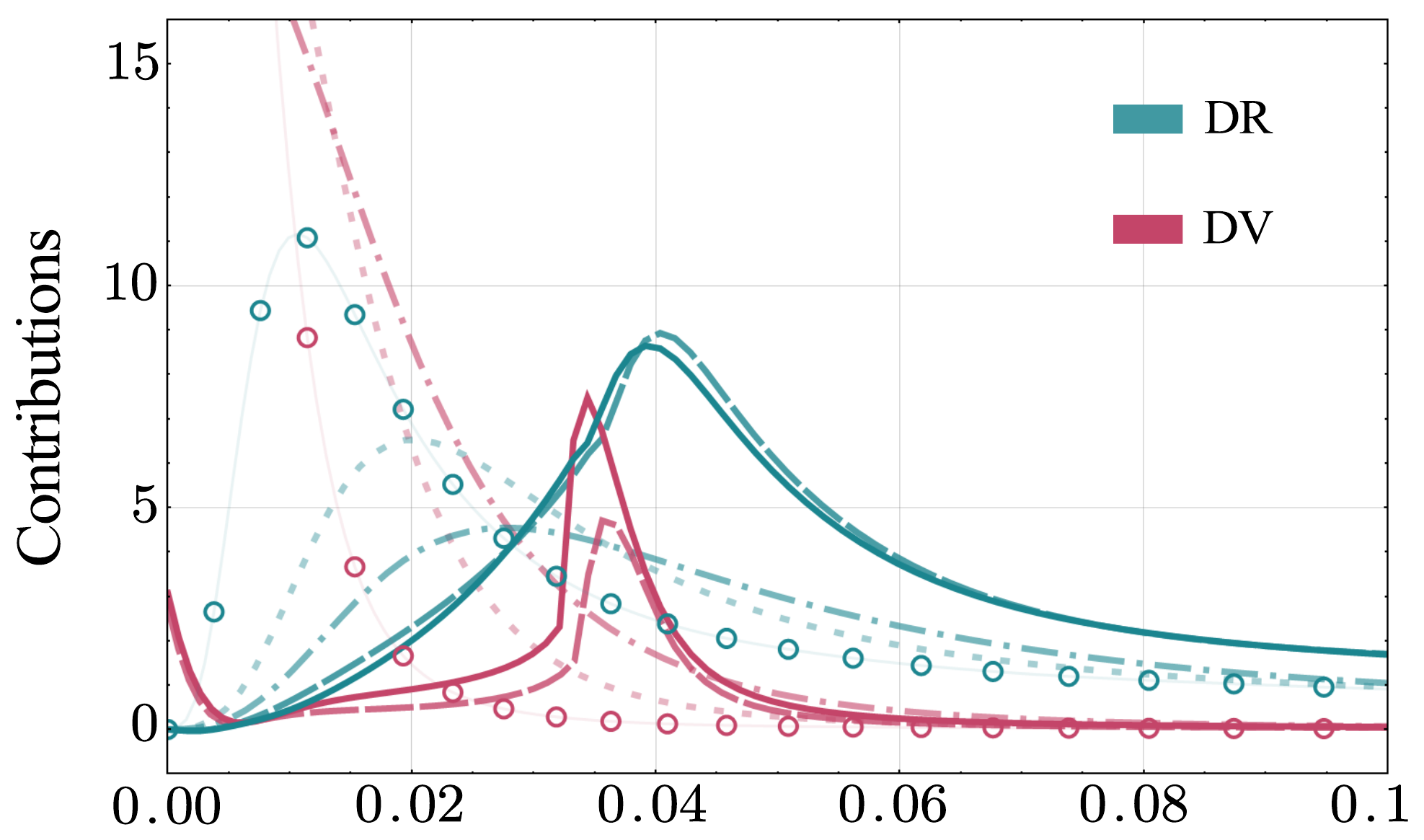}
\end{minipage}%
\hspace{0.04\textwidth}%
\begin{minipage}[t]{0.433\textwidth}
\hspace*{-1.2em}(\textit{b})\par
\vspace{-0.15em}
\centering
\includegraphics[width=\linewidth]{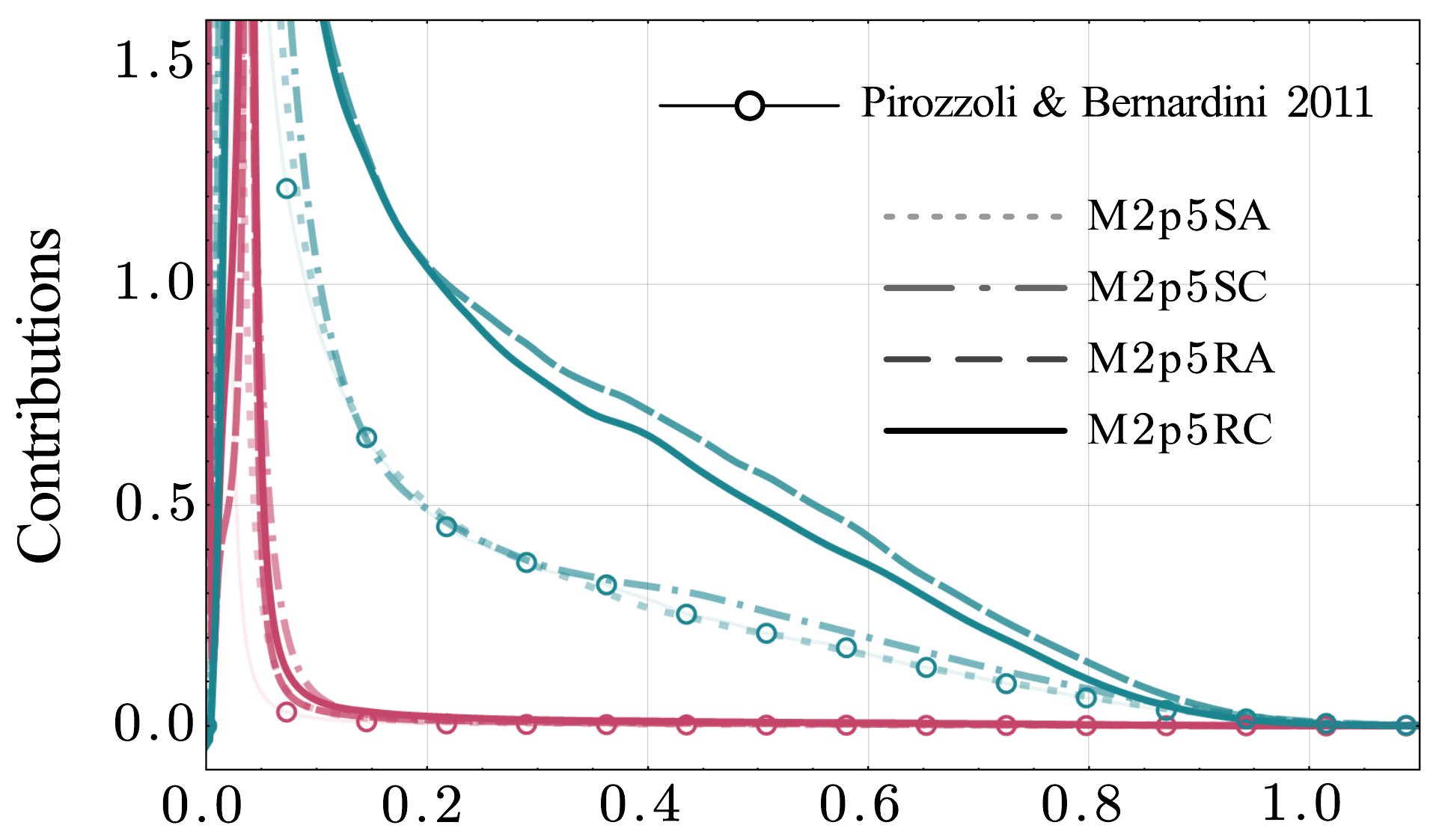}
\end{minipage}

\par\vspace{0.2cm}

\begin{minipage}[t]{0.428\textwidth}
\hspace*{-1.2em}(\textit{c})\par
\vspace{-0.15em}
\centering
\includegraphics[width=\linewidth]{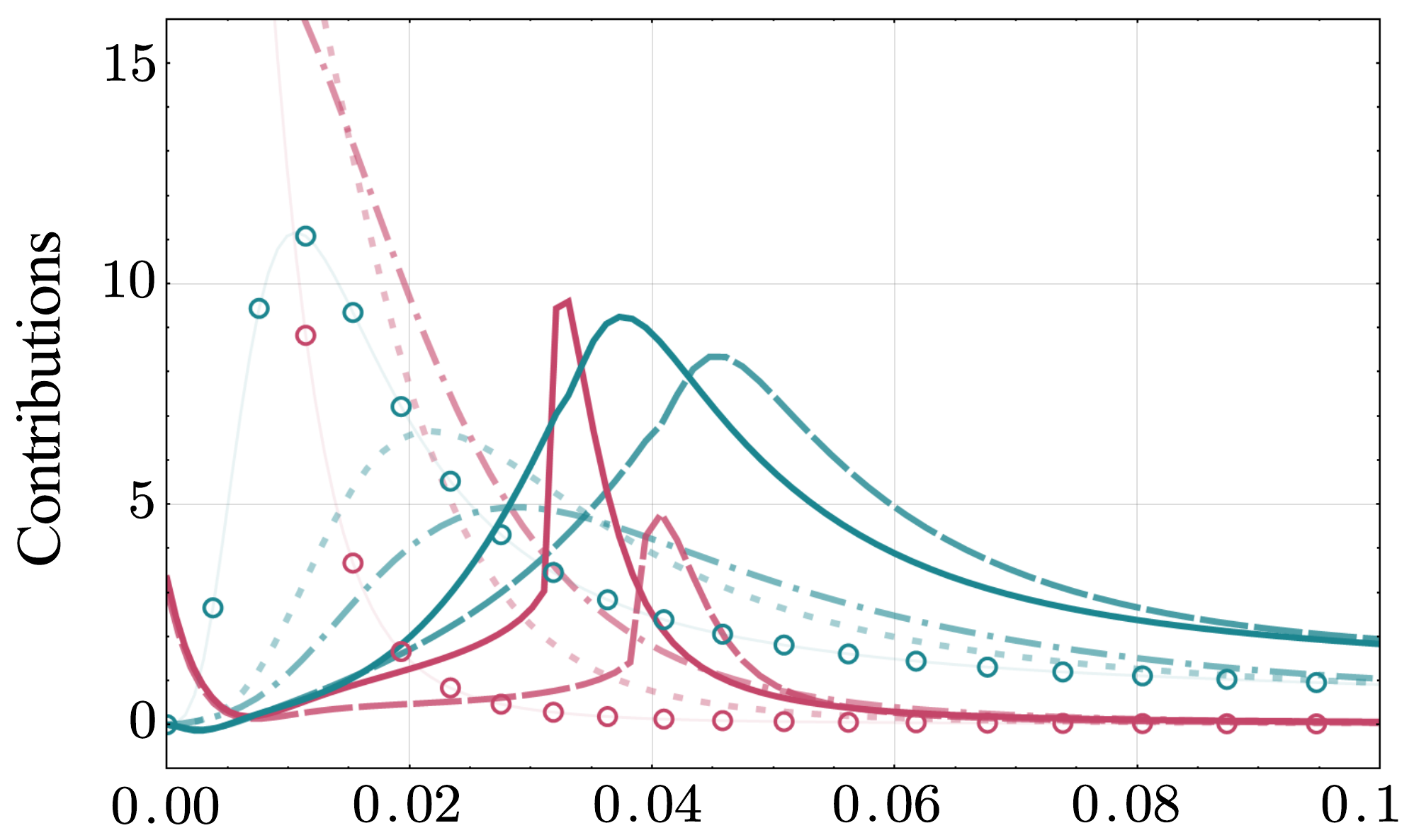}
\end{minipage}%
\hspace{0.04\textwidth}%
\begin{minipage}[t]{0.433\textwidth}
\hspace*{-1.2em}(\textit{d})\par
\vspace{-0.15em}
\centering
\includegraphics[width=\linewidth]{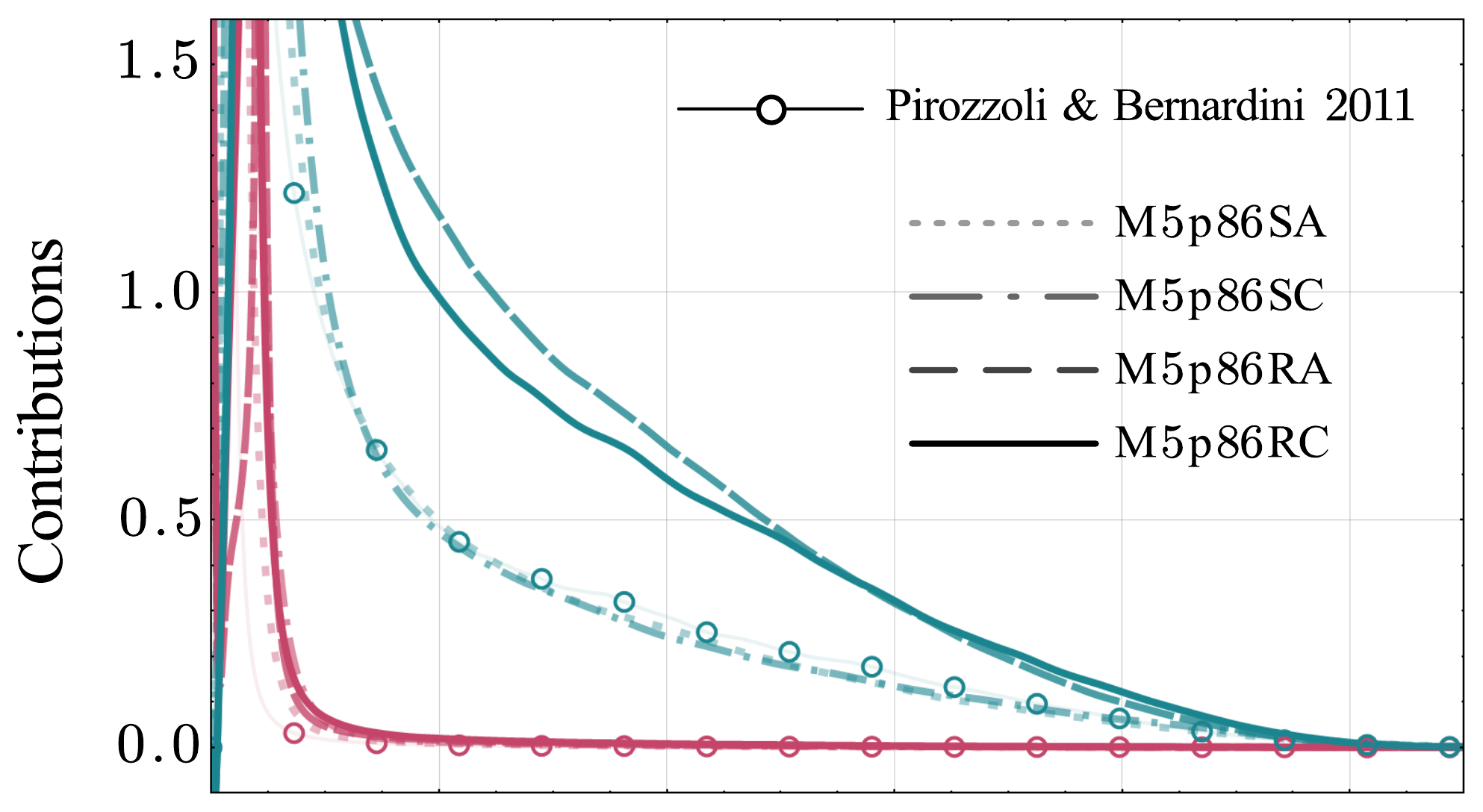}
\end{minipage}

\par\vspace{0.2cm}

\begin{minipage}[t]{0.428\textwidth}
\hspace*{-1.2em}(\textit{e})\par
\vspace{-0.15em}
\centering
\includegraphics[width=\linewidth]{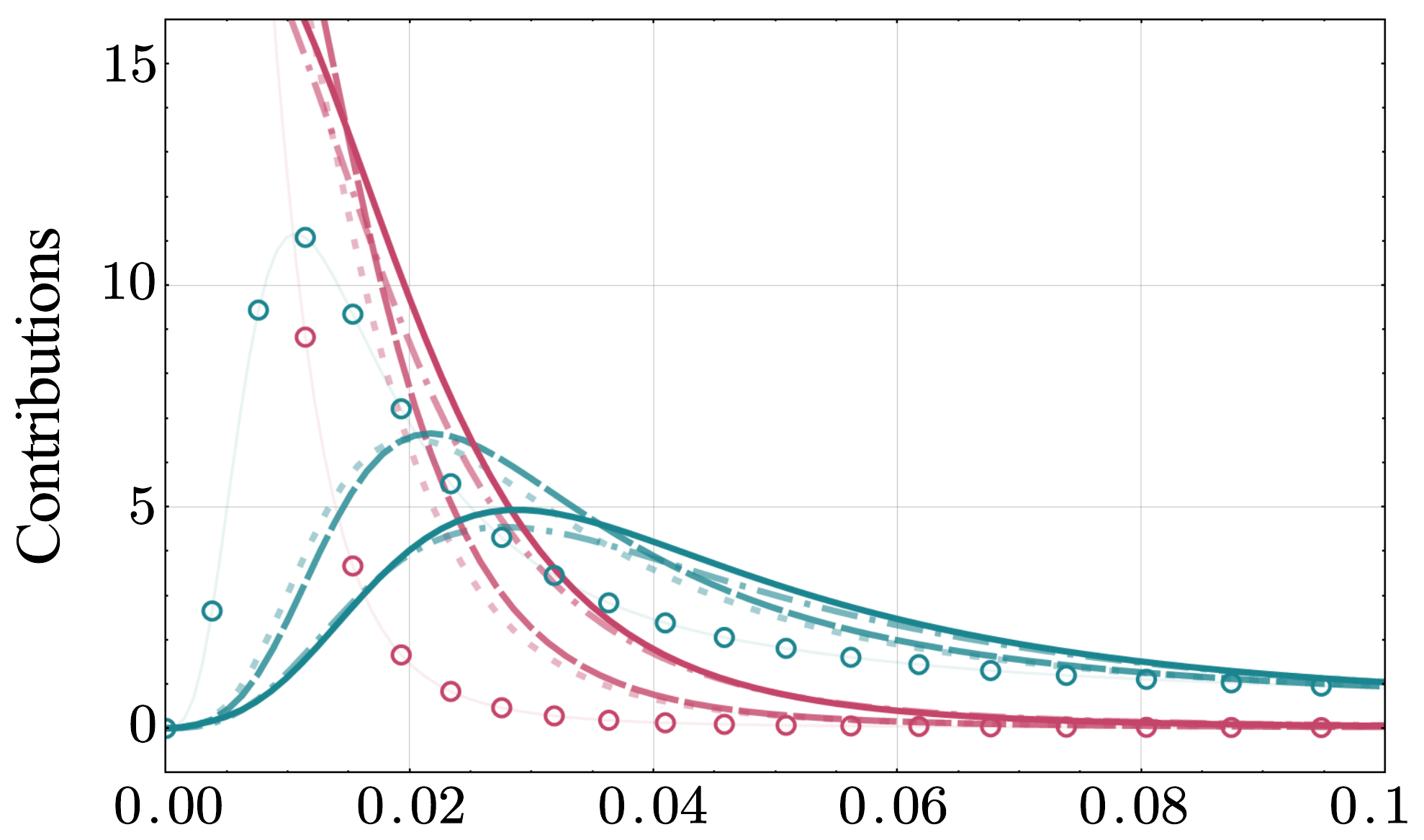}
\end{minipage}%
\hspace{0.04\textwidth}%
\begin{minipage}[t]{0.433\textwidth}
\hspace*{-1.2em}(\textit{f})\par
\vspace{-0.15em}
\centering
\includegraphics[width=\linewidth]{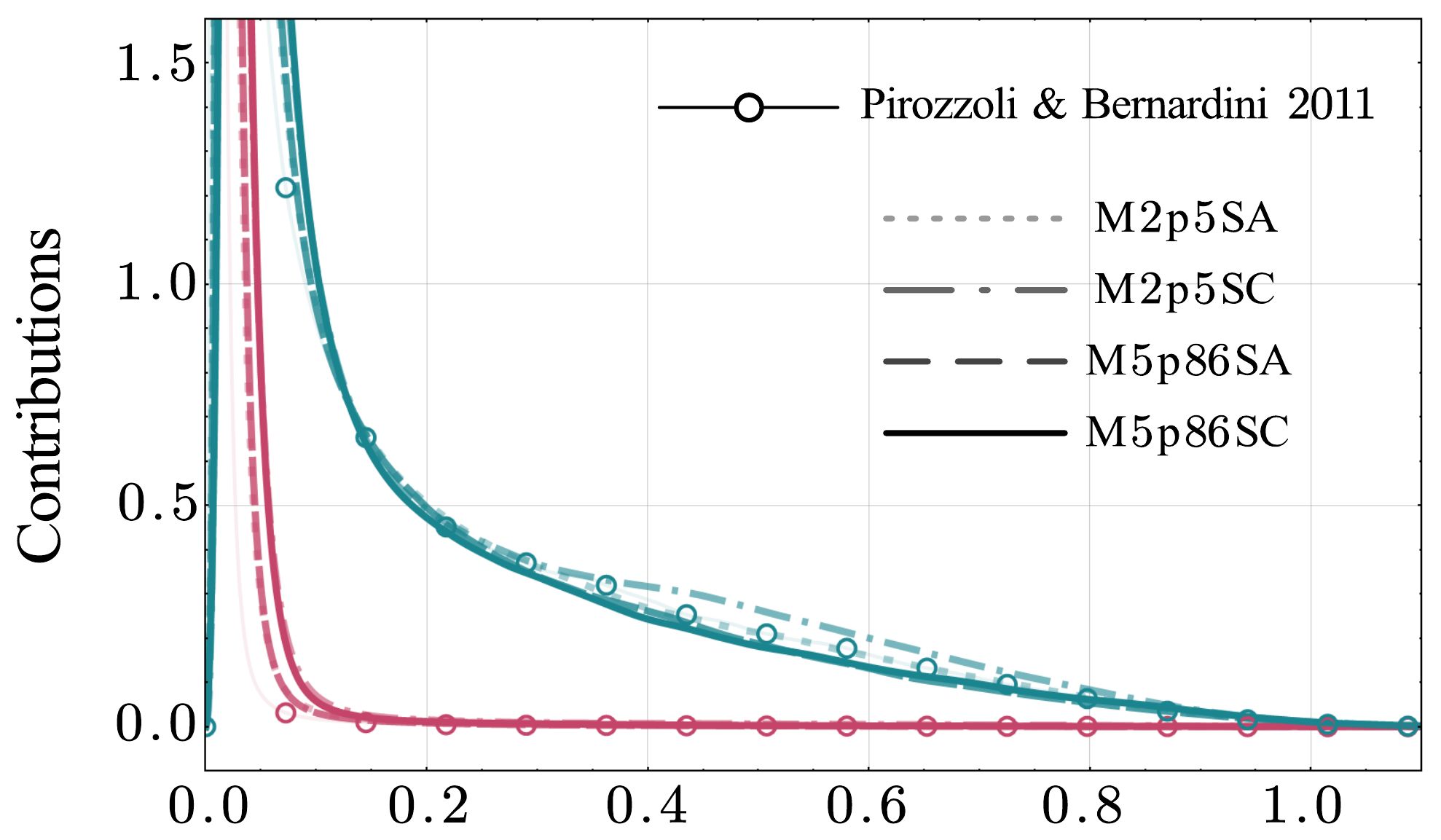}
\end{minipage}

\par\vspace{0.2cm}

\begin{minipage}[t]{0.428\textwidth}
\hspace*{-1.2em}(\textit{g})\par
\vspace{-0.15em}
\centering
\includegraphics[width=\linewidth]{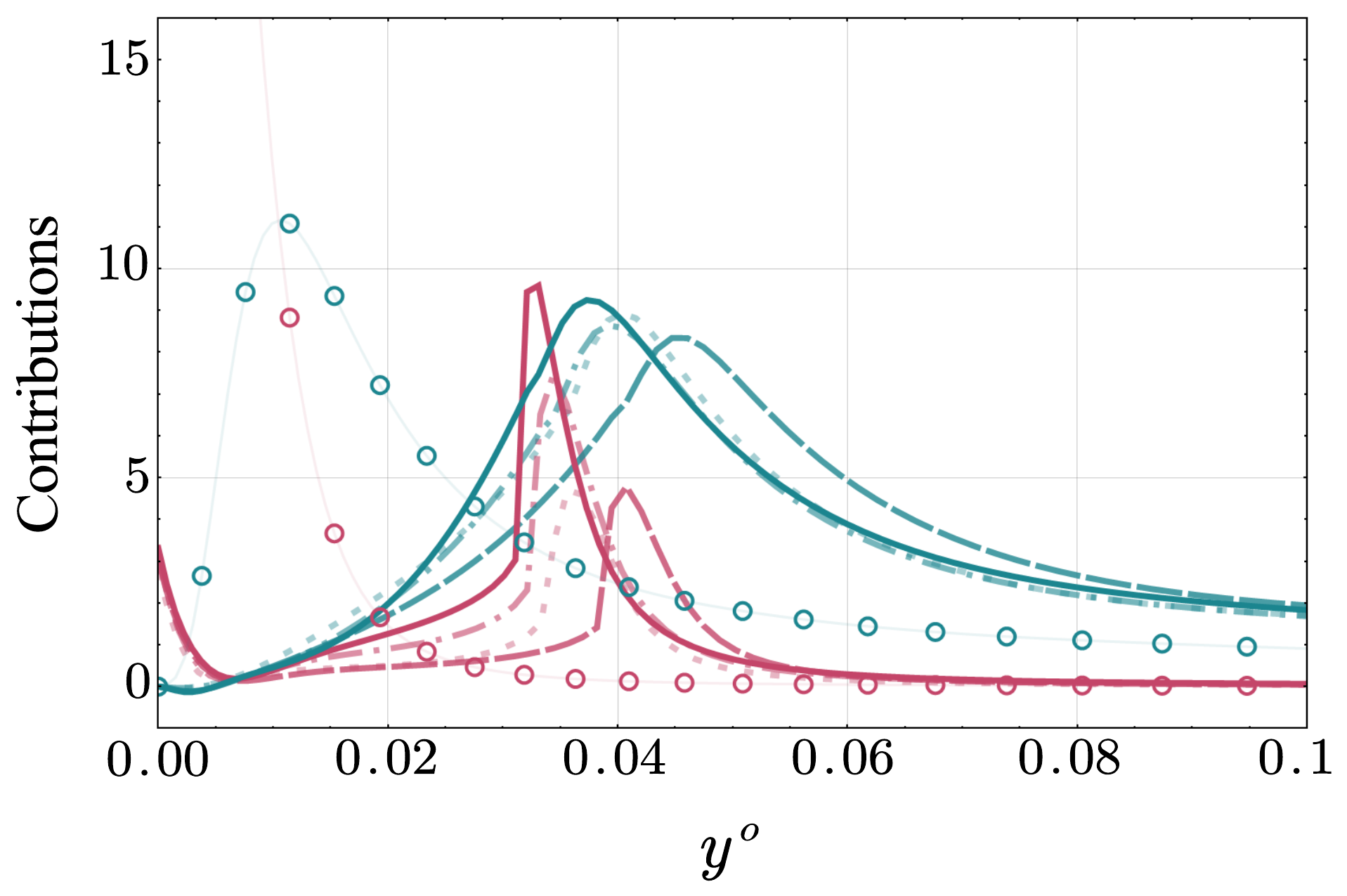}
\end{minipage}%
\hspace{0.04\textwidth}%
\begin{minipage}[t]{0.433\textwidth}
\hspace*{-1.2em}(\textit{h})\par
\vspace{-0.15em}
\centering
\includegraphics[width=\linewidth]{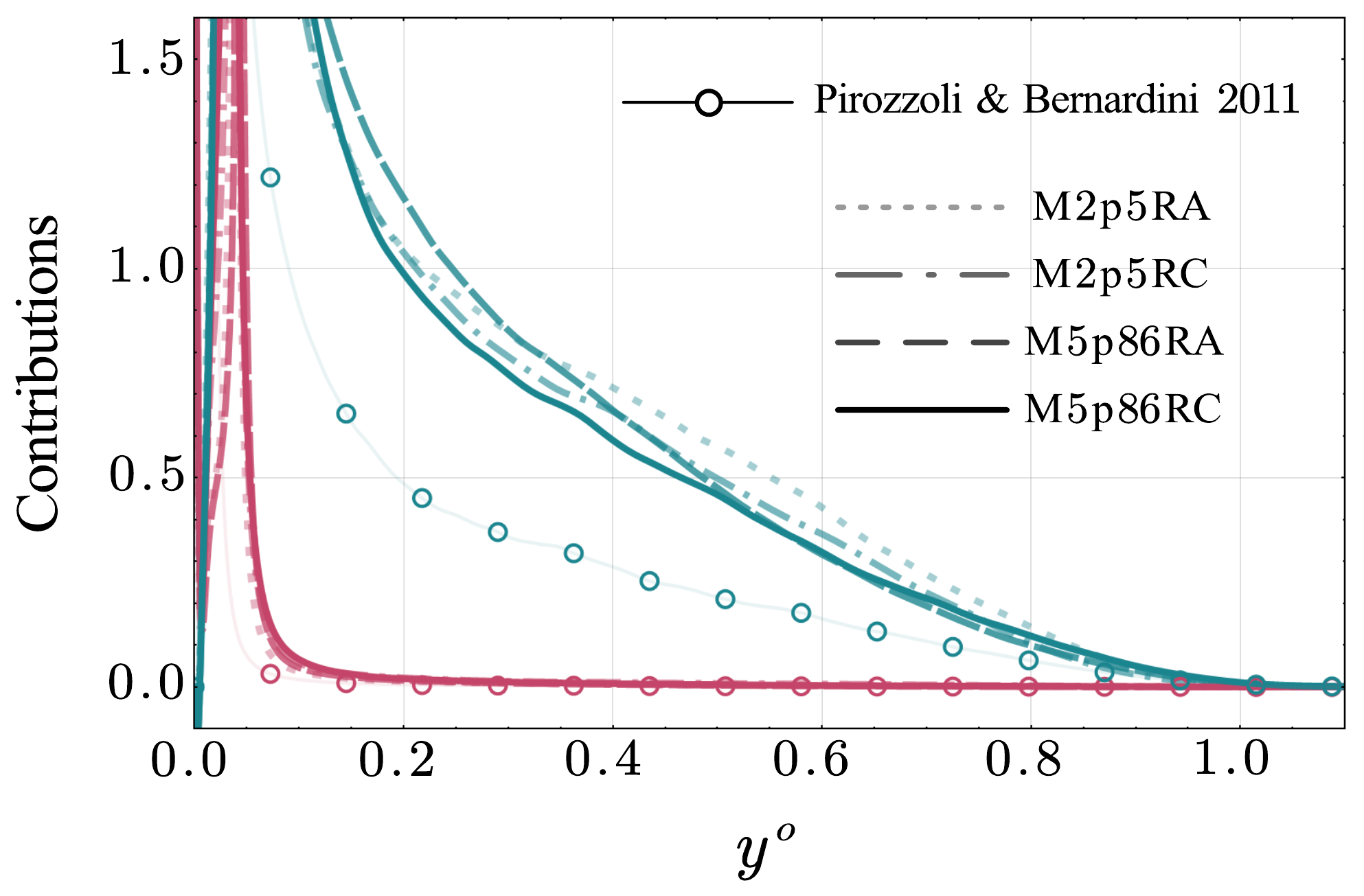}
\end{minipage}

\caption{Spatial distributions of the dissipation terms in the drag decomposition, where ``Contributions'' denotes the normalized contribution $\mathcal{C}_i^*$ defined by equation \eqref{eq:def_ciStar}. Each row compares the same four cases, with the left panel showing the near-wall distribution and the right panel showing the corresponding outer-layer distribution.}
\label{fig:GRD_D_ALL}
\end{figure}

\begin{figure}
\centering

\begin{minipage}[t]{0.428\textwidth}
\hspace*{-1.2em}(\textit{a})\par
\vspace{-0.15em}
\centering
\includegraphics[width=\linewidth]{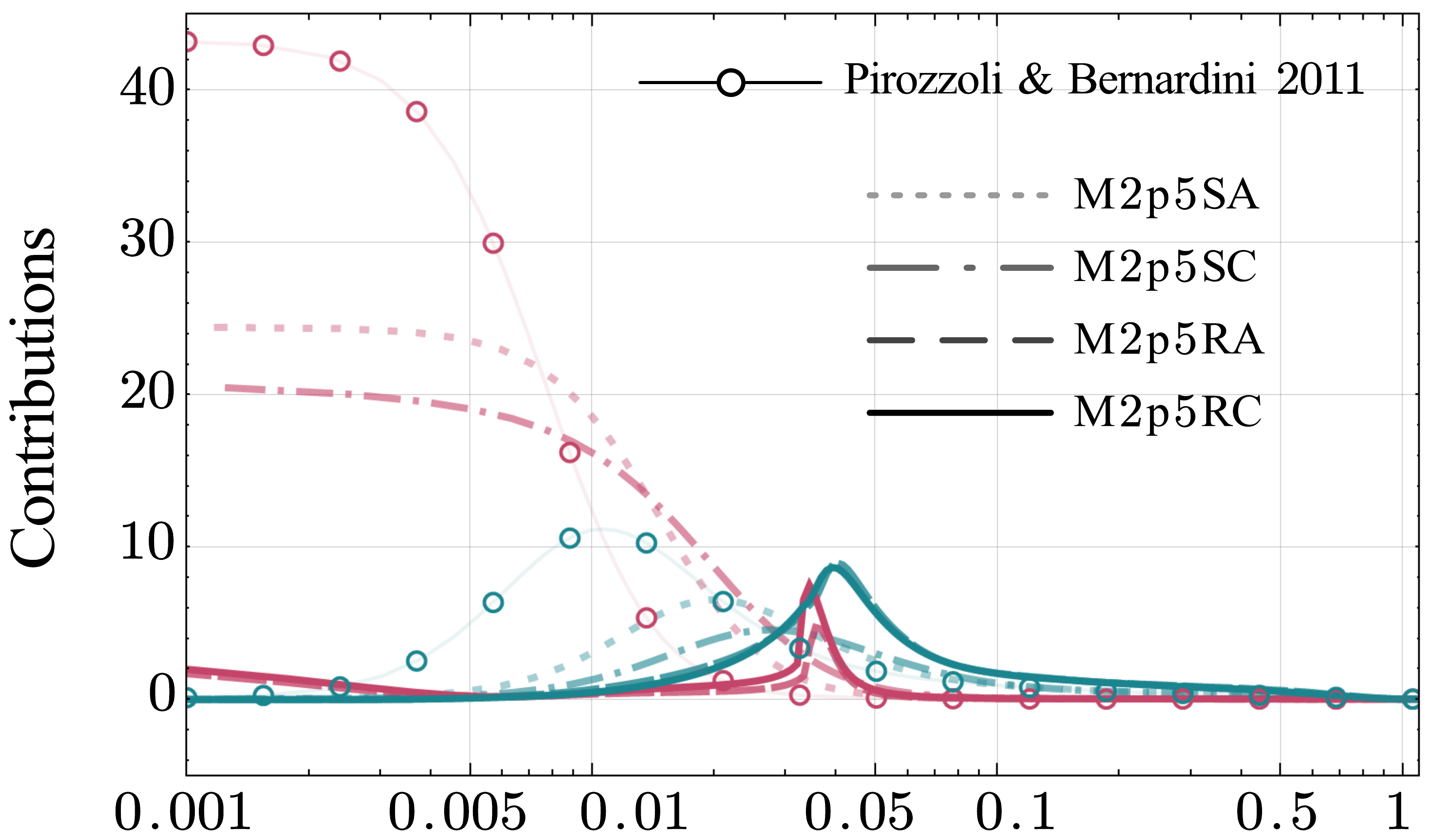}
\end{minipage}%
\hspace{0.04\textwidth}%
\begin{minipage}[t]{0.438\textwidth}
\hspace*{-1.2em}(\textit{b})\par
\vspace{-0.15em}
\centering
\includegraphics[width=\linewidth]{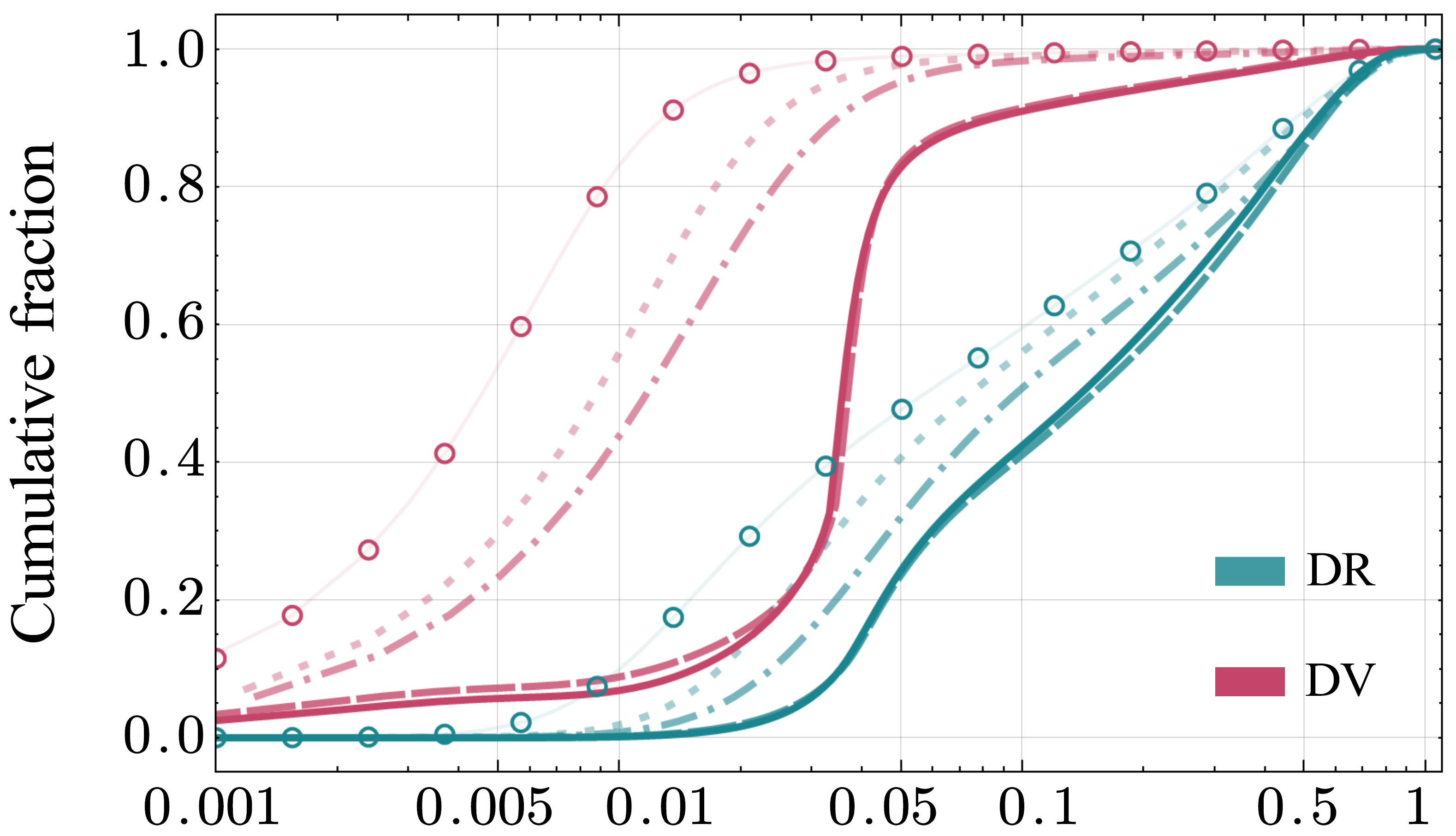}
\end{minipage}

\par\vspace{0.2cm}

\begin{minipage}[t]{0.428\textwidth}
\hspace*{-1.2em}(\textit{c})\par
\vspace{-0.15em}
\centering
\includegraphics[width=\linewidth]{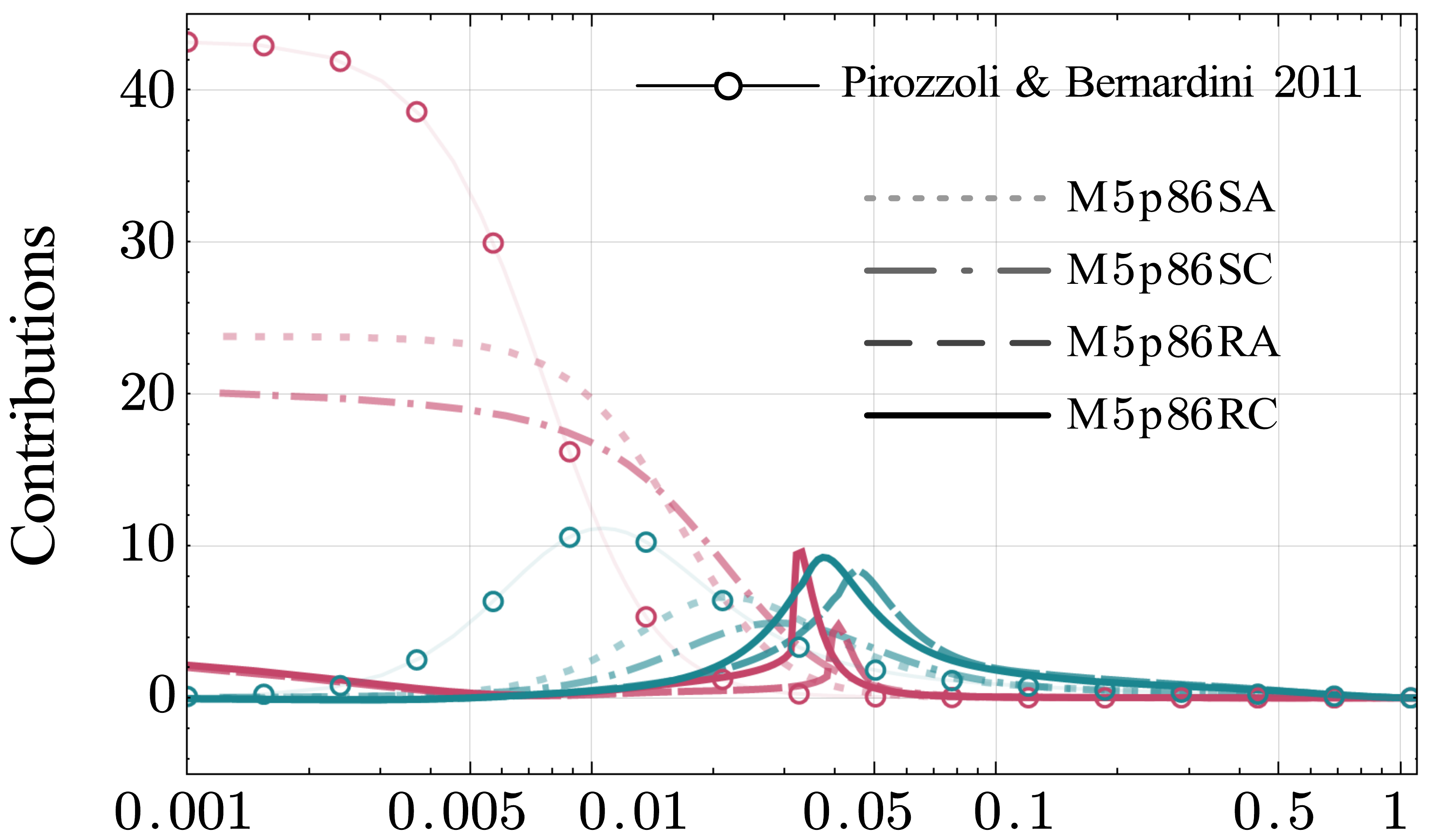}
\end{minipage}%
\hspace{0.04\textwidth}%
\begin{minipage}[t]{0.438\textwidth}
\hspace*{-1.2em}(\textit{d})\par
\vspace{-0.15em}
\centering
\includegraphics[width=\linewidth]{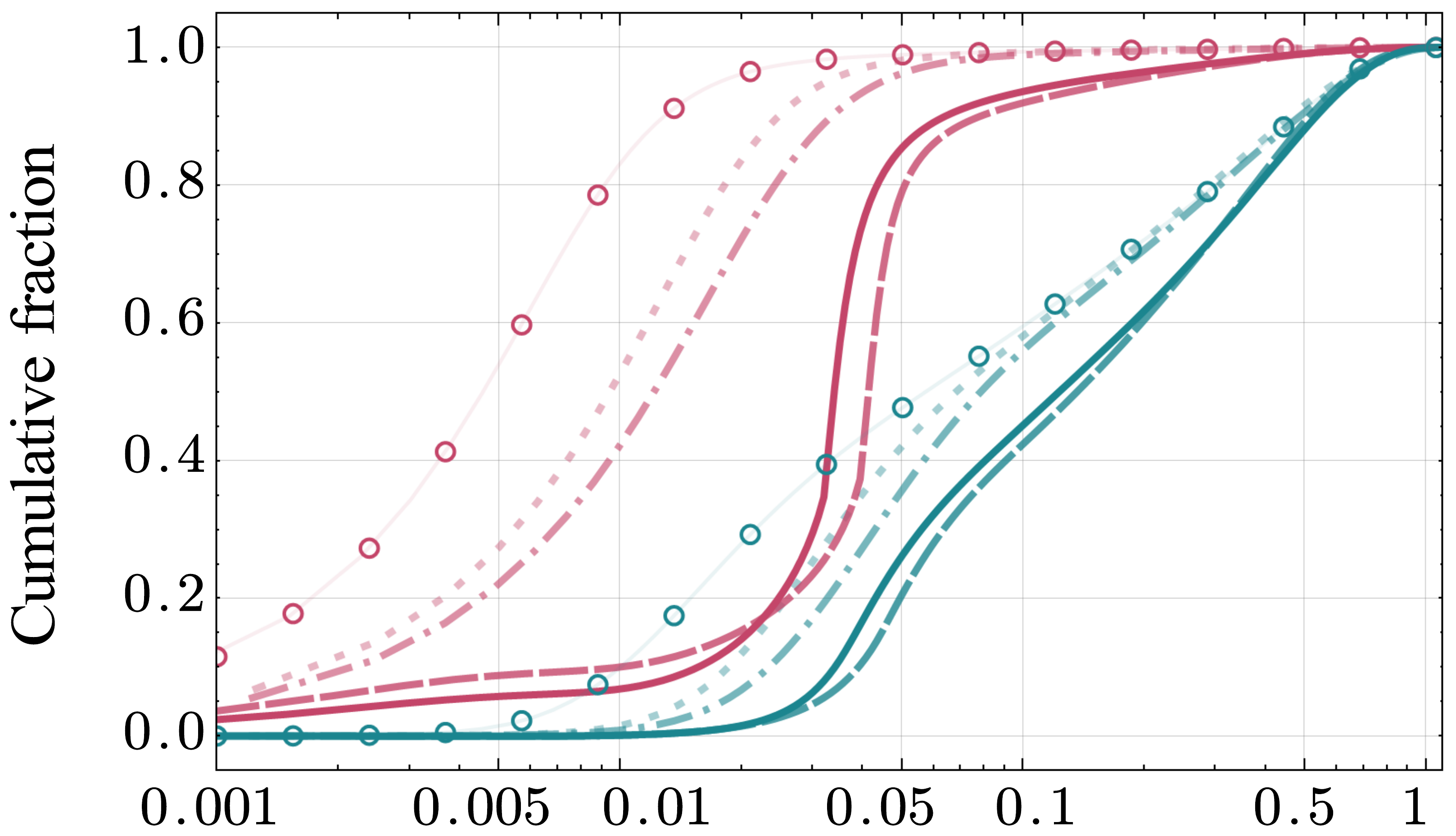}
\end{minipage}

\par\vspace{0.2cm}

\begin{minipage}[t]{0.428\textwidth}
\hspace*{-1.2em}(\textit{e})\par
\vspace{-0.15em}
\centering
\includegraphics[width=\linewidth]{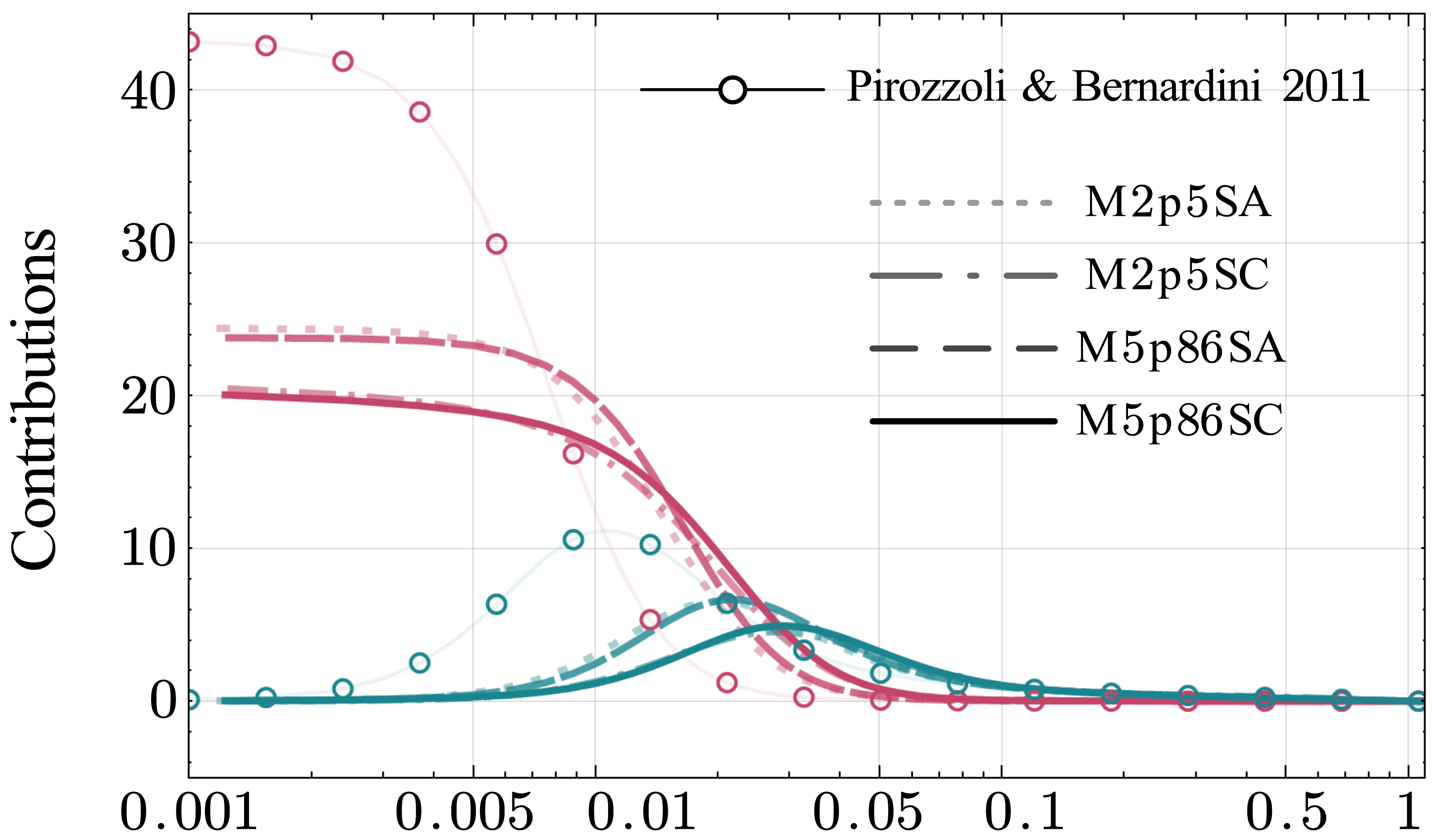}
\end{minipage}%
\hspace{0.04\textwidth}%
\begin{minipage}[t]{0.438\textwidth}
\hspace*{-1.2em}(\textit{f})\par
\vspace{-0.15em}
\centering
\includegraphics[width=\linewidth]{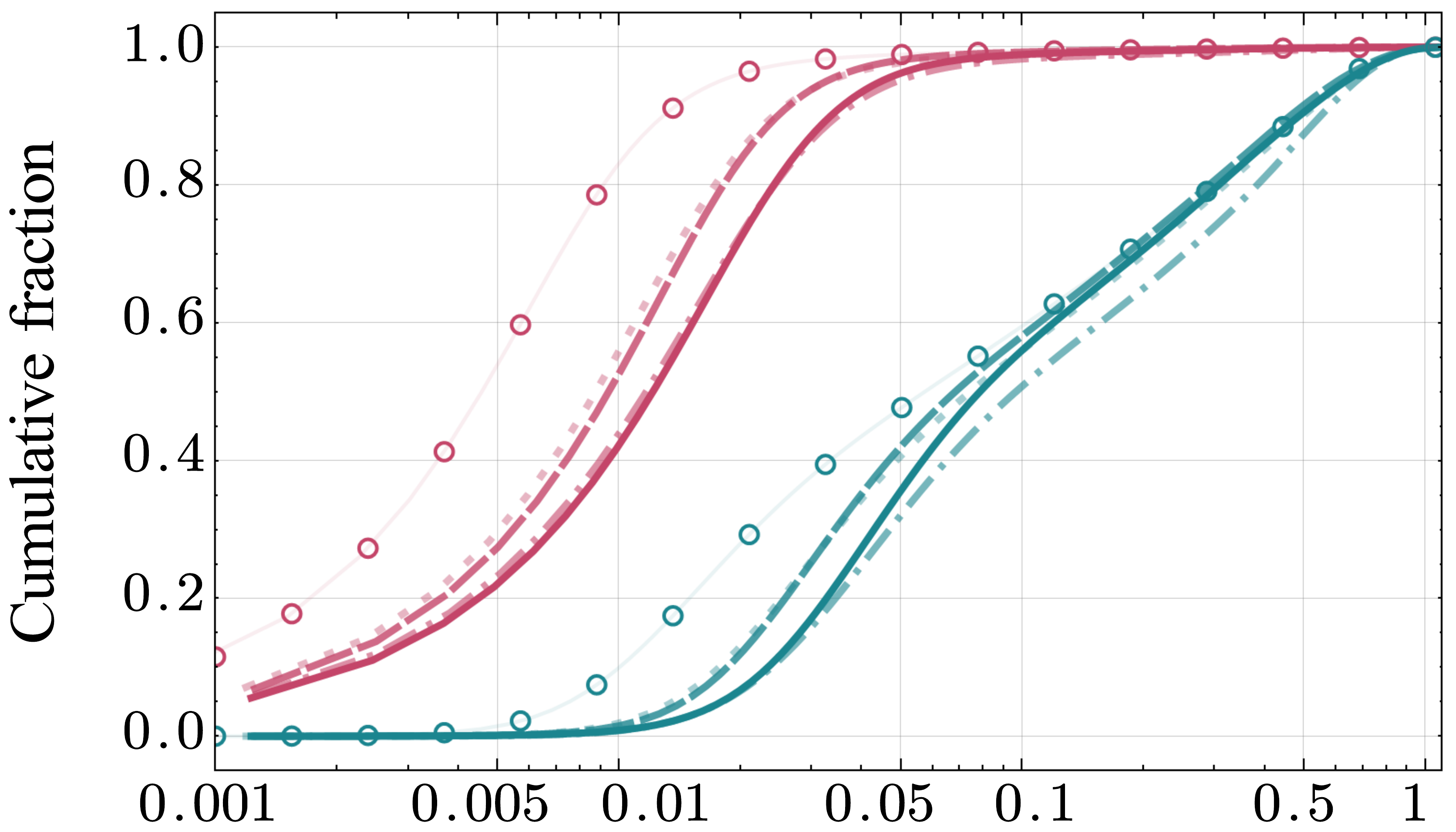}
\end{minipage}

\par\vspace{0.2cm}

\begin{minipage}[t]{0.428\textwidth}
\hspace*{-1.2em}(\textit{g})\par
\vspace{-0.15em}
\centering
\includegraphics[width=\linewidth]{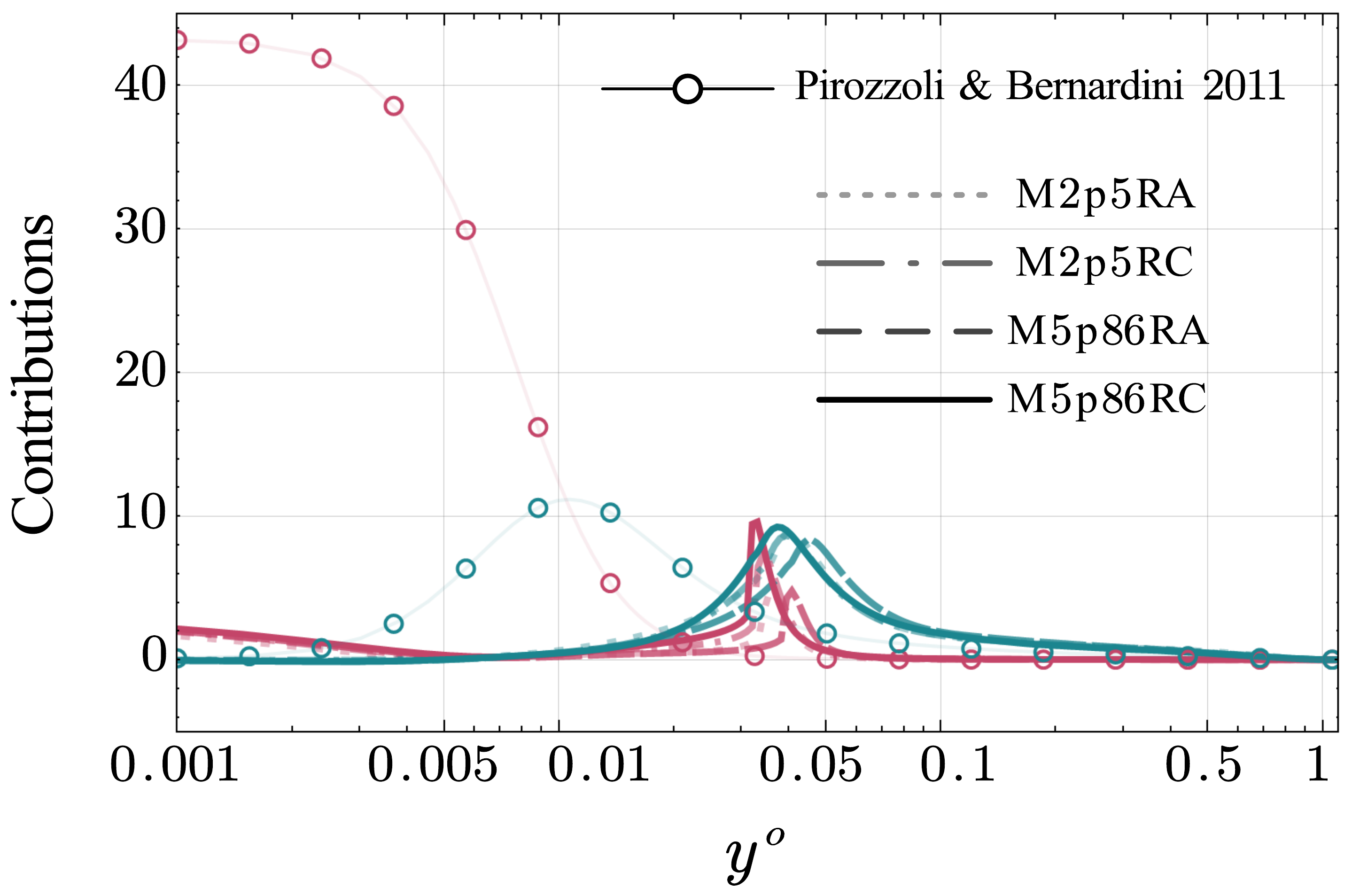}
\end{minipage}%
\hspace{0.04\textwidth}%
\begin{minipage}[t]{0.438\textwidth}
\hspace*{-1.2em}(\textit{h})\par
\vspace{-0.15em}
\centering
\includegraphics[width=\linewidth]{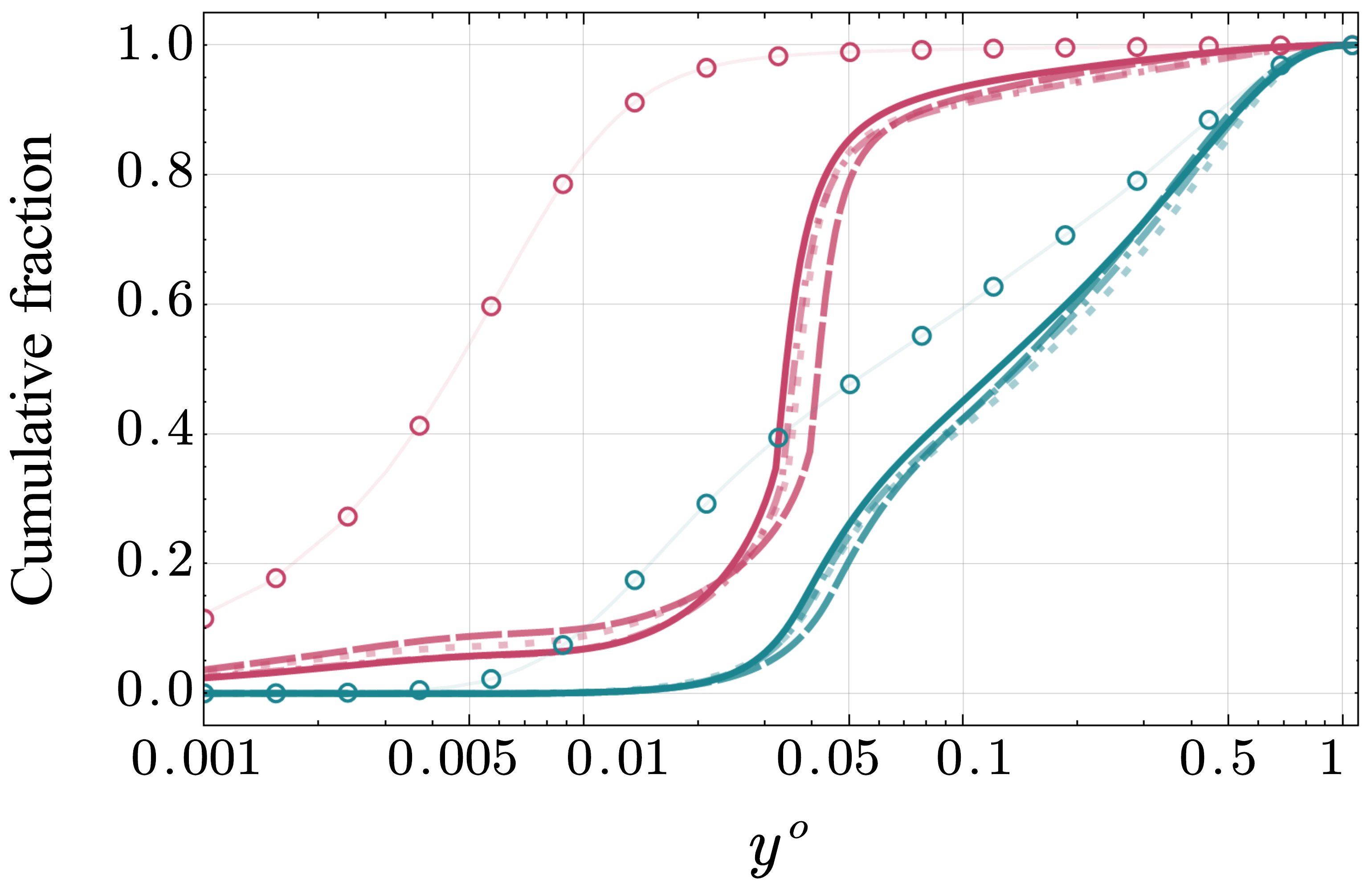}
\end{minipage}

\caption{Wall-normal distributions and cumulative contributions of the dissipation terms in the drag decomposition. Panels (a,c,e,g) show the distributions of the individual components on a logarithmic wall-normal scale. Panels (b,d,f,h) show the cumulative integral from the wall to a given height $y$, normalized by the total integral of the corresponding component, thereby quantifying the fraction of each contribution accumulated below $y$.}
\label{fig:GRD_D_CUM_ALL}
\end{figure}

The dominant wall-normal regions contributing to the individual dissipation components can be identified more directly from the right column of figure \ref{fig:GRD_D_CUM_ALL}. For the smooth-wall cases, the contribution of $DV$ is concentrated primarily within the inner layer. In contrast, for the rough-wall cases, the cumulative contribution below the roughness crest remains relatively small, accounting for only approximately $20\%$ of the total $DV$ contribution. Above the roughness crest, however, the cumulative fraction increases rapidly and reaches approximately $90\%$. It is also noteworthy that, unlike the smooth-wall cases, the $DV$ contribution for the rough-wall cases does not rapidly approach zero in the outer region, but instead exhibits a comparatively gradual decay, as shown in figure \ref{fig:GRD_D_CUM_ALL}(a, c). For the $DR$ component, the cumulative contribution at a given wall-normal location in the outer layer is consistently lower for the rough-wall cases than for the corresponding smooth-wall cases. This indicates that, for the smooth-wall flows, the contribution associated with the $DR$ peak emerging closer to the wall remains reflected in the cumulative distribution up to the edge of the boundary layer. The persistence of this difference throughout the outer layer suggests that roughness modifies the wall-normal organization of energy dissipation over a substantial portion of the boundary layer, rather than producing only a localized near-wall perturbation. Consequently, the drag characteristics of the present rough-wall flows cannot be represented solely through local wall modifications without accounting for the redistribution of dissipation across the boundary layer.

The distributions of the individual components of the dissipation term can be further examined from figures \ref{fig:GRD_D_ALL} and \ref{fig:GRD_D_CUM_ALL}. As shown in figure \ref{fig:GRD_D_CUM_ALL}(a,c), for the smooth-wall cases, the near-wall peak of $DT$ is contributed almost entirely by $DV$. With increasing wall-normal distance, $DV$ decreases rapidly, whereas $DR$ develops a distinct peak within the inner layer. Figure \ref{fig:GRD_D_ALL}(e) further shows that, as the friction Reynolds number increases, the $DR$ peak shifts towards the wall and increases in magnitude, whereas the Mach number has only a minor influence on its peak location. For the rough-wall cases, local peaks of $DV$ appear both at the bottom plane and near the roughness crest, followed by a rapid decay away from these locations. By contrast, the $DR$ peak is located above the roughness crest but remains below approximately twice the roughness height, as shown in figure \ref{fig:GRD_D_ALL}(a, c). The magnitude of the rough-wall $DV$ peak is also sensitive to the wall thermal condition, with the cold-wall cases exhibiting substantially higher peaks than their adiabatic counterparts. The corresponding $DR$ peaks, located slightly farther from the wall, show a much weaker dependence on wall temperature.

In the outer layer, $DR$ is the dominant contribution for all cases, and its wall-normal behavior is consistent with that observed for the total $DT$ term. As shown in figure \ref{fig:GRD_D_ALL}(f, h), the friction Reynolds number has only a minor influence on the $DR$ distribution for the smooth-wall cases. At comparable friction Reynolds numbers, however, clear differences are observed between the smooth- and rough-wall cases. In particular, $DR$ decays considerably more rapidly with increasing wall-normal distance for the smooth-wall cases than for the rough-wall cases.

\subsection{Regional decomposition of surface drag}\label{Regional_decomposition}
\label{sec:regional_drag}

The generalized RD decomposition provides a full-field energetic interpretation of the total drag by relating the drag work to the different transport and dissipation mechanisms within the boundary layer. The total drag itself, however, enters the derivation through the integral of the surface traction over the solid boundary. Consequently, the generalized RD decomposition does not directly distinguish whether the surface force originates from viscous friction or pressure differences across the roughness elements. To complement the energetic analysis, we therefore further examine the geometric origins of the surface drag by decomposing the wall force into contributions from different regions of a roughness unit.

\begin{figure}
    \centering
    \vspace{0.4cm}
    \includegraphics[width=0.95\textwidth]{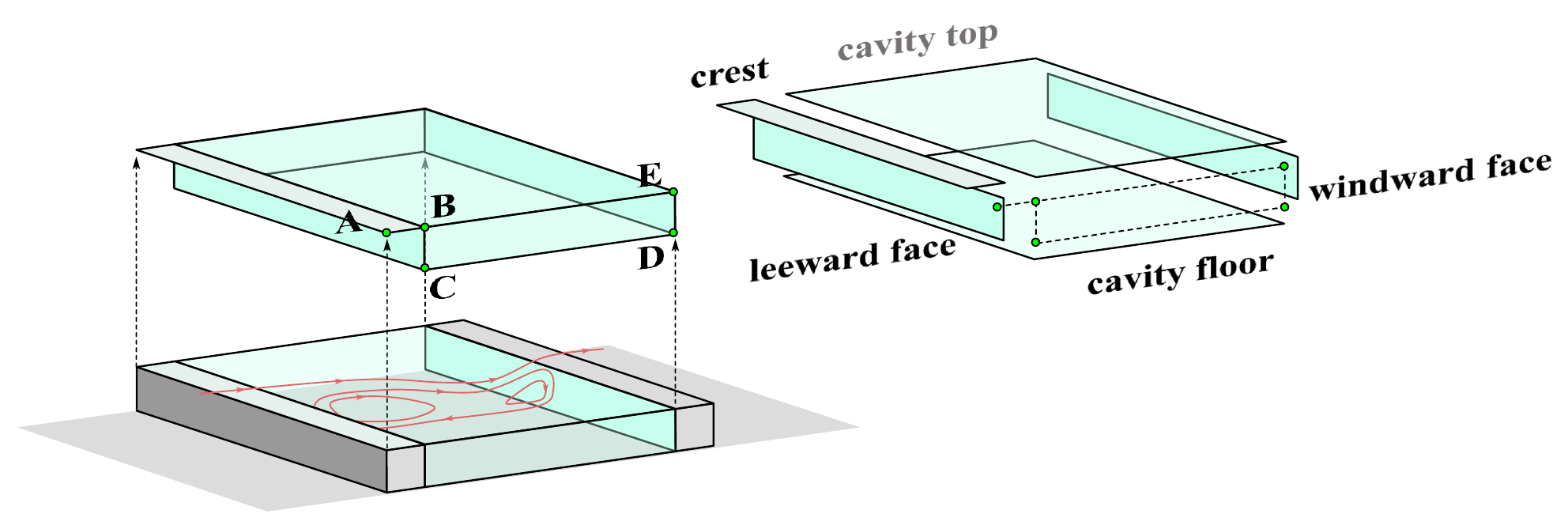}
    \vspace{0.2cm}
    \caption{Schematic illustration of the regional decomposition of roughness drag. The upper-left panel shows the cavity fluid region extracted from the same roughness unit shown below, while the upper-right panel provides an exploded view of the corresponding decomposition surfaces, including the roughness crest, cavity top, windward face, leeward face and cavity floor. The red curves in the lower panel schematically represent the streamlines within the cavity.}
    \label{fig:dragField}
\end{figure}

As illustrated in figure \ref{fig:dragField}, the drag acting on one roughness period can be separated into the viscous friction on the roughness crest and cavity floor, together with the pressure drag generated by the pressure difference between the windward and leeward faces. Since the present roughness geometry is homogeneous in the spanwise direction, the corresponding surface integrals can be reduced to two-dimensional line integrals per unit span. Denoting the total drag over one roughness period by $F_D$, the relative contributions from the roughness crest AB, the cavity floor CD and the pressure difference between the windward face ED and the leeward face BC are respectively defined as
\begin{equation}
\begin{aligned}
C_{f,AB}&=\frac{1}{F_D}\int_{AB}\mu\frac{\partial \overline{u}}{\partial y}\,\mathrm{d}x, \qquad
C_{f,CD}=\frac{1}{F_D}\int_{CD}\mu\frac{\partial \overline{u}}{\partial y}\,\mathrm{d}x,\\
C_p&=\frac{1}{F_D}\left(\int_{ED}\overline{p}\,\mathrm{d}y-\int_{BC}\overline{p}\,\mathrm{d}y\right)
\end{aligned}
\end{equation}
such that
\begin{equation}
C_{f,AB}+C_{f,CD}+C_p=1
\label{eq:regional_drag_balance}
\end{equation}
Here, the sign of each contribution is retained. In particular, the recirculating flow within the cavity can produce a negative skin-friction contribution over the cavity floor.

The momentum transfer between the outer flow and the cavity can be examined further by introducing a virtual plane BE across the cavity opening at the roughness-crest height. The streamwise momentum flux across this plane can be separated into viscous, turbulent and mean-flow contributions. Using the same normalization by $F_D$, these contributions are written as
\begin{equation}
\begin{aligned}
C_{f,BE}&=\frac{1}{F_D}\int_{BE}\mu(\frac{\partial \overline{u}}{\partial y} + \frac{\partial \overline{v}}{\partial x})\,\mathrm{d}x,\qquad
C_{t,BE}=\frac{1}{F_D}\int_{BE}-\overline{\rho}\,\widetilde{u''v''}\,\mathrm{d}x\\
C_{m,BE}&=\frac{1}{F_D}\int_{BE}-\overline{\rho}\,\widetilde{u}\widetilde{v}\,\mathrm{d}x
\end{aligned}
\end{equation}
where $C_{f,BE}$ represents viscous momentum transport, $C_{t,BE}$ the turbulent momentum transport associated with the Reynolds shear stress, and $C_{m,BE}$ the transport by the mean flow. Momentum conservation over the cavity control volume BCDE then gives
\begin{equation}
C_{f,BE}+C_{t,BE}+C_{m,BE}=C_p+C_{f,CD}
\label{eq:cavity_momentum_balance}
\end{equation}
Thus, the surface-based decomposition identifies the geometric origin of the drag, whereas the virtual-plane decomposition characterizes how the momentum required to sustain the cavity drag is transferred from the flow above the roughness crest into the cavity.

\begin{table}
    \centering
    \begin{tabular}{lcccccc}

        Case
        & $C_{f,AB}(\%)$
        & $C_p(\%)$
        & $C_{f,CD}(\%)$
        & $C_{f,BE}(\%)$
        & $C_{t,BE}(\%)$
        & $C_{m,BE}(\%)$ \\[3pt]

        M2p5RA  & 5.23  & 109.72 & -14.95 & 11.76 & 79.41 & 4.32 \\
        M2p5RC  & 8.66  & 108.50 & -17.16 & 18.81 & 67.21 & 3.70 \\
        M5p86RA & 6.82  & 108.51 & -15.23 & 12.27 & 78.85 & 3.79 \\
        M5p86RC & 11.20 & 106.08 & -17.28 & 21.82 & 63.58 & 2.39 \\

    \end{tabular}
    \caption{Regional decomposition of the roughness drag for the adiabatic and cold-wall cases at $Ma=2.5$ and $5.86$. The tabulated percentages indicate the relative contributions of the local tangential stresses, pressure force and momentum transport to the total drag within a single roughness element at $x_t$.}
    \label{tab:drag_decomposition}
\end{table}

The regional drag contributions are summarized in table \ref{tab:drag_decomposition}. The momentum balance between the virtual top (BE) and the physical cavity surfaces is satisfied within approximately $2\%$ for all cases, indicating good consistency of the present regional decomposition. The pressure drag is the dominant contribution, accounting for approximately $106$--$110\%$ of the total drag. The excess above $100\%$ is compensated primarily by the negative skin friction on the cavity floor, which contributes approximately $-15$--$-17\%$ owing to the recirculating flow within the cavity. The positive frictional contribution from the roughness crest is considerably smaller, ranging from approximately $5$ to $11\%$. Wall cooling substantially increases the relative frictional contribution from the roughness crest, from $5.23\%$ to $8.66\%$ at $Ma=2.5$ and from $6.82\%$ to $11.20\%$ at $Ma=5.86$. This trend is consistent with the generalized RD decomposition, in which the peak contribution of the $DV$ term is also enhanced under cold-wall conditions. At the virtual plane BE, the Reynolds shear stress is the dominant contribution to the momentum transfer across the cavity opening. The viscous contribution $C_{f,BE}$ also increases markedly under cold-wall conditions, whereas the Reynolds-shear-stress contribution $C_{t,BE}$ decreases from approximately $79\%$ to $67\%$ at $Ma=2.5$ and from $79\%$ to $64\%$ at $Ma=5.86$. The mean-flow contribution $C_{m,BE}$ remains small for all cases, accounting for only approximately $2$--$4\%$ of the total drag.

\section{Conclusions}\label{sec:Conclusions}

Eight direct numerical simulations of spatially developing compressible turbulent boundary layers at $Ma_\infty=2.5$ and $5.86$ compare smooth walls and transverse square-bar roughness with $\lambda_x/k=8$, under adiabatic conditions and wall cooling at $T_w/T_r=0.5$. The zero-plane displacement obtained by kinematic optimization, relative to the roughness height, increases with Mach number and cooling. The adiabatic roughness function is nearly independent of Mach number, whereas the cooled-wall value decreases with increasing Mach number. The Griffin--Fu--Moin transformation closely collapses the rough- and smooth-wall velocity-defect profiles at each Mach number, demonstrating outer-layer similarity at a given Mach number, although the transformed profiles still exhibit a clear Mach-number dependence. Regarding the velocity fluctuations, roughness suppresses the streamwise fluctuation peak. Differences in velocity fluctuations persist into the outer layer, where cooling reduces fluctuation intensities at $Ma_\infty=2.5$ but increases them at $Ma_\infty=5.86$.

The classical general Reynolds analogy exhibits near-wall discrepancies over rough walls, particularly under wall cooling. The previously developed rough-wall formulation (rGRA) gives good agreement with the DNS using a reference height of $h=k$ for both adiabatic cases and the cooled-wall case at $Ma_\infty=5.86$, but requires $h=2k$ for the cooled-wall case at $Ma_\infty=2.5$. This difference is consistent with the extent of near-wall thermal heterogeneity and the recovery of a quasi-parallel state above the roughness.

A generalized Renard--Deck (RD) decomposition applicable to rough-wall boundary layers is developed in the present study, which decomposes the total drag work into contributions associated with different transport and dissipation mechanisms distributed throughout the flow field. Mean-flow viscous dissipation and turbulent kinetic-energy production together account for approximately $90\%$ of the drag work, with the convective transport of kinetic energy accounting for most of the remainder. The streamwise-development, normal-displacement and additional compressibility terms have small contributions. Viscous dissipation and turbulent production are comparable in the present smooth-wall cases, whereas turbulent production accounts for approximately $80\%$ of the rough-wall drag work. Cooling increases the rough-wall viscous contribution from approximately $7\%$ to $10$--$12\%$, while reducing the turbulent-production share.

The rough-wall turbulent-production peak is located between the roughness crest and approximately $2k$, while its contribution decays more slowly towards the outer layer than in the smooth-wall cases, even at comparable friction Reynolds numbers. Meanwhile, the convection peak shifts from the inner region over smooth walls to approximately half the boundary-layer thickness over rough walls. The regional force balance further shows that the surface drag is dominated by pressure differences, which contribute approximately $106$--$110\%$ of the total drag and are partly offset by the negative cavity-floor friction of approximately $-15$ to $-17\%$. Reynolds shear stress dominates the momentum transfer across the cavity opening, while cooling increases the relative viscous contributions at both the roughness crest and the cavity opening. Taken together, for the present roughness configuration, the two decompositions show that the surface drag is generated predominantly by pressure forces on the roughness elements, whereas the corresponding drag work is associated mainly with energy transfer from the mean flow to turbulence. They further indicate that the energetic influence of roughness is not confined to the roughness sublayer, but extends over a substantial fraction of the boundary layer.

\section*{Acknowledgements}\label{sec:Acknowledgements}
This work was supported by National Natural Science Foundation of China under Grant Nos. U2570214, 12425206 and 12388101.

\section*{Declaration of Interests}\label{sec:Declaration}
The authors report no conflict of interest.

\appendix
\section{Grid distribution and resolution assessment}\label{app:grid_resolution}

The grid distributions in both the streamwise and wall-normal directions are carefully designed to resolve the turbulent flow structures, particularly in the vicinity of the roughness elements. In the streamwise direction, local grid refinement is employed around each roughness element to capture the separated shear layers and wake structures. For the $Ma_\infty=2.5$ cases, each square bar is resolved by $21\times37$ and $21\times35$ grid points in the streamwise and wall-normal directions for the adiabatic and cooled-wall conditions, respectively, with corresponding streamwise grid spacings of $\Delta x^+=1.81$ and $1.74$. For the $Ma_\infty=5.86$ cases, the corresponding resolutions are $21\times38$ and $21\times40$, with $\Delta x^+=1.89$ and $1.54$, respectively. These streamwise grid spacings are comparable to those adopted by \citet{modesti2022direct} and \citet{cogo2025surface}, who employed $\Delta x^+=2.0$ and $2.8$, respectively, in simulations of cube roughness. Furthermore, the maximum streamwise grid spacing in the rough-wall cases is maintained below that of the corresponding smooth-wall cases to adequately resolve roughness-induced flow structures.

In the wall-normal direction, a non-uniform grid distribution is employed, with grid points clustered near the wall and gradually stretched towards the free stream. For the $Ma_\infty=2.5$ cases, the maximum wall-normal grid spacing within the boundary layer is limited to $\Delta y=0.034\delta_{in}$ up to $y=4.9\delta_{in}$, beyond which the grid is further stretched towards the upper boundary. This arrangement provides sufficient resolution for the turbulent structures within the thickened rough-wall boundary layers. For the $Ma_\infty=5.86$ cases, the corresponding maximum wall-normal spacing is $\Delta y=0.0203\delta_{in}$, maintained up to $y=4.3\delta_{in}$, followed by further stretching towards the free stream.

For rough-wall flows, the conventional assessment of grid resolution based on inner scaling requires particular care. Since pressure drag constitutes the dominant contribution to the total drag stress, the friction velocity defined from the total drag is substantially larger than that associated with viscous friction alone. Consequently, grid spacings normalized by the inner scale may appear relatively large without necessarily indicating insufficient resolution of the relevant turbulent structures. In particular, turbulence within the roughness sublayer is strongly influenced by separated shear layers and wake structures, rather than the near-wall viscous mechanisms characteristic of smooth-wall flows. The local Kolmogorov length scale, $\eta=(\nu^3/\varepsilon)^{1/4}$, therefore provides an additional measure for assessing grid resolution. As shown in table~\ref{tab:grid_resolution}, the Kolmogorov scales evaluated at $y=2k$ are comparable between the rough- and smooth-wall cases. The maximum ratio of wall-normal grid spacing to the Kolmogorov length scale is observed in case M2p5RA, while the corresponding smooth-wall case exhibits a value approximately $7\%$ lower. The grid spacings satisfy the DNS resolution criteria of $\Delta x/\eta=7.5$ and $\Delta z/\eta=4.4$ reported by \cite{moin1998direct}. These results indicate that the present grid resolution is sufficient to capture the turbulent structures within the roughness boundary layers.

\begin{table}
    \begin{center}
    \def~{\hphantom{0}}
    \setlength{\tabcolsep}{2.5pt}
    \begin{tabular}{lcccccc}

        Case & $\Delta x^+$ & $\Delta y^+$ & $\Delta z^+$ & \mbox{$\Delta x/\eta \mid_{2k}$} & \mbox{$\Delta y/\eta \mid_{2k}$} & \mbox{$\Delta z/\eta \mid_{2k}$} \\[3pt]

        M2p5SA  & 8.21       & 0.602--9.45 & 5.28 & 4.23        & 0.600 & 2.72 \\
        M2p5SC  & 8.82       & 0.647--10.1 & 5.67 & 3.40        & 0.458 & 2.18 \\
        M2p5RA  & 1.81--10.2 & 0.916--14.3 & 8.03 & 0.660--3.72 & 0.647 & 2.93 \\
        M2p5RC  & 1.74--9.90 & 0.945--14.7 & 8.28 & 0.518--2.95 & 0.518 & 2.47 \\[3pt]

        M5p86SA & 8.48       & 0.56--5.82  & 5.45 & 4.33        & 0.580 & 2.79 \\
        M5p86SC & 8.31       & 0.55--5.70  & 5.35 & 3.45        & 0.477 & 2.22 \\
        M5p86RA & 1.89--10.7 & 0.92--9.55  & 8.96 & 0.545--3.09 & 0.540 & 2.59 \\
        M5p86RC & 1.54--8.70 & 0.70--7.30  & 6.84 & 0.408--2.30 & 0.390 & 1.81 \\

    \end{tabular}
    \caption{Grid resolution of the present DNS cases at the target streamwise stations $x_t$. The grid spacings $\Delta x^+$, $\Delta y^+$ and $\Delta z^+$ are expressed in inner scaling. The ratios $\Delta x/\eta$, $\Delta y/\eta$ and $\Delta z/\eta$ characterize the local resolution relative to the Kolmogorov length scale $\eta$, evaluated at $y=2k$ for the rough-wall cases and at the same wall-normal locations for the corresponding smooth-wall cases under identical thermal conditions.}
    \label{tab:grid_resolution}
    \end{center}
\end{table}

\section{Detailed derivation of the generalized Renard--Deck decomposition}\label{app:RD_derivation}

Starting from the three-term integral relation in \eqref{RD_original_three}, the three contributions are considered separately.For the first term, since the phase indicator $\gamma$ is unity in the fluid and zero in the solid, the volume integral may be restricted to $\Omega_f$. Application of the divergence theorem gives
\begin{equation}
\iiint_{\Omega}\gamma\nabla\cdot\left[(\overline{\boldsymbol U}-\boldsymbol U_\infty)\cdot\overline{\boldsymbol{\mathcal T}}\right]\,\mathrm{d}V
=\iint_{\partial\Omega_f}(\overline{\boldsymbol U}-\boldsymbol U_\infty)\cdot\overline{\boldsymbol{\mathcal T}}\cdot\mathrm{d}\boldsymbol A
\end{equation}
With $\partial\Omega_f=L\cup R\cup T\cup W$ defined in figure \ref{fig:RDField}, the contribution from $W$ is $U_\infty F_D$. At the upper boundary, the free-stream flow is approximately uniform and the shear stresses are negligible, leaving only the normal-pressure contribution
\begin{equation}
\iint_T(\overline{\boldsymbol U}-\boldsymbol U_\infty)\cdot\overline{\boldsymbol{\mathcal T}}\cdot\mathrm{d}\boldsymbol A=-\iint_T\overline p\,\overline v\,\mathrm{d}A
\end{equation}
Using the no-penetration condition at the wall, this contribution can be rewritten as
\begin{equation}
-\iint_T\overline p\,\overline v\,\mathrm{d}A=-\iiint_\Omega\gamma\frac{\partial(\overline p\,\overline v)}{\partial y}\,\mathrm{d}V
\end{equation}
and hence the first term becomes
\begin{equation}
\begin{aligned}
\iiint_{\Omega}\gamma\nabla\cdot\left[(\overline{\boldsymbol U}-\boldsymbol U_\infty)\cdot\overline{\boldsymbol{\mathcal T}}\right]\,\mathrm{d}V
={}&\left.\iint_S\left[(\overline{\boldsymbol U}-\boldsymbol U_\infty)\cdot\overline{\boldsymbol{\mathcal T}}\right]_x\,\mathrm{d}A\right|_L^R\\
&-\iiint_\Omega\gamma\frac{\partial(\overline p\,\overline v)}{\partial y}\,\mathrm{d}V+U_\infty F_D
\end{aligned}
\label{eq:RD_first_term_app}
\end{equation}

For the second term, noting that $\boldsymbol U_\infty$ is spatially uniform, the contraction between the mean velocity gradient and the total stress can be expanded as
\begin{equation}
\frac{\partial\overline U_j}{\partial x_k}\overline{\mathcal T}_{jk}
=-\overline p\,\overline\Theta-\frac{2}{3}\mu\overline\Theta^2+\mu\frac{\partial\overline U_j}{\partial x_k}\frac{\partial\overline U_k}{\partial x_j}+\Phi_\nu+\Phi_R
\end{equation}
The cross-gradient term satisfies
\begin{equation}
\mu\frac{\partial\overline U_j}{\partial x_k}\frac{\partial\overline U_k}{\partial x_j}
=\mu\nabla\cdot\left[(\overline{\boldsymbol U}\cdot\nabla)\overline{\boldsymbol U}\right]-\mu\overline{\boldsymbol U}\cdot\nabla\overline\Theta
\end{equation}
and, for variable viscosity,
\begin{equation}
\mu\nabla\cdot\left[(\overline{\boldsymbol U}\cdot\nabla)\overline{\boldsymbol U}\right]
=\nabla\cdot\left[\mu(\overline{\boldsymbol U}\cdot\nabla)\overline{\boldsymbol U}\right]-\nabla\mu\cdot\left[(\overline{\boldsymbol U}\cdot\nabla)\overline{\boldsymbol U}\right]
\end{equation}
The divergence term is converted into a surface integral. The wall contribution vanishes because of the no-slip condition, while the velocity gradients vanish at the upper boundary, leaving only the upstream and downstream cross-sections. The second term therefore becomes
\begin{equation}
\begin{aligned}
\iiint_{\Omega}\gamma\nabla\overline{\boldsymbol U}:\overline{\boldsymbol{\mathcal T}}\,\mathrm{d}V
={}&\left.\iint_S\mu\left[(\overline{\boldsymbol U}\cdot\nabla)\overline{\boldsymbol U}\right]_x\,\mathrm{d}A\right|_L^R
+\iiint_\Omega\gamma\left(-\overline p\,\overline\Theta\right)\,\mathrm{d}V\\
&+\iiint_\Omega\gamma\left[-\frac{2}{3}\mu\overline\Theta^2-\nabla\mu\cdot\left((\overline{\boldsymbol U}\cdot\nabla)\overline{\boldsymbol U}\right)-\mu\overline{\boldsymbol U}\cdot\nabla\overline\Theta\right]\,\mathrm{d}V\\
&+\iiint_\Omega\gamma\left(\Phi_\nu+\Phi_R\right)\,\mathrm{d}V
\end{aligned}
\label{eq:RD_second_term_app}
\end{equation}

For the third term, the mean continuity equation
\begin{equation}
\nabla\cdot\left(\overline\rho\,\widetilde{\boldsymbol U}\right)=0
\end{equation}
gives
\begin{equation}
\nabla\cdot\left(\overline\rho\,\widetilde{\boldsymbol U}\widetilde{\boldsymbol U}\right)
=\overline\rho\left(\widetilde{\boldsymbol U}\cdot\nabla\right)\widetilde{\boldsymbol U}
\end{equation}
The velocity difference is decomposed as
\begin{equation}
\overline{\boldsymbol U}-\boldsymbol U_\infty
=(\overline{\boldsymbol U}-\widetilde{\boldsymbol U})+(\widetilde{\boldsymbol U}-\boldsymbol U_\infty)
\end{equation}
where the first term accounts for the difference between the Reynolds- and Favre-averaged velocities. The contribution associated with the second term can be written as
\begin{equation}
\begin{aligned}
(\widetilde{\boldsymbol U}-\boldsymbol U_\infty)\cdot\overline\rho(\widetilde{\boldsymbol U}\cdot\nabla)\widetilde{\boldsymbol U}
={}&\nabla\cdot\left[\frac{1}{2}\overline\rho\,\widetilde{\boldsymbol U}\left|\widetilde{\boldsymbol U}-\boldsymbol U_\infty\right|^2\right]\\
&-\frac{1}{2}\left|\widetilde{\boldsymbol U}-\boldsymbol U_\infty\right|^2\nabla\cdot\left(\overline\rho\,\widetilde{\boldsymbol U}\right)
\end{aligned}
\end{equation}
The second term on the right-hand side vanishes by continuity. Applying the divergence theorem to the remaining term, the contributions from $W$ and $T$ vanish, giving
\begin{equation}
\begin{aligned}
&\iiint_\Omega\gamma(\overline{\boldsymbol U}-\boldsymbol U_\infty)\cdot\nabla\cdot\left(\overline\rho\,\widetilde{\boldsymbol U}\widetilde{\boldsymbol U}\right)\,\mathrm{d}V\\
={}&\left.\iint_S\frac{1}{2}\overline\rho\,\widetilde U_x\left|\widetilde{\boldsymbol U}-\boldsymbol U_\infty\right|^2\,\mathrm{d}A\right|_L^R
+\iiint_\Omega\gamma(\overline{\boldsymbol U}-\widetilde{\boldsymbol U})\cdot\overline\rho(\widetilde{\boldsymbol U}\cdot\nabla)\widetilde{\boldsymbol U}\,\mathrm{d}V
\end{aligned}
\label{eq:RD_third_term_app}
\end{equation}

Substituting \eqref{eq:RD_first_term_app}, \eqref{eq:RD_second_term_app} and \eqref{eq:RD_third_term_app} into \eqref{RD_original_three} gives
\begin{equation}
\begin{aligned}
U_\infty F_D
={}&-\left.\iint_S\left[(\overline{\boldsymbol U}-\boldsymbol U_\infty)\cdot\overline{\boldsymbol{\mathcal T}}\right]_x\,\mathrm{d}A\right|_L^R
+\iiint_\Omega\gamma\frac{\partial(\overline p\,\overline v)}{\partial y}\,\mathrm{d}V\\
&+\left.\iint_S\mu\left[(\overline{\boldsymbol U}\cdot\nabla)\overline{\boldsymbol U}\right]_x\,\mathrm{d}A\right|_L^R
+\iiint_\Omega\gamma(-\overline p\,\overline\Theta)\,\mathrm{d}V\\
&+\iiint_\Omega\gamma\left[-\frac{2}{3}\mu\overline\Theta^2-\nabla\mu\cdot\left((\overline{\boldsymbol U}\cdot\nabla)\overline{\boldsymbol U}\right)-\mu\overline{\boldsymbol U}\cdot\nabla\overline\Theta\right]\,\mathrm{d}V\\
&+\iiint_\Omega\gamma(\Phi_\nu+\Phi_R)\,\mathrm{d}V
+\left.\iint_S\frac{1}{2}\overline\rho\,\widetilde U_x\left|\widetilde{\boldsymbol U}-\boldsymbol U_\infty\right|^2\,\mathrm{d}A\right|_L^R\\
&+\iiint_\Omega\gamma(\overline{\boldsymbol U}-\widetilde{\boldsymbol U})\cdot\overline\rho(\widetilde{\boldsymbol U}\cdot\nabla)\widetilde{\boldsymbol U}\,\mathrm{d}V
\end{aligned}
\label{eq:RD_generalized_drag_app}
\end{equation}

The two pressure-related volume terms can be combined as
\begin{equation}
\frac{\partial(\overline p\,\overline v)}{\partial y}-\overline p\,\overline\Theta
=\overline v\frac{\partial\overline p}{\partial y}-\overline p\left(\frac{\partial\overline u}{\partial x}+\frac{\partial\overline w}{\partial z}\right)
\label{eq:RD_pressure_rearrangement}
\end{equation}
which separates the pressure contribution into the streamwise-development, wall-normal-displacement and spanwise-inhomogeneity terms. Grouping the remaining terms according to their physical origin recovers the generalized RD decomposition given in \eqref{eq:RD_generalized_final}.

\bibliographystyle{jfm}
\bibliography{jfm-instructions}

\end{document}